\documentclass[
 reprint,
 superscriptaddress,
 bibnotes,
 amsmath,
 amssymb,
 aps,
 prb,
 citeautoscript,
 floatfix,
]{revtex4-2}

\usepackage{graphicx}
\usepackage{dcolumn}
\usepackage{bm}
\usepackage[dvipsnames]{xcolor}
\usepackage[colorlinks=true, citecolor={blue!80!black}, urlcolor={blue!50!black}, linkcolor = {blue!80!black}]{hyperref}
\usepackage{lipsum}  
\usepackage{upgreek}
\usepackage{siunitx}
\usepackage{mathtools}
\usepackage{ulem}
\usepackage{braket}

\usepackage{nameref,zref-xr}
\zxrsetup{toltxlabel}

\newcommand{\Ej}{$E_{\rm J}$}

\newcommand{\flux}{$\phi_{\rm ext}$}

\makeatletter
\newcommand*{\addFileDependency}[1]{
  \typeout{(#1)}
  \@addtofilelist{#1}
  \IfFileExists{#1}{}{\typeout{No file #1.}}
}
\makeatother

\begin{document}
\preprint{APS/123-QED}

\title{Direct manipulation of a superconducting spin qubit strongly coupled to a transmon qubit}

\author{Marta Pita-Vidal}
\thanks{These two authors contributed equally.}
\affiliation{QuTech and Kavli Institute of Nanoscience, Delft University of Technology, 2600 GA Delft, The Netherlands}

\author{Arno Bargerbos}
\thanks{These two authors contributed equally.}
\affiliation{QuTech and Kavli Institute of Nanoscience, Delft University of Technology, 2600 GA Delft, The Netherlands}

\author{Rok Žitko}
\affiliation{Jožef Stefan Institute, Jamova 39, SI-1000 Ljubljana, Slovenia}
\affiliation{Faculty of Mathematics and Physics, University of Ljubljana, Jadranska 19, SI-1000 Ljubljana, Slovenia}

\author{Lukas J. Splitthoff}
\affiliation{QuTech and Kavli Institute of Nanoscience, Delft University of Technology, 2600 GA Delft, The Netherlands}

\author{Lukas Grünhaupt}
\affiliation{QuTech and Kavli Institute of Nanoscience, Delft University of Technology, 2600 GA Delft, The Netherlands}

\author{Jaap J. Wesdorp}
\affiliation{QuTech and Kavli Institute of Nanoscience, Delft University of Technology, 2600 GA Delft, The Netherlands}

\author{Yu Liu}
\affiliation{Center for Quantum Devices, Niels Bohr Institute, University of Copenhagen, 2100 Copenhagen, Denmark}

\author{Leo P. Kouwenhoven}
\affiliation{QuTech and Kavli Institute of Nanoscience, Delft University of Technology, 2600 GA Delft, The Netherlands}

\author{Ramón Aguado}
\affiliation{Instituto de Ciencia de Materiales de Madrid (ICMM),
Consejo Superior de Investigaciones Cientificas (CSIC), Sor Juana Ines de la Cruz 3, 28049 Madrid, Spain}

\author{Bernard van Heck}
\affiliation{Leiden Institute of Physics, Leiden University, Niels Bohrweg 2, 2333 CA Leiden, The Netherlands}

\author{Angela Kou}
\affiliation{Department of Physics and Frederick Seitz Materials Research Laboratory,
University of Illinois Urbana-Champaign, Urbana, IL 61801, USA}

\author{Christian Kraglund Andersen}
\affiliation{QuTech and Kavli Institute of Nanoscience, Delft University of Technology, 2600 GA Delft, The Netherlands}

\date{\today}

\begin{abstract}
Spin qubits in semiconductors are currently one of the most promising architectures for quantum computing. However, they face challenges in realizing multi-qubit interactions over extended distances. Superconducting spin qubits provide a promising alternative by encoding a qubit in the spin degree of freedom of an Andreev level. Such an Andreev spin qubit could leverage the advantages of circuit quantum electrodynamic, enabled by an intrinsic spin-supercurrent coupling.
The first realization of an Andreev spin qubit encoded the qubit in the excited states of a semiconducting weak-link, leading to frequent decay out of the computational subspace. Additionally, rapid qubit manipulation was hindered by the need for indirect Raman transitions.
Here, we exploit a different qubit subspace, using the spin-split doublet ground state of an electrostatically-defined quantum dot Josephson junction with large charging energy. Additionally, we use a magnetic field to enable direct spin manipulation over a frequency range of \SI{10}{GHz}. Using an all-electric microwave drive we achieve Rabi frequencies exceeding \SI{200}{MHz}. We furthermore embed the Andreev spin qubit in a superconducting transmon qubit, demonstrating strong coherent qubit-qubit coupling. 
These results are a crucial step towards a hybrid architecture that combines the beneficial aspects of both superconducting and semiconductor qubits.
\end{abstract}

\maketitle



Spin qubits in semiconductors \cite{diVincenzo1998, Hanson2007} and transmon qubits in superconducting circuits \cite{Koch2007} are currently two of the most promising platforms for quantum computing. Spin qubits are promising from a scalability standpoint due to their small footprint and compatibility with industrial semiconductor processing \cite{Zwerver2022}. However, despite encouraging progress in recent years \cite{Mi2018, Samkharadze2018, Landig2018, Borjans2020, HarveyCollard2022}, spin qubit architectures face challenges in realizing multi-qubit interactions over extended distances.
Transmon-based circuits currently boast some of the largest numbers of qubits on a single device~\cite{Arute2019, ibmquantum}, and are readily controlled, read out, and coupled over long distances due to the use of circuit quantum electrodynamics (QED) techniques \cite{Blais2004, Wallraff2004, Blais2021a}. However, transmon qubits have a small anharmonicity, limiting the speed of qubit operations, and they are relatively large (typically 0.01 to $\SI{1}{mm^2}$ per qubit), which leads to large chip sizes and makes transmons susceptible to cross-coupling with distant control elements.

A potential route to leverage the benefits of both superconducting qubits and spin qubits is to encode a qubit in the spin degree of freedom of a quasi-particle occupying an Andreev bound state in a Josephson junction \cite{Chtchelkatchev2003, Padurariu2010}. These states are confined by Andreev reflections at the superconducting interfaces and, thus, are localized in a small and well-defined region, similarly to conventional spin qubits. Furthermore, in the presence of spin-orbit interaction (SOI), the supercurrent across the Josephson junction becomes spin-dependent \cite{Chtchelkatchev2003, Beri2008}, which allows for interfacing with superconducting circuit elements. Such a superconducting spin qubit has recently been realized in the weak link of a superconductor-semiconductor hybrid nanowire \cite{Hays2021}, where it was named the Andreev spin qubit (ASQ). This first implementation showed that an ASQ can be efficiently read out using standard circuit QED techniques. However, qubit control was hindered by frequent leakage out of the computational subspace of the qubit, formed by higher energy Andreev levels of the junction. Additionally, due to the selection rules of the system in the absence of a magnetic field \cite{Park2017}, the ASQ required virtual driving of auxiliary states to induce qubit transitions.

In this work, we utilize previous insights from semiconducting spin-orbit qubits (SOQ) \cite{Nowack2007, NadjPerge2010} to construct an ASQ using a electrostatic gate-defined quantum dot within a Josephson junction~\cite{Bargerbos2022, Bargerbos2022b}.
Our implementation gives rise to qubit states of a different character than the non-interacting Andreev levels that form in the extended weak-link 
of Ref.~\cite{Hays2021}. The Coulomb interaction in the quantum dot leads to a charging energy, which can be exploited to deterministically prepare the quantum dot into a doublet phase~\cite{Bargerbos2022} with well-defined spin-split states.
As a consequence, the computational subspace of the qubit is now formed by the lowest energy states of the junction in the doublet phase, while the charging energy of the quantum dot provides additional protection against leakage resulting from parity switches of the junction \cite{Padurariu2010}. 
Furthermore, this design allows for fast and direct qubit control through spin-orbit mediated electric dipole spin resonance (EDSR) \cite{Golovach2006, Nowack2007, NadjPerge2010, vandenBerg2013}. 
We additionally demonstrate the magnetic field tunability of the qubit transition frequency over a frequency range of more than \SI{10}{GHz}. At elevated qubit frequencies, this results in a suppression of the population of the excited state, facilitating qubit manipulation and readout. Finally, the intrinsic coupling between the spin degree of freedom and the supercurrent facilitates strong coherent coupling between the ASQ and a transmon qubit.


\section{ANDREEV SPIN QUBIT}

\begin{figure}[t!]
    \centering
    \includegraphics[scale=1.0]{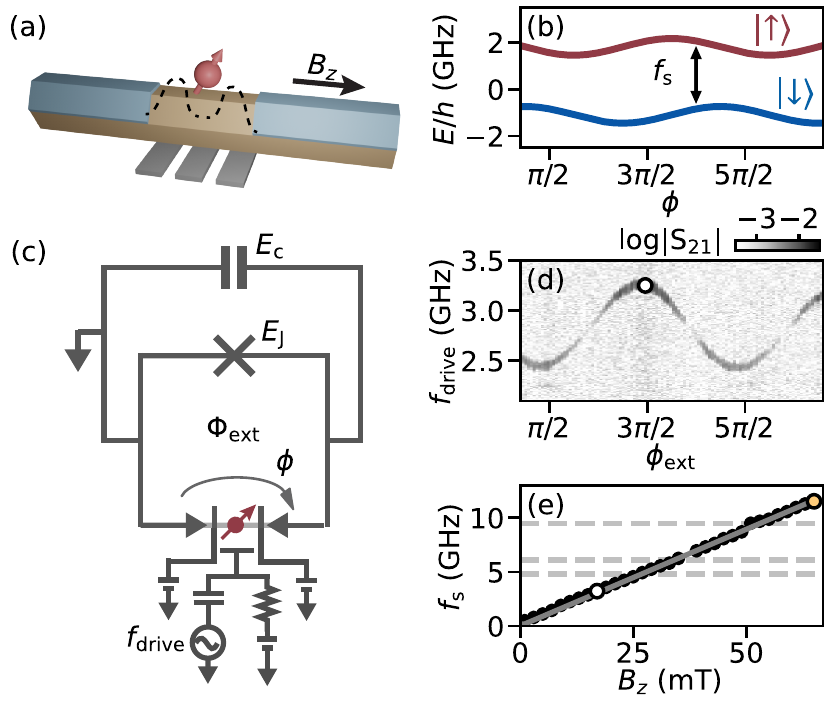}
    \caption{(a) Schematic depiction of an Andreev spin qubit in a hybrid superconductor-semiconductor nanowire. The qubit is formed in a gate-defined quantum dot with an odd number of electrons and is coupled to superconducting leads. (b) Eigenenergies of the qubit levels as a function of the phase difference $\phi$, as described by the effective model of Eq.~\ref{eq:H_ASQ}. The frequency of the qubit spin-flip transition $\ket{\downarrow} \leftrightarrow \ket{\uparrow}$ is denoted by $f_{\rm s}$. In this panel the component of the Zeeman energy parallel to the zero-field spin-polarization direction is $E_{\rm Z}^\parallel = \SI{2.9}{GHz}$. (c) Circuit model of the Andreev spin qubit embedded in a transmon circuit. The spin state is manipulated by a microwave drive, at frequency $f_{\rm drive}$, applied to the central gate electrode. The transmon island, with charging energy $E_{\rm c}$, is connected to ground by a SQUID formed by the parallel combination of the ASQ and a reference Josephson junction. Here, $\phi$ denotes the superconducting phase difference across the quantum dot junction, while $\Phi_\textrm{ext}$ is the externally applied magnetic flux through the SQUID loop. (d) Transmission through the readout circuit \cite{Supplement} as a function of the external flux and the applied drive frequency, measured at a magnetic field $B_{\rm {z}}=\SI{17}{mT}$ parallel to the nanowire. (e) Extracted qubit frequency $f_{\rm s}$ versus $B_z$ (markers), measured at $\phi_{\rm ext} = 3\pi/2$.  The data is fitted with a linear dependence (solid line), resulting in an effective Landé $g$-factor of $g^* = 12.7 \pm 0.2$. Horizontal dashed lines denote, from bottom to top, the first transmon frequency, the readout resonator frequency and the second transmon frequency. 
    }
    \label{fig:intro}
\end{figure}

We implement the ASQ in a quantum dot Josephson junction formed in a hybrid InAs/Al semiconducting-superconducting nanowire, see Fig.~\ref{fig:intro}(a). The quantum dot is electrostatically defined by three gate electrodes under an uncovered InAs section of the nanowire and tunnel-coupled to the superconducting segments \cite{Bargerbos2022}. In the presence of a magnetic field, the ASQ can be described by the effective Hamiltonian \cite{Bargerbos2022b, Padurariu2010}
\begin{equation} \label{eq:H_ASQ}
H_{\rm s} = E_0\cos\left(\phi\right) - E_{\rm SO}\, \vec{\sigma} \cdot \vec{n}\,\sin\left(\phi\right) +\frac{1}{2} \vec{E}_{\rm Z} \cdot\vec{\sigma} ,
\end{equation}
where $\phi$ is the phase difference across the junction, $\vec{\sigma}$ is the spin operator, $\vec{n}$ is a unit vector along the zero-field spin-polarization direction, set by the SOI, and $\vec{E}_{\rm Z}$ is a Zeeman field arising in the presence of an external magnetic field. 
$E_0$ denotes the effective Josephson energy of the quantum dot junction, common for both spin states. We note that the the term proportional to $E_0$ has a minimum at $\phi=\pi$, originating from the odd occupancy of the quantum dot junction~\cite{Bargerbos2022b}. In turn, $E_{\rm SO}$ denotes the spin-dependent contribution to the energy of the junction. The spin-dependent potential energy originates from the occurrence of electron co-tunneling accompanied by a spin flip, and it is finite only if SOI is present and multiple levels of the quantum dot are involved in the co-tunneling sequence \cite{Padurariu2010}. The difference between the energies of the $\ket{\downarrow}$ and $\ket{\uparrow}$ eigenstates of Eq.~\eqref{eq:H_ASQ} determines the ASQ qubit frequency $f_{\rm s}=E_\uparrow - E_\downarrow$, as depicted in Fig.~\ref{fig:intro}(b). 
For readout and control, we embed the ASQ into a superconducting transmon circuit, as illustrated in Fig.~\ref{fig:intro}(c). The transmon circuit consists of a capacitor, with charging energy $E_{\rm c}$, shunting a superconducting quantum interference device (SQUID) formed by the parallel combination of a gate-tunable Josephson junction with Josephson energy \Ej, and the quantum dot Josephson junction hosting the ASQ. We operate in the regime $E_{\rm J}/\sqrt{E_0^2+E_{\rm SO}^2} > 20$ so that the phase difference $\phi$ across the quantum dot Josephson junction can be controlled by the magnetic flux through the SQUID loop $\Phi_{\rm ext}=\phi_{\rm ext}\Phi_0/(2\pi)$, where $\Phi_0 = h/2e$ is the magnetic flux quantum. Due to the presence of the $E_{\rm SO}$ term, the transmon frequency $f_{\rm t}$ becomes spin-dependent and can be distinguished by probing a readout resonator capacitively coupled to the transmon using standard circuit QED techniques~\cite{Bargerbos2022b}. 
Finally, the spin-flipping qubit transition can be directly driven, while maintaining the transmon in its ground state, by applying a microwave tone on the central quantum dot gate~\cite{Metzger2021, Bargerbos2022b, Wesdorp2022}. Such microwave drive allows for all-electrical manipulation through EDSR \cite{Nowack2007, NadjPerge2010}. For further details about the device implementation and setup, see the Supplementary Information \cite{Supplement}.

Following the gate-tuning strategy described in Ref.~\cite{Bargerbos2022}, we prepare the quantum dot junction in a regime where it is occupied by an odd number of electrons, with $E_0/h = \SI{211}{MHz}$ and $E_{\rm SO}/h = \SI{305}{MHz}$ \cite{Supplement}. In this regime, the qubit states $\ket{\uparrow}$ and $\ket{\downarrow}$ are the lowest energy levels of the system, and the qubit subspace is separated from higher lying states by a frequency gap of at least \SI{20}{GHz} \cite{Supplement}. 
After fixing the gate voltages of the quantum dot, we investigate the tunability of the spin-flip transition $f_{\rm s}$ by applying a microwave tone at frequency $f_{\rm drive}$ and performing dispersive readout of the transmon qubit. As shown in Fig.~\ref{fig:intro}(d), we can finely control $f_{\rm s}$ by applying a magnetic flux through the SQUID loop, although the visibility of the measurement signal is reduced around $\phi_{\rm ext}=0,\pi$, where the spin-dependent transmon frequencies are degenerate \cite{Bargerbos2022b}.  
By applying an external magnetic field along the nanowire $B_z$ of up to \SI{65}{mT}, the qubit frequency can be varied from \SI{250}{MHz} to \SI{12}{GHz}, see Fig.~\ref{fig:intro}(e). The magnetic field direction is chosen to maximize the magnetic field compatibility of the Al shell of the nanowire and is generally not aligned with the spin-orbit direction $\vec{n}$ \cite{Han2022, Bargerbos2022b}.


\section{QUBIT COHERENCE}
\begin{figure}[t]
    \centering
        \includegraphics[scale=1.0]{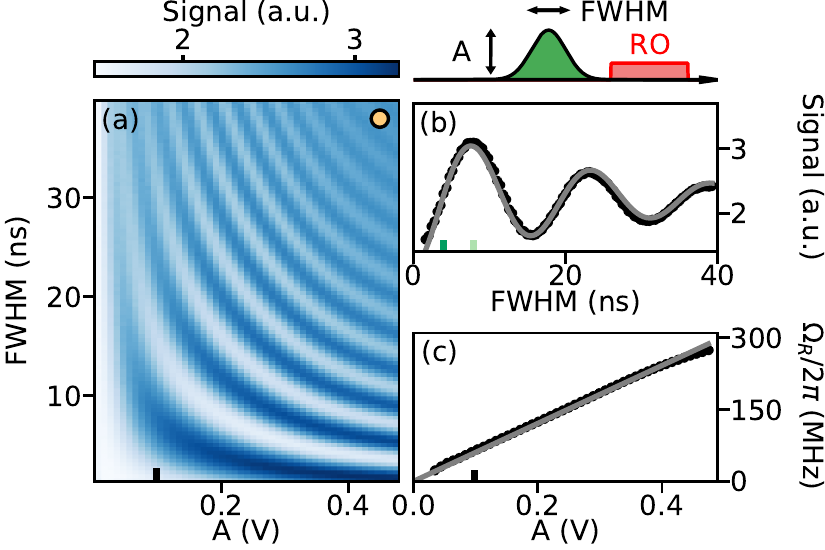}    \caption{Coherent manipulation of the Andreev spin qubit for $f_{\rm s} = \SI{11.5}{GHz}$ at $B_z = \SI{65}{mT}$. (a) Rabi oscillations for a range of Gaussian pulses characterized by their amplitude $\rm A$ at the waveform generator output and their full width at half maximum (FWHM), see pulse sequence. As also indicated in the pulse sequence, the Rabi pulse is immediately followed by a readout (RO) pulse (red, not to scale). (b) Rabi oscillation corresponding to $A=\SI{0.1}{V}$, fit with $a \cos{\left(t \Omega_R\right)} \exp{\left(\rm{t}/t_{\rm d}\right)}$ (solid line). The fit yields a decay time $t_{\rm d}=\SI{27}{ns}$. (c) Extracted Rabi frequencies versus pulse amplitude, fit with a linear equation (solid line).}  
    \label{fig:Rabi}
\end{figure}

To perform coherent manipulation of the spin states we fix $B_z$~=~\SI{65}{mT} and \flux~=~$3\pi/2$, setting $f_{\rm s} = \SI{11.5}{GHz}$, where the residual population of the excited state is suppressed to less than 5~\% \cite{Supplement}, facilitating qubit manipulation and readout. We observe Rabi oscillations between the qubit states $\ket{\uparrow}$ and $\ket{\downarrow}$ by applying a Gaussian microwave pulse with a carrier frequency at the spin-flip transition frequency $f_{\rm drive} = f_{\rm s}$, see Fig.~\ref{fig:Rabi}. Here, the Gaussian pulses are truncated so that the total pulse length is 2.5 times the Gaussian full width at half maximum (FWHM). As shown in Fig.~\ref{fig:Rabi}(a), we resolve up to 10 oscillations by varying the amplitude and duration of the pulse envelope.
The population transfer between the spin states, as measured by the dispersive readout scheme with a readout time of \SI{2}{\micro \second}, follows the expected time-dependence of a standard Rabi oscillation, as shown in Fig.~\ref{fig:Rabi}(b), from which we extract the Rabi frequency for each pulse amplitude. For a fixed Rabi frequency, we calibrate the FWHM needed for $\pi$ and $\pi/2$ pulses for single qubit manipulation. 

As expected for a two-level system, the Rabi frequency is linear over a wide range of pulse amplitudes and only starts to deviate from this linear dependence for strong drive amplitudes, see Fig.~\ref{fig:Rabi}(c). This deviation is due to saturation of the maximum power provided by the room-temperature electronics. We measure Rabi frequencies larger than \SI{200}{MHz}, exceeding the largest Rabi frequencies achieved in SOQ \cite{vandenBerg2013} and more than an order of magnitude faster than previous results for the ASQ \cite{Hays2021}. We observe that the Rabi frequency is approaching the anharmonicity of typical transmon qubits, with no indications of higher order levels being driven. The two-level nature of the ASQ thus intrinsically supports faster single qubit gates than standard transmon qubits \cite{Werninghaus2021}.

\begin{figure}[t]
    \centering
    \includegraphics[scale=1.0]{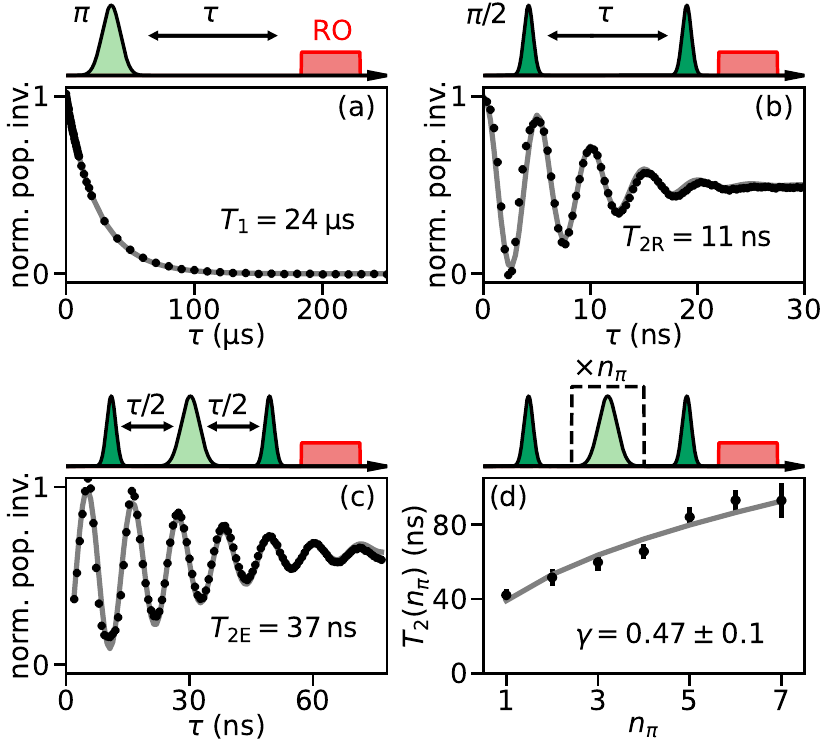} \caption{Coherence of the Andreev spin qubit at the same setpoint as Fig. \ref{fig:Rabi}. (a) Qubit lifetime, (b) Ramsey, (c) Hahn-echo and (d) CP experiments. Solid lines indicate fits to the data. For (b-d) oscillations are introduced into the decay by adding a phase proportional to the delay time for the final $\pi/2$-pulse. The data of (a-c) is obtained using a $\pi$-pulse ($\pi/2$-pulse) of $\rm{FWHM}=\SI{8}{ns}$ ($\SI{4}{ns}$), while for (d) this is \SI{4}{ns} (\SI{2}{ns}). For (a-c) we plot the normalized population inversion, where each sub-panel is individually normalized to the resulting fit. }
    \label{fig:coherence}
\end{figure}

Next, we characterize the lifetime of the ASQ by applying a $\pi$ pulse and reading out the qubit state after a delay time $\tau$. We obtain an exponential decay with a characteristic time $T_1$~=~24.4~$\pm$~\SI{0.5}{\micro s} at $B_z$~=~\SI{65}{mT}, see Fig.~\ref{fig:coherence}(a). As a function of magnetic field, $T_1$ varies between 10 and \SI{40}{\micro s} for qubit frequencies above the transmon frequency. We conjecture that the observed lifetime is limited by Purcell-like decay from coupling to the transmon, given the short transmon lifetime of around \SI{250}{ns}. For $B_z$ closer to zero, $T_1$ drops down to around \SI{1}{\micro s} \cite{Supplement}. This low lifetime is in contrast to the near-zero-field lifetimes found in previous ASQ experiments \cite{Hays2020, Hays2021}, which were in the range of $\SIrange{10}{50}{\micro s}$. The cause of this discrepancy is unknown, but a potential reason is an enhanced resonant exchange with the nuclear spins~\cite{Stockill2016} due to stronger strain in the InAs nanowire, which may differ for different nanowires depending on the exact growth conditions.

To characterize the coherence time of the qubit, we apply two $\pi/2$~pulses separated by a delay time, after which we read out the qubit state. From this experiment we extract a Ramsey coherence time of $T_{\rm 2R}$~=~11~$\pm$~\SI{1}{ns}, see Fig.~\ref{fig:coherence}(b), much smaller than $T_1$, and thus indicative of strong dephasing. Dephasing that originates from slow noise compared to the spin dynamics can be partially cancelled using a Hahn-echo sequence \cite{Hahn1950}, which introduces a $\pi$ pulse halfway between the two $\pi/2$ pulses. This echo sequence increases the measured coherence time by more than three times, to $T_{\rm 2E}$~=~37~$\pm$~\SI{4}{ns}, see Fig.~\ref{fig:coherence}(c).

The coherence time of the qubit can be further enhanced by using dynamical-decoupling pulse sequences, which serve to filter out faster environmental fluctuations. We apply Carr–Purcell (CP) sequences \cite{Carr1954, Barthel2010, Bylander2011}, interleaving a varying number of equidistant $\pi$ pulses, $n_\pi$, in between two $\pi/2$ pulses. As $n_\pi$ increases, higher frequency noise is cancelled out, extending the decoherence times. We reach $T_2$ times up to more than \SI{90}{ns} for $n_\pi=7$, at which stage we are most likely limited by decoherence during the $\pi$ pulses, see Fig.~\ref{fig:coherence}(d) \cite{Supplement}. We subsequently fit the $n_\pi$ dependence of $T_2$ with a power law $T_2(n_\pi) \propto n_{\pi}^\gamma$. Assuming a noise power spectral density of the form $f^{1/\beta}$, we expect the relation $\beta = \gamma/(1-\gamma)$ \cite{Cywinski2008, Bylander2011, Medford2012}. The observed scaling with $\gamma = 0.47 \pm 0.1$ therefore suggests that the decoherence is governed by noise with a $1/f$ spectral density in the frequency range 25 to 100~MHz. 

There are several potential sources of dephasing that are compatible with a $1/f$ noise spectral density, such as flux noise through the SQUID loop and charge noise \cite{Schreier2008, Braumueller2020}. We exclude the former, as we do not observe an increase of coherence times at the flux sweet spots \cite{Supplement}. Similarly, no consistent trend is observed when varying the gate voltages, nor when increasing the magnetic field strength. The latter indicates that charge noise is likely not the dominant contributor to dephasing, given that EDSR becomes more effective at coupling charge noise to the qubit at elevated fields. Additionally, based on the evolution of the Rabi decay time with increasing pulse amplitudes \cite{Malinowski2017}, the size of the charge fluctuations required to cause the observed amount of dephasing is estimated to be \SI{0.25}{mV}, significantly larger than what is expected to originate from the gate lines \cite{Supplement}.  However, the contribution of charge fluctuations originating elsewhere, such as in the dielectric material on the device, could still be contributing to the dephasing. Given that the sensitivity to fluctuations in environmental offset charge on the transmon island is suppressed by the large $E_{\rm J}/E_{\rm c}> 30 $ ratio, it is furthermore unlikely that the ASQ dephasing originates from offset-charge-dependent fluctuations of the transmon frequency qubit \cite{Koch2007}.

Another potential source of dephasing originates from the dynamics of the spinful nuclei in the nanowire, which may couple to the ASQ as a result of the hyperfine interaction. It has previously been shown that these dynamics can lead to longitudinal Overhauser field fluctuations with a $1/f$ spectral density \cite{Malinowski2017b}. Moreover, this effect is expected to be particularly strong in InAs due to the large nuclear spin of indium ($I = 9/2$) and should not be strongly affected by magnetic field in the $B_z$ range investigated here, which is not enough to polarize the nuclear spins. Corroborated by the fact that the extracted $T_{\rm 2R}$ and $T_{\rm 2E}$ times are strikingly similar to those found for the weak-link InAs ASQ \cite{Hays2021}, the InAs SOQ \cite{NadjPerge2010} and the InSb SOQ \cite{vandenBerg2013}, we conjecture that the nuclear environment provides a significant contribution to the decoherence of the ASQ.


\section{ASQ-TRANSMON COUPLING}
One of the main characteristics of the ASQ is the intrinsic coupling between the spin degree of freedom and the supercurrent across the quantum dot Josephson junction. We have so far only exploited this coupling for read-out of the qubit state using circuit QED techniques. Now, we demonstrate the observation of coherent coupling of the ASQ with the transmon qubit. 

\begin{figure}[h!]
    \centering
        \includegraphics[scale=1.0]{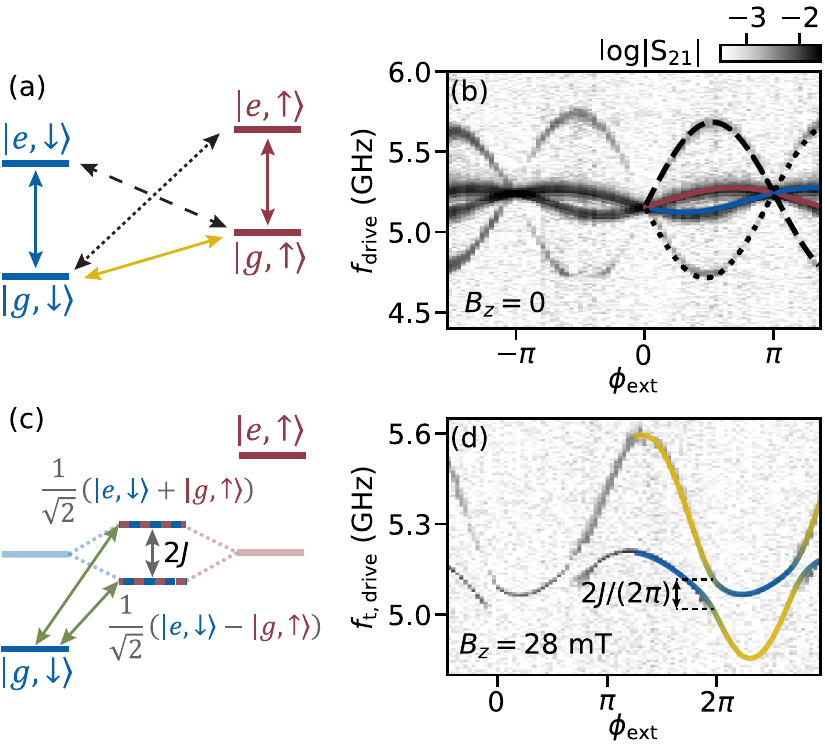}    \caption{Coherent ASQ-transmon coupling. (a) Frequency diagram of the joint ASQ-transmon circuit of Fig.~\ref{fig:intro}(c) at large detuning between ASQ and transmon qubit energy levels. In addition to the two spin-conserving transmon transitions (solid red and blue) and the transmon-conserving spin qubit transition (solid yellow), two additional transitions involving both qubits can take place in the presence of coherent coupling between them (dashed and dotted black). (b) Two-tone spectroscopy of the joint two-qubit system at $B_z = 0$. In addition to the two spin-dependent branches of the transmon qubit frequency \cite{Bargerbos2022b}, two additional transitions appear. Overlaid are transition frequencies obtained from the  model of Eq~\ref{eq:H_total}. (c) Frequency diagram of the joint ASQ-transmon circuit for $\ket{e,\downarrow} =\ket{g,\uparrow}$. In the presence of coherent coupling, the two qubits hybridize into states with a frequency splitting of $2J$. Green arrows denote the transitions from ground to the two hybridized states. (d) Two-tone spectroscopy versus  external flux at $B_z=\SI{28}{mT}$, where $f_{\rm s} \approx f_{\rm t}$. This results in avoided crossings between the two qubit frequencies. Overlaid are the transition frequencies obtained from the model of Eq.~\ref{eq:H_total}. Their colors denote the expectation value of the spin degree of freedom of the excited state and go from  $\ket{\downarrow}$  (blue) for the transmon transition to  $\ket{\uparrow}$ (yellow) for the spin-flip transition. $f_{\rm t, drive}$ denotes the frequency of the second tone, sent through the readout resonator.}
    \label{fig:coupling}
\end{figure}

A first signature of a coherent coupling is the presence of transitions that involve both qubits, in addition to the single-qubit transitions, see Fig.~\ref{fig:coupling}(a). At zero applied magnetic field, we spectroscopically detect two of such transitions at $f_{\rm t}+f_{\rm s}$ and $f_{\rm t}-f_{\rm s}$, where $f_{\rm t}$ is the transmon frequency, see Fig.~\ref{fig:coupling}(b). We classify them based on a fit with the joint Hamiltonian of the total ASQ-transmon circuit of Fig.~\ref{fig:intro}(c), given by
\begin{equation}\label{eq:H_total}
H_{\rm tot} = -4 E_{\rm c} \partial_\phi^2 - E_{\rm J} \cos{ ( \phi-\phi_{\rm ext} )} + H_{\rm s}(\phi).
\end{equation}
We identify the additional observed resonances as the double excitation $\ket{g \downarrow} \leftrightarrow \ket{e \uparrow}$ and the $\ket{g \uparrow} \leftrightarrow \ket{e \downarrow}$ SWAP transitions, where $\ket{g}$ and $\ket{e}$ denote the ground and first excited transmon states, respectively. These transitions could be used to construct entanglement and two qubit gates between the two different qubit platforms, provided the transitions can be driven at a faster rate than the decoherence rates of either qubit. 
 
Additionally, one of the hallmarks of strong coherent coupling is the appearance of an avoided level crossing when both qubit frequencies are made equal, $f_{\rm t} \approx f_{\rm s}$. In this case the $\ket{e,\downarrow}$ and $\ket{g,\uparrow}$ states are expected to hybridize into superposition states with a frequency splitting of $2J$, see Fig.~\ref{fig:coupling}(c). At $B_z$~=~\SI{28}{mT} this splitting can be readily observed in the experiment. By varying the external flux \flux~such that the ASQ frequency $f_{\rm s}$ crosses the transmon frequency $f_{\rm t}$, we find avoided crossings with a minimum frequency splitting $2J/(2\pi)= 2\times 52$~MHz, as shown in Fig.~\ref{fig:coupling}(d). As $J$ is four times larger than the decoherence rate of the ASQ, $1/T_{\rm 2R} \approx 14\times 2\pi\,$MHz and one order of magnitude larger than the decoherence rate of the transmon, $\approx1.2\times 2\pi\,$MHz, the coupling between the two qubits falls into the strong coupling regime. This result establishes the first realization of a direct strong coupling between a spin qubit and a superconducting qubit, in contrast to the results of Ref.~\cite{Landig2019}, where a high-impedance bus resonator was required to mediate the coupling between spin and transmon qubit through virtual photons.

Analytical estimates predict that the coupling  $J \propto E_{\rm SO} \phi_{\rm zpf} \sin(\theta)$, where $\phi_{\rm zpf}$ is the magnitude of zero-point fluctuation of the transmon phase, and $\theta$ is the angle between the Zeeman field and the spin-orbit direction $\vec{n}$ \cite{Supplement}. This expression suggests that by choosing a resonance with a larger $E_{\rm SO}$ \cite{Bargerbos2022b} and by aligning the magnetic field perpendicular to the spin-orbit direction, coupling rates of hundreds of MHz can be achieved, which would enable rapid two-qubit gates between the transmon and the ASQ and potentially allow for the study of light-matter interactions in the ultrastrong coupling regime \cite{FornDiaz2019, Scarlino2022}.


\section{TOWARDS NEW PLATFORMS AND MULTIPLE ASQ}
We have implemented an Andreev spin qubit, where the spin degree of freedom of a quasi-particle in a quantum dot with superconducting leads encodes the qubit state. The qubit subspace is stabilized by the charging energy of the quantum dot and direct microwave driving of the transitions is possible without the requirement of auxiliary levels. The qubit coherence was found to be comparable to previous results for qubits implemented in InAs or InSb nanowires \cite{Hays2021, NadjPerge2010, vandenBerg2013}. Our results therefore suggest that the nuclear environment contributes strongly to the ASQ decoherence, although the contribution of charge noise can not be fully neglected. This limitation motivates future investigation of alternative material platforms for ASQs, such as superconductor-proximitized nuclear-spin-free semiconductors \cite{Leon2021}, e.g. isotopically purified germanium \cite{Hendrickx2018, Scappucci2021, Tosato2022}.

We furthermore observed direct strong coherent coupling between the ASQ and a transmon qubit. Such strong coupling showcases the advantage of the intrinsic spin-supercurrent coupling, allowing the ASQ to be readily integrated into a circuit QED architecture. Our results open avenues towards multi-qubit devices: we propose to leverage the fact that transmon qubits can be readily coupled together using capacitive coupling, useful for mediating interactions between multiple ASQ. Furthermore, our results are a crucial step towards the coupling of distant Andreev spin qubits through bus resonators or a shared inductance \cite{Padurariu2010}, as well as short-distance coupling through wavefunction overlap \cite{Spethmann2022}. 


\begin{acknowledgments}
We acknowledge fruitful discussion with Menno Veldhorst, Maximillian Russ, Filip Malinowski, Valla Fatemi, and Yuli Nazarov. We further thank Peter Krogstrup for guidance in the material growth. This research was inspired by prior work by co-author J.J.W. where the spin-flip transition in an InAs/Al nanowire weak-link was directly observed in spectroscopy under the application of a magnetic field~\cite{Wesdorp2022}. This research is co-funded by the allowance for Top consortia for Knowledge and Innovation (TKI’s) from the Dutch Ministry of Economic Affairs, research project {\it Scalable circuits of Majorana qubits with topological protection} (i39, SCMQ) with project number 14SCMQ02, from the Dutch Research Council (NWO), and the Microsoft Quantum initiative. R. \v{Z}. acknowledges the support of the Slovenian Research agency (ARRS) under P1-0416 and J1-3008. R. A. acknowledges support from the Spanish Ministry of Science and Innovation through Grant PGC2018-097018-B-I00 and from the CSIC Research Platform on Quantum Technologies PTI-001. B.v.H. and C.K.A. acknowledge support from the Dutch Research Council (NWO).
\end{acknowledgments}

\section*{Data avalability}
The data and analysis code that support the findings of this study will be made available in 4TU.ResearchData before final publication.\\

\section*{Author contributions}

A.B., M.P.V., and A.K. conceived the experiment.
Y.L. developed and provided the nanowire materials.
A.B., M.P.V., L.S., L.G. and J.J.W prepared the experimental setup and data acquisition tools.
L.S. deposited the nanowires.
A.B. and M.P.V. designed and fabricated the device, performed the measurements and analysed the data, with continuous feedback from L.S., L.G., J.J.W, B.v.H, A.K. and C.K.A.
R.A., B.v.H. and R.Z. provided theory support during and after the measurements.
A.B., M.P.V. and B.v.H. wrote the code to compute the circuit energy levels and extract experimental parameters.
L.P.K., R.A., B.v.H., A.K. and  C.K.A. supervised the work.
A.B., M.P.V., and C.K.A. wrote the manuscript with feedback from all authors.

\bibliography{ms.bib}

\begin{thebibliography}{50}%
\makeatletter
\providecommand \@ifxundefined [1]{%
 \@ifx{#1\undefined}
}%
\providecommand \@ifnum [1]{%
 \ifnum #1\expandafter \@firstoftwo
 \else \expandafter \@secondoftwo
 \fi
}%
\providecommand \@ifx [1]{%
 \ifx #1\expandafter \@firstoftwo
 \else \expandafter \@secondoftwo
 \fi
}%
\providecommand \natexlab [1]{#1}%
\providecommand \enquote  [1]{``#1''}%
\providecommand \bibnamefont  [1]{#1}%
\providecommand \bibfnamefont [1]{#1}%
\providecommand \citenamefont [1]{#1}%
\providecommand \href@noop [0]{\@secondoftwo}%
\providecommand \href [0]{\begingroup \@sanitize@url \@href}%
\providecommand \@href[1]{\@@startlink{#1}\@@href}%
\providecommand \@@href[1]{\endgroup#1\@@endlink}%
\providecommand \@sanitize@url [0]{\catcode `\\12\catcode `\$12\catcode
  `\&12\catcode `\#12\catcode `\^12\catcode `\_12\catcode `\%12\relax}%
\providecommand \@@startlink[1]{}%
\providecommand \@@endlink[0]{}%
\providecommand \url  [0]{\begingroup\@sanitize@url \@url }%
\providecommand \@url [1]{\endgroup\@href {#1}{\urlprefix }}%
\providecommand \urlprefix  [0]{URL }%
\providecommand \Eprint [0]{\href }%
\providecommand \doibase [0]{https://doi.org/}%
\providecommand \selectlanguage [0]{\@gobble}%
\providecommand \bibinfo  [0]{\@secondoftwo}%
\providecommand \bibfield  [0]{\@secondoftwo}%
\providecommand \translation [1]{[#1]}%
\providecommand \BibitemOpen [0]{}%
\providecommand \bibitemStop [0]{}%
\providecommand \bibitemNoStop [0]{.\EOS\space}%
\providecommand \EOS [0]{\spacefactor3000\relax}%
\providecommand \BibitemShut  [1]{\csname bibitem#1\endcsname}%
\let\auto@bib@innerbib\@empty
\bibitem [{\citenamefont {Loss}\ and\ \citenamefont
  {DiVincenzo}(1998)}]{diVincenzo1998}%
  \BibitemOpen
  \bibfield  {author} {\bibinfo {author} {\bibfnamefont {D.}~\bibnamefont
  {Loss}}\ and\ \bibinfo {author} {\bibfnamefont {D.~P.}\ \bibnamefont
  {DiVincenzo}},\ }\bibfield  {title} {\bibinfo {title} {Quantum computation
  with quantum dots},\ }\bibfield  {journal} {\bibinfo  {journal} {Phys Rev A}\
  }\textbf {\bibinfo {volume} {57}},\ \href
  {https://doi.org/10.1103/PhysRevA.57.120} {10.1103/PhysRevA.57.120} (\bibinfo
  {year} {1998})\BibitemShut {NoStop}%
\bibitem [{\citenamefont {Hanson}\ \emph {et~al.}(2007)\citenamefont {Hanson},
  \citenamefont {Kouwenhoven}, \citenamefont {Petta}, \citenamefont {Tarucha},\
  and\ \citenamefont {Vandersypen}}]{Hanson2007}%
  \BibitemOpen
  \bibfield  {author} {\bibinfo {author} {\bibfnamefont {R.}~\bibnamefont
  {Hanson}}, \bibinfo {author} {\bibfnamefont {L.~P.}\ \bibnamefont
  {Kouwenhoven}}, \bibinfo {author} {\bibfnamefont {J.~R.}\ \bibnamefont
  {Petta}}, \bibinfo {author} {\bibfnamefont {S.}~\bibnamefont {Tarucha}},\
  and\ \bibinfo {author} {\bibfnamefont {L.~M.~K.}\ \bibnamefont
  {Vandersypen}},\ }\bibfield  {title} {\bibinfo {title} {Spins in few-electron
  quantum dots},\ }\bibfield  {journal} {\bibinfo  {journal} {Rev Mod Phys}\
  }\textbf {\bibinfo {volume} {79}},\ \href
  {https://doi.org/10.1103/RevModPhys.79.1217} {10.1103/RevModPhys.79.1217}
  (\bibinfo {year} {2007})\BibitemShut {NoStop}%
\bibitem [{\citenamefont {Koch}\ \emph {et~al.}(2007)\citenamefont {Koch},
  \citenamefont {Yu}, \citenamefont {Gambetta}, \citenamefont {Houck},
  \citenamefont {Schuster}, \citenamefont {Majer}, \citenamefont {Blais},
  \citenamefont {Devoret}, \citenamefont {Girvin},\ and\ \citenamefont
  {Schoelkopf}}]{Koch2007}%
  \BibitemOpen
  \bibfield  {author} {\bibinfo {author} {\bibfnamefont {J.}~\bibnamefont
  {Koch}}, \bibinfo {author} {\bibfnamefont {T.~M.}\ \bibnamefont {Yu}},
  \bibinfo {author} {\bibfnamefont {J.}~\bibnamefont {Gambetta}}, \bibinfo
  {author} {\bibfnamefont {A.~A.}\ \bibnamefont {Houck}}, \bibinfo {author}
  {\bibfnamefont {D.~I.}\ \bibnamefont {Schuster}}, \bibinfo {author}
  {\bibfnamefont {J.}~\bibnamefont {Majer}}, \bibinfo {author} {\bibfnamefont
  {A.}~\bibnamefont {Blais}}, \bibinfo {author} {\bibfnamefont {M.~H.}\
  \bibnamefont {Devoret}}, \bibinfo {author} {\bibfnamefont {S.~M.}\
  \bibnamefont {Girvin}},\ and\ \bibinfo {author} {\bibfnamefont {R.~J.}\
  \bibnamefont {Schoelkopf}},\ }\bibfield  {title} {\bibinfo {title}
  {Charge-insensitive qubit design derived from the cooper pair box},\
  }\bibfield  {journal} {\bibinfo  {journal} {Physical Review A}\ }\textbf
  {\bibinfo {volume} {76}},\ \href {https://doi.org/10.1103/physreva.76.042319}
  {10.1103/physreva.76.042319} (\bibinfo {year} {2007})\BibitemShut {NoStop}%
\bibitem [{\citenamefont {Zwerver}\ \emph {et~al.}(2022)\citenamefont
  {Zwerver}, \citenamefont {Kr{\"a}henmann}, \citenamefont {Watson},
  \citenamefont {Lampert}, \citenamefont {George}, \citenamefont
  {Pillarisetty}, \citenamefont {Bojarski}, \citenamefont {Amin}, \citenamefont
  {Amitonov}, \citenamefont {Boter}, \citenamefont {Caudillo}, \citenamefont
  {Correas-Serrano}, \citenamefont {Dehollain}, \citenamefont {Droulers},
  \citenamefont {Henry}, \citenamefont {Kotlyar}, \citenamefont {Lodari},
  \citenamefont {L{\"u}thi}, \citenamefont {Michalak}, \citenamefont {Mueller},
  \citenamefont {Neyens}, \citenamefont {Roberts}, \citenamefont {Samkharadze},
  \citenamefont {Zheng}, \citenamefont {Zietz}, \citenamefont {Scappucci},
  \citenamefont {Veldhorst}, \citenamefont {Vandersypen},\ and\ \citenamefont
  {Clarke}}]{Zwerver2022}%
  \BibitemOpen
  \bibfield  {author} {\bibinfo {author} {\bibfnamefont {A.~M.~J.}\
  \bibnamefont {Zwerver}}, \bibinfo {author} {\bibfnamefont {T.}~\bibnamefont
  {Kr{\"a}henmann}}, \bibinfo {author} {\bibfnamefont {T.~F.}\ \bibnamefont
  {Watson}}, \bibinfo {author} {\bibfnamefont {L.}~\bibnamefont {Lampert}},
  \bibinfo {author} {\bibfnamefont {H.~C.}\ \bibnamefont {George}}, \bibinfo
  {author} {\bibfnamefont {R.}~\bibnamefont {Pillarisetty}}, \bibinfo {author}
  {\bibfnamefont {S.~A.}\ \bibnamefont {Bojarski}}, \bibinfo {author}
  {\bibfnamefont {P.}~\bibnamefont {Amin}}, \bibinfo {author} {\bibfnamefont
  {S.~V.}\ \bibnamefont {Amitonov}}, \bibinfo {author} {\bibfnamefont {J.~M.}\
  \bibnamefont {Boter}}, \bibinfo {author} {\bibfnamefont {R.}~\bibnamefont
  {Caudillo}}, \bibinfo {author} {\bibfnamefont {D.}~\bibnamefont
  {Correas-Serrano}}, \bibinfo {author} {\bibfnamefont {J.~P.}\ \bibnamefont
  {Dehollain}}, \bibinfo {author} {\bibfnamefont {G.}~\bibnamefont {Droulers}},
  \bibinfo {author} {\bibfnamefont {E.~M.}\ \bibnamefont {Henry}}, \bibinfo
  {author} {\bibfnamefont {R.}~\bibnamefont {Kotlyar}}, \bibinfo {author}
  {\bibfnamefont {M.}~\bibnamefont {Lodari}}, \bibinfo {author} {\bibfnamefont
  {F.}~\bibnamefont {L{\"u}thi}}, \bibinfo {author} {\bibfnamefont {D.~J.}\
  \bibnamefont {Michalak}}, \bibinfo {author} {\bibfnamefont {B.~K.}\
  \bibnamefont {Mueller}}, \bibinfo {author} {\bibfnamefont {S.}~\bibnamefont
  {Neyens}}, \bibinfo {author} {\bibfnamefont {J.}~\bibnamefont {Roberts}},
  \bibinfo {author} {\bibfnamefont {N.}~\bibnamefont {Samkharadze}}, \bibinfo
  {author} {\bibfnamefont {G.}~\bibnamefont {Zheng}}, \bibinfo {author}
  {\bibfnamefont {O.~K.}\ \bibnamefont {Zietz}}, \bibinfo {author}
  {\bibfnamefont {G.}~\bibnamefont {Scappucci}}, \bibinfo {author}
  {\bibfnamefont {M.}~\bibnamefont {Veldhorst}}, \bibinfo {author}
  {\bibfnamefont {L.~M.~K.}\ \bibnamefont {Vandersypen}},\ and\ \bibinfo
  {author} {\bibfnamefont {J.~S.}\ \bibnamefont {Clarke}},\ }\bibfield  {title}
  {\bibinfo {title} {Qubits made by advanced semiconductor manufacturing},\
  }\bibfield  {journal} {\bibinfo  {journal} {Nature Electronics}\ }\textbf
  {\bibinfo {volume} {5}},\ \href {https://doi.org/10.1038/s41928-022-00727-9}
  {10.1038/s41928-022-00727-9} (\bibinfo {year} {2022})\BibitemShut {NoStop}%
\bibitem [{\citenamefont {Mi}\ \emph {et~al.}(2018)\citenamefont {Mi},
  \citenamefont {Benito}, \citenamefont {Putz}, \citenamefont {Zajac},
  \citenamefont {Taylor}, \citenamefont {Burkard},\ and\ \citenamefont
  {Petta}}]{Mi2018}%
  \BibitemOpen
  \bibfield  {author} {\bibinfo {author} {\bibfnamefont {X.}~\bibnamefont
  {Mi}}, \bibinfo {author} {\bibfnamefont {M.}~\bibnamefont {Benito}}, \bibinfo
  {author} {\bibfnamefont {S.}~\bibnamefont {Putz}}, \bibinfo {author}
  {\bibfnamefont {D.~M.}\ \bibnamefont {Zajac}}, \bibinfo {author}
  {\bibfnamefont {J.~M.}\ \bibnamefont {Taylor}}, \bibinfo {author}
  {\bibfnamefont {G.}~\bibnamefont {Burkard}},\ and\ \bibinfo {author}
  {\bibfnamefont {J.~R.}\ \bibnamefont {Petta}},\ }\bibfield  {title} {\bibinfo
  {title} {A coherent spin--photon interface in silicon},\ }\bibfield
  {journal} {\bibinfo  {journal} {Nature}\ }\textbf {\bibinfo {volume} {555}},\
  \href {https://doi.org/10.1038/nature25769} {10.1038/nature25769} (\bibinfo
  {year} {2018})\BibitemShut {NoStop}%
\bibitem [{\citenamefont {Samkharadze}\ \emph {et~al.}(2018)\citenamefont
  {Samkharadze}, \citenamefont {Zheng}, \citenamefont {Kalhor}, \citenamefont
  {Brousse}, \citenamefont {Sammak}, \citenamefont {Mendes}, \citenamefont
  {Blais}, \citenamefont {Scappucci},\ and\ \citenamefont
  {Vandersypen}}]{Samkharadze2018}%
  \BibitemOpen
  \bibfield  {author} {\bibinfo {author} {\bibfnamefont {N.}~\bibnamefont
  {Samkharadze}}, \bibinfo {author} {\bibfnamefont {G.}~\bibnamefont {Zheng}},
  \bibinfo {author} {\bibfnamefont {N.}~\bibnamefont {Kalhor}}, \bibinfo
  {author} {\bibfnamefont {D.}~\bibnamefont {Brousse}}, \bibinfo {author}
  {\bibfnamefont {A.}~\bibnamefont {Sammak}}, \bibinfo {author} {\bibfnamefont
  {U.~C.}\ \bibnamefont {Mendes}}, \bibinfo {author} {\bibfnamefont
  {A.}~\bibnamefont {Blais}}, \bibinfo {author} {\bibfnamefont
  {G.}~\bibnamefont {Scappucci}},\ and\ \bibinfo {author} {\bibfnamefont
  {L.~M.~K.}\ \bibnamefont {Vandersypen}},\ }\bibfield  {title} {\bibinfo
  {title} {Strong spin-photon coupling in silicon},\ }\bibfield  {journal}
  {\bibinfo  {journal} {Science}\ }\textbf {\bibinfo {volume} {359}},\ \href
  {https://doi.org/10.1126/science.aar4054} {10.1126/science.aar4054} (\bibinfo
  {year} {2018})\BibitemShut {NoStop}%
\bibitem [{\citenamefont {Landig}\ \emph {et~al.}(2018)\citenamefont {Landig},
  \citenamefont {Koski}, \citenamefont {Scarlino}, \citenamefont {Mendes},
  \citenamefont {Blais}, \citenamefont {Reichl}, \citenamefont {Wegscheider},
  \citenamefont {Wallraff}, \citenamefont {Ensslin},\ and\ \citenamefont
  {Ihn}}]{Landig2018}%
  \BibitemOpen
  \bibfield  {author} {\bibinfo {author} {\bibfnamefont {A.~J.}\ \bibnamefont
  {Landig}}, \bibinfo {author} {\bibfnamefont {J.~V.}\ \bibnamefont {Koski}},
  \bibinfo {author} {\bibfnamefont {P.}~\bibnamefont {Scarlino}}, \bibinfo
  {author} {\bibfnamefont {U.~C.}\ \bibnamefont {Mendes}}, \bibinfo {author}
  {\bibfnamefont {A.}~\bibnamefont {Blais}}, \bibinfo {author} {\bibfnamefont
  {C.}~\bibnamefont {Reichl}}, \bibinfo {author} {\bibfnamefont
  {W.}~\bibnamefont {Wegscheider}}, \bibinfo {author} {\bibfnamefont
  {A.}~\bibnamefont {Wallraff}}, \bibinfo {author} {\bibfnamefont
  {K.}~\bibnamefont {Ensslin}},\ and\ \bibinfo {author} {\bibfnamefont
  {T.}~\bibnamefont {Ihn}},\ }\bibfield  {title} {\bibinfo {title} {Coherent
  spin--photon coupling using a resonant exchange qubit},\ }\bibfield
  {journal} {\bibinfo  {journal} {Nature}\ }\textbf {\bibinfo {volume} {560}},\
  \href {https://doi.org/10.1038/s41586-018-0365-y} {10.1038/s41586-018-0365-y}
  (\bibinfo {year} {2018})\BibitemShut {NoStop}%
\bibitem [{\citenamefont {Borjans}\ \emph {et~al.}(2020)\citenamefont
  {Borjans}, \citenamefont {Croot}, \citenamefont {Mi}, \citenamefont
  {Gullans},\ and\ \citenamefont {Petta}}]{Borjans2020}%
  \BibitemOpen
  \bibfield  {author} {\bibinfo {author} {\bibfnamefont {F.}~\bibnamefont
  {Borjans}}, \bibinfo {author} {\bibfnamefont {X.~G.}\ \bibnamefont {Croot}},
  \bibinfo {author} {\bibfnamefont {X.}~\bibnamefont {Mi}}, \bibinfo {author}
  {\bibfnamefont {M.~J.}\ \bibnamefont {Gullans}},\ and\ \bibinfo {author}
  {\bibfnamefont {J.~R.}\ \bibnamefont {Petta}},\ }\bibfield  {title} {\bibinfo
  {title} {Resonant microwave-mediated interactions between distant electron
  spins},\ }\bibfield  {journal} {\bibinfo  {journal} {Nature}\ }\textbf
  {\bibinfo {volume} {577}},\ \href {https://doi.org/10.1038/s41586-019-1867-y}
  {10.1038/s41586-019-1867-y} (\bibinfo {year} {2020})\BibitemShut {NoStop}%
\bibitem [{\citenamefont {Harvey-Collard}\ \emph {et~al.}(2022)\citenamefont
  {Harvey-Collard}, \citenamefont {Dijkema}, \citenamefont {Zheng},
  \citenamefont {Sammak}, \citenamefont {Scappucci},\ and\ \citenamefont
  {Vandersypen}}]{HarveyCollard2022}%
  \BibitemOpen
  \bibfield  {author} {\bibinfo {author} {\bibfnamefont {P.}~\bibnamefont
  {Harvey-Collard}}, \bibinfo {author} {\bibfnamefont {J.}~\bibnamefont
  {Dijkema}}, \bibinfo {author} {\bibfnamefont {G.}~\bibnamefont {Zheng}},
  \bibinfo {author} {\bibfnamefont {A.}~\bibnamefont {Sammak}}, \bibinfo
  {author} {\bibfnamefont {G.}~\bibnamefont {Scappucci}},\ and\ \bibinfo
  {author} {\bibfnamefont {L.~M.}\ \bibnamefont {Vandersypen}},\ }\bibfield
  {title} {\bibinfo {title} {Coherent spin-spin coupling mediated by virtual
  microwave photons},\ }\bibfield  {journal} {\bibinfo  {journal} {Phys. Rev.
  X}\ }\textbf {\bibinfo {volume} {12}},\ \href
  {https://doi.org/10.1103/physrevx.12.021026} {10.1103/physrevx.12.021026}
  (\bibinfo {year} {2022})\BibitemShut {NoStop}%
\bibitem [{\citenamefont {Arute}\ \emph {et~al.}(2019)\citenamefont {Arute},
  \citenamefont {Arya}, \citenamefont {Babbush}, \citenamefont {Bacon},
  \citenamefont {Bardin}, \citenamefont {Barends}, \citenamefont {Biswas},
  \citenamefont {Boixo}, \citenamefont {Brandao}, \citenamefont {Buell},
  \citenamefont {Burkett}, \citenamefont {Chen}, \citenamefont {Chen},
  \citenamefont {Chiaro}, \citenamefont {Collins}, \citenamefont {Courtney},
  \citenamefont {Dunsworth}, \citenamefont {Farhi}, \citenamefont {Foxen},
  \citenamefont {Fowler}, \citenamefont {Gidney}, \citenamefont {Giustina},
  \citenamefont {Graff}, \citenamefont {Guerin}, \citenamefont {Habegger},
  \citenamefont {Harrigan}, \citenamefont {Hartmann}, \citenamefont {Ho},
  \citenamefont {Hoffmann}, \citenamefont {Huang}, \citenamefont {Humble},
  \citenamefont {Isakov}, \citenamefont {Jeffrey}, \citenamefont {Jiang},
  \citenamefont {Kafri}, \citenamefont {Kechedzhi}, \citenamefont {Kelly},
  \citenamefont {Klimov}, \citenamefont {Knysh}, \citenamefont {Korotkov},
  \citenamefont {Kostritsa}, \citenamefont {Landhuis}, \citenamefont
  {Lindmark}, \citenamefont {Lucero}, \citenamefont {Lyakh}, \citenamefont
  {Mandr{\`{a}}}, \citenamefont {McClean}, \citenamefont {McEwen},
  \citenamefont {Megrant}, \citenamefont {Mi}, \citenamefont {Michielsen},
  \citenamefont {Mohseni}, \citenamefont {Mutus}, \citenamefont {Naaman},
  \citenamefont {Neeley}, \citenamefont {Neill}, \citenamefont {Niu},
  \citenamefont {Ostby}, \citenamefont {Petukhov}, \citenamefont {Platt},
  \citenamefont {Quintana}, \citenamefont {Rieffel}, \citenamefont {Roushan},
  \citenamefont {Rubin}, \citenamefont {Sank}, \citenamefont {Satzinger},
  \citenamefont {Smelyanskiy}, \citenamefont {Sung}, \citenamefont
  {Trevithick}, \citenamefont {Vainsencher}, \citenamefont {Villalonga},
  \citenamefont {White}, \citenamefont {Yao}, \citenamefont {Yeh},
  \citenamefont {Zalcman}, \citenamefont {Neven},\ and\ \citenamefont
  {Martinis}}]{Arute2019}%
  \BibitemOpen
  \bibfield  {author} {\bibinfo {author} {\bibfnamefont {F.}~\bibnamefont
  {Arute}}, \bibinfo {author} {\bibfnamefont {K.}~\bibnamefont {Arya}},
  \bibinfo {author} {\bibfnamefont {R.}~\bibnamefont {Babbush}}, \bibinfo
  {author} {\bibfnamefont {D.}~\bibnamefont {Bacon}}, \bibinfo {author}
  {\bibfnamefont {J.~C.}\ \bibnamefont {Bardin}}, \bibinfo {author}
  {\bibfnamefont {R.}~\bibnamefont {Barends}}, \bibinfo {author} {\bibfnamefont
  {R.}~\bibnamefont {Biswas}}, \bibinfo {author} {\bibfnamefont
  {S.}~\bibnamefont {Boixo}}, \bibinfo {author} {\bibfnamefont {F.~G. S.~L.}\
  \bibnamefont {Brandao}}, \bibinfo {author} {\bibfnamefont {D.~A.}\
  \bibnamefont {Buell}}, \bibinfo {author} {\bibfnamefont {B.}~\bibnamefont
  {Burkett}}, \bibinfo {author} {\bibfnamefont {Y.}~\bibnamefont {Chen}},
  \bibinfo {author} {\bibfnamefont {Z.}~\bibnamefont {Chen}}, \bibinfo {author}
  {\bibfnamefont {B.}~\bibnamefont {Chiaro}}, \bibinfo {author} {\bibfnamefont
  {R.}~\bibnamefont {Collins}}, \bibinfo {author} {\bibfnamefont
  {W.}~\bibnamefont {Courtney}}, \bibinfo {author} {\bibfnamefont
  {A.}~\bibnamefont {Dunsworth}}, \bibinfo {author} {\bibfnamefont
  {E.}~\bibnamefont {Farhi}}, \bibinfo {author} {\bibfnamefont
  {B.}~\bibnamefont {Foxen}}, \bibinfo {author} {\bibfnamefont
  {A.}~\bibnamefont {Fowler}}, \bibinfo {author} {\bibfnamefont
  {C.}~\bibnamefont {Gidney}}, \bibinfo {author} {\bibfnamefont
  {M.}~\bibnamefont {Giustina}}, \bibinfo {author} {\bibfnamefont
  {R.}~\bibnamefont {Graff}}, \bibinfo {author} {\bibfnamefont
  {K.}~\bibnamefont {Guerin}}, \bibinfo {author} {\bibfnamefont
  {S.}~\bibnamefont {Habegger}}, \bibinfo {author} {\bibfnamefont {M.~P.}\
  \bibnamefont {Harrigan}}, \bibinfo {author} {\bibfnamefont {M.~J.}\
  \bibnamefont {Hartmann}}, \bibinfo {author} {\bibfnamefont {A.}~\bibnamefont
  {Ho}}, \bibinfo {author} {\bibfnamefont {M.}~\bibnamefont {Hoffmann}},
  \bibinfo {author} {\bibfnamefont {T.}~\bibnamefont {Huang}}, \bibinfo
  {author} {\bibfnamefont {T.~S.}\ \bibnamefont {Humble}}, \bibinfo {author}
  {\bibfnamefont {S.~V.}\ \bibnamefont {Isakov}}, \bibinfo {author}
  {\bibfnamefont {E.}~\bibnamefont {Jeffrey}}, \bibinfo {author} {\bibfnamefont
  {Z.}~\bibnamefont {Jiang}}, \bibinfo {author} {\bibfnamefont
  {D.}~\bibnamefont {Kafri}}, \bibinfo {author} {\bibfnamefont
  {K.}~\bibnamefont {Kechedzhi}}, \bibinfo {author} {\bibfnamefont
  {J.}~\bibnamefont {Kelly}}, \bibinfo {author} {\bibfnamefont {P.~V.}\
  \bibnamefont {Klimov}}, \bibinfo {author} {\bibfnamefont {S.}~\bibnamefont
  {Knysh}}, \bibinfo {author} {\bibfnamefont {A.}~\bibnamefont {Korotkov}},
  \bibinfo {author} {\bibfnamefont {F.}~\bibnamefont {Kostritsa}}, \bibinfo
  {author} {\bibfnamefont {D.}~\bibnamefont {Landhuis}}, \bibinfo {author}
  {\bibfnamefont {M.}~\bibnamefont {Lindmark}}, \bibinfo {author}
  {\bibfnamefont {E.}~\bibnamefont {Lucero}}, \bibinfo {author} {\bibfnamefont
  {D.}~\bibnamefont {Lyakh}}, \bibinfo {author} {\bibfnamefont
  {S.}~\bibnamefont {Mandr{\`{a}}}}, \bibinfo {author} {\bibfnamefont {J.~R.}\
  \bibnamefont {McClean}}, \bibinfo {author} {\bibfnamefont {M.}~\bibnamefont
  {McEwen}}, \bibinfo {author} {\bibfnamefont {A.}~\bibnamefont {Megrant}},
  \bibinfo {author} {\bibfnamefont {X.}~\bibnamefont {Mi}}, \bibinfo {author}
  {\bibfnamefont {K.}~\bibnamefont {Michielsen}}, \bibinfo {author}
  {\bibfnamefont {M.}~\bibnamefont {Mohseni}}, \bibinfo {author} {\bibfnamefont
  {J.}~\bibnamefont {Mutus}}, \bibinfo {author} {\bibfnamefont
  {O.}~\bibnamefont {Naaman}}, \bibinfo {author} {\bibfnamefont
  {M.}~\bibnamefont {Neeley}}, \bibinfo {author} {\bibfnamefont
  {C.}~\bibnamefont {Neill}}, \bibinfo {author} {\bibfnamefont {M.~Y.}\
  \bibnamefont {Niu}}, \bibinfo {author} {\bibfnamefont {E.}~\bibnamefont
  {Ostby}}, \bibinfo {author} {\bibfnamefont {A.}~\bibnamefont {Petukhov}},
  \bibinfo {author} {\bibfnamefont {J.~C.}\ \bibnamefont {Platt}}, \bibinfo
  {author} {\bibfnamefont {C.}~\bibnamefont {Quintana}}, \bibinfo {author}
  {\bibfnamefont {E.~G.}\ \bibnamefont {Rieffel}}, \bibinfo {author}
  {\bibfnamefont {P.}~\bibnamefont {Roushan}}, \bibinfo {author} {\bibfnamefont
  {N.~C.}\ \bibnamefont {Rubin}}, \bibinfo {author} {\bibfnamefont
  {D.}~\bibnamefont {Sank}}, \bibinfo {author} {\bibfnamefont {K.~J.}\
  \bibnamefont {Satzinger}}, \bibinfo {author} {\bibfnamefont {V.}~\bibnamefont
  {Smelyanskiy}}, \bibinfo {author} {\bibfnamefont {K.~J.}\ \bibnamefont
  {Sung}}, \bibinfo {author} {\bibfnamefont {M.~D.}\ \bibnamefont
  {Trevithick}}, \bibinfo {author} {\bibfnamefont {A.}~\bibnamefont
  {Vainsencher}}, \bibinfo {author} {\bibfnamefont {B.}~\bibnamefont
  {Villalonga}}, \bibinfo {author} {\bibfnamefont {T.}~\bibnamefont {White}},
  \bibinfo {author} {\bibfnamefont {Z.~J.}\ \bibnamefont {Yao}}, \bibinfo
  {author} {\bibfnamefont {P.}~\bibnamefont {Yeh}}, \bibinfo {author}
  {\bibfnamefont {A.}~\bibnamefont {Zalcman}}, \bibinfo {author} {\bibfnamefont
  {H.}~\bibnamefont {Neven}},\ and\ \bibinfo {author} {\bibfnamefont {J.~M.}\
  \bibnamefont {Martinis}},\ }\bibfield  {title} {\bibinfo {title} {Quantum
  supremacy using a programmable superconducting processor},\ }\bibfield
  {journal} {\bibinfo  {journal} {Nature}\ }\textbf {\bibinfo {volume} {574}},\
  \href {https://doi.org/10.1038/s41586-019-1666-5} {10.1038/s41586-019-1666-5}
  (\bibinfo {year} {2019})\BibitemShut {NoStop}%
\bibitem [{ibm()}]{ibmquantum}%
  \BibitemOpen
  \href@noop {} {\bibinfo {title} {{IBM} {Q}uantum {S}ervices}},\ \bibinfo
  {howpublished}
  {\url{https://quantum-computing.ibm.com/services?services=systems}},\
  \bibinfo {note} {{A}ccessed: 2022-06-13}\BibitemShut {NoStop}%
\bibitem [{\citenamefont {Blais}\ \emph {et~al.}(2004)\citenamefont {Blais},
  \citenamefont {Huang}, \citenamefont {Wallraff}, \citenamefont {Girvin},\
  and\ \citenamefont {Schoelkopf}}]{Blais2004}%
  \BibitemOpen
  \bibfield  {author} {\bibinfo {author} {\bibfnamefont {A.}~\bibnamefont
  {Blais}}, \bibinfo {author} {\bibfnamefont {R.-S.}\ \bibnamefont {Huang}},
  \bibinfo {author} {\bibfnamefont {A.}~\bibnamefont {Wallraff}}, \bibinfo
  {author} {\bibfnamefont {S.~M.}\ \bibnamefont {Girvin}},\ and\ \bibinfo
  {author} {\bibfnamefont {R.~J.}\ \bibnamefont {Schoelkopf}},\ }\bibfield
  {title} {\bibinfo {title} {Cavity quantum electrodynamics for superconducting
  electrical circuits: An architecture for quantum computation},\ }\bibfield
  {journal} {\bibinfo  {journal} {Phys Rev A}\ }\textbf {\bibinfo {volume}
  {69}},\ \href {https://doi.org/10.1103/PhysRevA.69.062320}
  {10.1103/PhysRevA.69.062320} (\bibinfo {year} {2004})\BibitemShut {NoStop}%
\bibitem [{\citenamefont {Wallraff}\ \emph {et~al.}(2004)\citenamefont
  {Wallraff}, \citenamefont {Schuster}, \citenamefont {Blais}, \citenamefont
  {Frunzio}, \citenamefont {Huang}, \citenamefont {Majer}, \citenamefont
  {Kumar}, \citenamefont {Girvin},\ and\ \citenamefont
  {Schoelkopf}}]{Wallraff2004}%
  \BibitemOpen
  \bibfield  {author} {\bibinfo {author} {\bibfnamefont {A.}~\bibnamefont
  {Wallraff}}, \bibinfo {author} {\bibfnamefont {D.~I.}\ \bibnamefont
  {Schuster}}, \bibinfo {author} {\bibfnamefont {A.}~\bibnamefont {Blais}},
  \bibinfo {author} {\bibfnamefont {L.}~\bibnamefont {Frunzio}}, \bibinfo
  {author} {\bibfnamefont {R.-.~S.}\ \bibnamefont {Huang}}, \bibinfo {author}
  {\bibfnamefont {J.}~\bibnamefont {Majer}}, \bibinfo {author} {\bibfnamefont
  {S.}~\bibnamefont {Kumar}}, \bibinfo {author} {\bibfnamefont {S.~M.}\
  \bibnamefont {Girvin}},\ and\ \bibinfo {author} {\bibfnamefont {R.~J.}\
  \bibnamefont {Schoelkopf}},\ }\bibfield  {title} {\bibinfo {title} {Strong
  coupling of a single photon to a superconducting qubit using circuit quantum
  electrodynamics},\ }\bibfield  {journal} {\bibinfo  {journal} {Nature}\
  }\textbf {\bibinfo {volume} {431}},\ \href
  {https://doi.org/10.1038/nature02851} {10.1038/nature02851} (\bibinfo {year}
  {2004})\BibitemShut {NoStop}%
\bibitem [{\citenamefont {Blais}\ \emph {et~al.}(2021)\citenamefont {Blais},
  \citenamefont {Grimsmo}, \citenamefont {Girvin},\ and\ \citenamefont
  {Wallraff}}]{Blais2021a}%
  \BibitemOpen
  \bibfield  {author} {\bibinfo {author} {\bibfnamefont {A.}~\bibnamefont
  {Blais}}, \bibinfo {author} {\bibfnamefont {A.~L.}\ \bibnamefont {Grimsmo}},
  \bibinfo {author} {\bibfnamefont {S.}~\bibnamefont {Girvin}},\ and\ \bibinfo
  {author} {\bibfnamefont {A.}~\bibnamefont {Wallraff}},\ }\bibfield  {title}
  {\bibinfo {title} {Circuit quantum electrodynamics},\ }\bibfield  {journal}
  {\bibinfo  {journal} {Rev Mod Phys}\ }\textbf {\bibinfo {volume} {93}},\
  \href {https://doi.org/10.1103/revmodphys.93.025005}
  {10.1103/revmodphys.93.025005} (\bibinfo {year} {2021})\BibitemShut {NoStop}%
\bibitem [{\citenamefont {Chtchelkatchev}\ and\ \citenamefont
  {Nazarov}(2003)}]{Chtchelkatchev2003}%
  \BibitemOpen
  \bibfield  {author} {\bibinfo {author} {\bibfnamefont {N.~M.}\ \bibnamefont
  {Chtchelkatchev}}\ and\ \bibinfo {author} {\bibfnamefont {Y.~V.}\
  \bibnamefont {Nazarov}},\ }\bibfield  {title} {\bibinfo {title} {{A}ndreev
  quantum dots for spin manipulation},\ }\bibfield  {journal} {\bibinfo
  {journal} {Phys Rev Lett}\ }\textbf {\bibinfo {volume} {90}},\ \href
  {https://doi.org/10.1103/PhysRevLett.90.226806}
  {10.1103/PhysRevLett.90.226806} (\bibinfo {year} {2003})\BibitemShut
  {NoStop}%
\bibitem [{\citenamefont {Padurariu}\ and\ \citenamefont
  {Nazarov}(2010)}]{Padurariu2010}%
  \BibitemOpen
  \bibfield  {author} {\bibinfo {author} {\bibfnamefont {C.}~\bibnamefont
  {Padurariu}}\ and\ \bibinfo {author} {\bibfnamefont {Y.~V.}\ \bibnamefont
  {Nazarov}},\ }\bibfield  {title} {\bibinfo {title} {Theoretical proposal for
  superconducting spin qubits},\ }\bibfield  {journal} {\bibinfo  {journal}
  {Phys Rev B}\ }\textbf {\bibinfo {volume} {81}},\ \href
  {https://doi.org/10.1103/PhysRevB.81.144519} {10.1103/PhysRevB.81.144519}
  (\bibinfo {year} {2010})\BibitemShut {NoStop}%
\bibitem [{\citenamefont {B{\'{e}}ri}\ \emph {et~al.}(2008)\citenamefont
  {B{\'{e}}ri}, \citenamefont {Bardarson},\ and\ \citenamefont
  {Beenakker}}]{Beri2008}%
  \BibitemOpen
  \bibfield  {author} {\bibinfo {author} {\bibfnamefont {B.}~\bibnamefont
  {B{\'{e}}ri}}, \bibinfo {author} {\bibfnamefont {J.~H.}\ \bibnamefont
  {Bardarson}},\ and\ \bibinfo {author} {\bibfnamefont {C.~W.~J.}\ \bibnamefont
  {Beenakker}},\ }\bibfield  {title} {\bibinfo {title} {Splitting of {Andreev}
  levels in a {Josephson} junction by spin-orbit coupling},\ }\bibfield
  {journal} {\bibinfo  {journal} {Phys Rev B}\ }\textbf {\bibinfo {volume}
  {77}},\ \href {https://doi.org/10.1103/physrevb.77.045311}
  {10.1103/physrevb.77.045311} (\bibinfo {year} {2008})\BibitemShut {NoStop}%
\bibitem [{\citenamefont {Hays}\ \emph {et~al.}(2021)\citenamefont {Hays},
  \citenamefont {Fatemi}, \citenamefont {Bouman}, \citenamefont {Cerrillo},
  \citenamefont {Diamond}, \citenamefont {Serniak}, \citenamefont {Connolly},
  \citenamefont {Krogstrup}, \citenamefont {Nyg{\aa}rd}, \citenamefont
  {Levy~Yeyati}, \citenamefont {Geresdi},\ and\ \citenamefont
  {Devoret}}]{Hays2021}%
  \BibitemOpen
  \bibfield  {author} {\bibinfo {author} {\bibfnamefont {M.}~\bibnamefont
  {Hays}}, \bibinfo {author} {\bibfnamefont {V.}~\bibnamefont {Fatemi}},
  \bibinfo {author} {\bibfnamefont {D.}~\bibnamefont {Bouman}}, \bibinfo
  {author} {\bibfnamefont {J.}~\bibnamefont {Cerrillo}}, \bibinfo {author}
  {\bibfnamefont {S.}~\bibnamefont {Diamond}}, \bibinfo {author} {\bibfnamefont
  {K.}~\bibnamefont {Serniak}}, \bibinfo {author} {\bibfnamefont
  {T.}~\bibnamefont {Connolly}}, \bibinfo {author} {\bibfnamefont
  {P.}~\bibnamefont {Krogstrup}}, \bibinfo {author} {\bibfnamefont
  {J.}~\bibnamefont {Nyg{\aa}rd}}, \bibinfo {author} {\bibfnamefont
  {A.}~\bibnamefont {Levy~Yeyati}}, \bibinfo {author} {\bibfnamefont
  {A.}~\bibnamefont {Geresdi}},\ and\ \bibinfo {author} {\bibfnamefont {M.~H.}\
  \bibnamefont {Devoret}},\ }\bibfield  {title} {\bibinfo {title} {Coherent
  manipulation of an {A}ndreev spin qubit},\ }\bibfield  {journal} {\bibinfo
  {journal} {Science}\ }\textbf {\bibinfo {volume} {373}},\ \href
  {https://doi.org/10.1126/science.abf0345} {10.1126/science.abf0345} (\bibinfo
  {year} {2021})\BibitemShut {NoStop}%
\bibitem [{\citenamefont {Park}\ and\ \citenamefont {Yeyati}(2017)}]{Park2017}%
  \BibitemOpen
  \bibfield  {author} {\bibinfo {author} {\bibfnamefont {S.}~\bibnamefont
  {Park}}\ and\ \bibinfo {author} {\bibfnamefont {A.~L.}\ \bibnamefont
  {Yeyati}},\ }\bibfield  {title} {\bibinfo {title} {Andreev spin qubits in
  multichannel {Rashba} nanowires},\ }\bibfield  {journal} {\bibinfo  {journal}
  {Phys Rev B}\ }\textbf {\bibinfo {volume} {96}},\ \href
  {https://doi.org/10.1103/physrevb.96.125416} {10.1103/physrevb.96.125416}
  (\bibinfo {year} {2017})\BibitemShut {NoStop}%
\bibitem [{\citenamefont {Nowack}\ \emph {et~al.}(2007)\citenamefont {Nowack},
  \citenamefont {Koppens}, \citenamefont {Nazarov},\ and\ \citenamefont
  {Vandersypen}}]{Nowack2007}%
  \BibitemOpen
  \bibfield  {author} {\bibinfo {author} {\bibfnamefont {K.~C.}\ \bibnamefont
  {Nowack}}, \bibinfo {author} {\bibfnamefont {F.~H.~L.}\ \bibnamefont
  {Koppens}}, \bibinfo {author} {\bibfnamefont {Y.~V.}\ \bibnamefont
  {Nazarov}},\ and\ \bibinfo {author} {\bibfnamefont {L.~M.~K.}\ \bibnamefont
  {Vandersypen}},\ }\bibfield  {title} {\bibinfo {title} {Coherent control of a
  single electron spin with electric fields},\ }\bibfield  {journal} {\bibinfo
  {journal} {Science}\ }\textbf {\bibinfo {volume} {318}},\ \href
  {https://doi.org/10.1126/science.1148092} {10.1126/science.1148092} (\bibinfo
  {year} {2007})\BibitemShut {NoStop}%
\bibitem [{\citenamefont {Nadj-Perge}\ \emph {et~al.}(2010)\citenamefont
  {Nadj-Perge}, \citenamefont {Frolov}, \citenamefont {Bakkers},\ and\
  \citenamefont {Kouwenhoven}}]{NadjPerge2010}%
  \BibitemOpen
  \bibfield  {author} {\bibinfo {author} {\bibfnamefont {S.}~\bibnamefont
  {Nadj-Perge}}, \bibinfo {author} {\bibfnamefont {S.~M.}\ \bibnamefont
  {Frolov}}, \bibinfo {author} {\bibfnamefont {E.~P. A.~M.}\ \bibnamefont
  {Bakkers}},\ and\ \bibinfo {author} {\bibfnamefont {L.~P.}\ \bibnamefont
  {Kouwenhoven}},\ }\bibfield  {title} {\bibinfo {title}
  {Spin{\textendash}orbit qubit in a semiconductor nanowire},\ }\bibfield
  {journal} {\bibinfo  {journal} {Nature}\ }\textbf {\bibinfo {volume} {468}},\
  \href {https://doi.org/10.1038/nature09682} {10.1038/nature09682} (\bibinfo
  {year} {2010})\BibitemShut {NoStop}%
\bibitem [{\citenamefont {Bargerbos}\ \emph {et~al.}(2022)\citenamefont
  {Bargerbos}, \citenamefont {Pita-Vidal}, \citenamefont
  {\ifmmode~\check{Z}\else \v{Z}\fi{}itko}, \citenamefont {\'Avila},
  \citenamefont {Splitthoff}, \citenamefont {Gr\"unhaupt}, \citenamefont
  {Wesdorp}, \citenamefont {Andersen}, \citenamefont {Liu}, \citenamefont
  {Kouwenhoven}, \citenamefont {Aguado}, \citenamefont {Kou},\ and\
  \citenamefont {van Heck}}]{Bargerbos2022}%
  \BibitemOpen
  \bibfield  {author} {\bibinfo {author} {\bibfnamefont {A.}~\bibnamefont
  {Bargerbos}}, \bibinfo {author} {\bibfnamefont {M.}~\bibnamefont
  {Pita-Vidal}}, \bibinfo {author} {\bibfnamefont {R.}~\bibnamefont
  {\ifmmode~\check{Z}\else \v{Z}\fi{}itko}}, \bibinfo {author} {\bibfnamefont
  {J.}~\bibnamefont {\'Avila}}, \bibinfo {author} {\bibfnamefont {L.~J.}\
  \bibnamefont {Splitthoff}}, \bibinfo {author} {\bibfnamefont
  {L.}~\bibnamefont {Gr\"unhaupt}}, \bibinfo {author} {\bibfnamefont {J.~J.}\
  \bibnamefont {Wesdorp}}, \bibinfo {author} {\bibfnamefont {C.~K.}\
  \bibnamefont {Andersen}}, \bibinfo {author} {\bibfnamefont {Y.}~\bibnamefont
  {Liu}}, \bibinfo {author} {\bibfnamefont {L.~P.}\ \bibnamefont
  {Kouwenhoven}}, \bibinfo {author} {\bibfnamefont {R.}~\bibnamefont {Aguado}},
  \bibinfo {author} {\bibfnamefont {A.}~\bibnamefont {Kou}},\ and\ \bibinfo
  {author} {\bibfnamefont {B.}~\bibnamefont {van Heck}},\ }\bibfield  {title}
  {\bibinfo {title} {Singlet-doublet transitions of a quantum dot josephson
  junction detected in a transmon circuit},\ }\href
  {https://doi.org/10.1103/PRXQuantum.3.030311} {\bibfield  {journal} {\bibinfo
   {journal} {PRX Quantum}\ }\textbf {\bibinfo {volume} {3}},\ \bibinfo {pages}
  {030311} (\bibinfo {year} {2022})}\BibitemShut {NoStop}%
\bibitem [{\citenamefont {{Bargerbos}}\ \emph {et~al.}(2022)\citenamefont
  {{Bargerbos}}, \citenamefont {{Pita-Vidal}}, \citenamefont {{{\v{Z}}itko}},
  \citenamefont {{Splitthoff}}, \citenamefont {{Gr{\"u}nhaupt}}, \citenamefont
  {{Wesdorp}}, \citenamefont {{Liu}}, \citenamefont {{Kouwenhoven}},
  \citenamefont {{Aguado}}, \citenamefont {{Kraglund Andersen}}, \citenamefont
  {{Kou}},\ and\ \citenamefont {{van Heck}}}]{Bargerbos2022b}%
  \BibitemOpen
  \bibfield  {author} {\bibinfo {author} {\bibfnamefont {A.}~\bibnamefont
  {{Bargerbos}}}, \bibinfo {author} {\bibfnamefont {M.}~\bibnamefont
  {{Pita-Vidal}}}, \bibinfo {author} {\bibfnamefont {R.}~\bibnamefont
  {{{\v{Z}}itko}}}, \bibinfo {author} {\bibfnamefont {L.~J.}\ \bibnamefont
  {{Splitthoff}}}, \bibinfo {author} {\bibfnamefont {L.}~\bibnamefont
  {{Gr{\"u}nhaupt}}}, \bibinfo {author} {\bibfnamefont {J.~J.}\ \bibnamefont
  {{Wesdorp}}}, \bibinfo {author} {\bibfnamefont {Y.}~\bibnamefont {{Liu}}},
  \bibinfo {author} {\bibfnamefont {L.~P.}\ \bibnamefont {{Kouwenhoven}}},
  \bibinfo {author} {\bibfnamefont {R.}~\bibnamefont {{Aguado}}}, \bibinfo
  {author} {\bibfnamefont {C.}~\bibnamefont {{Kraglund Andersen}}}, \bibinfo
  {author} {\bibfnamefont {A.}~\bibnamefont {{Kou}}},\ and\ \bibinfo {author}
  {\bibfnamefont {B.}~\bibnamefont {{van Heck}}},\ }\bibfield  {title}
  {\bibinfo {title} {{Spectroscopy of spin-split Andreev levels in a quantum
  dot with superconducting leads}},\ }\href@noop {} {\bibfield  {journal}
  {\bibinfo  {journal} {arXiv e-prints}\ ,\ \bibinfo {eid} {arXiv:2208.09314}}
  (\bibinfo {year} {2022})},\ \Eprint {https://arxiv.org/abs/2208.09314}
  {arXiv:2208.09314} \BibitemShut {NoStop}%
\bibitem [{\citenamefont {Golovach}\ \emph {et~al.}(2006)\citenamefont
  {Golovach}, \citenamefont {Borhani},\ and\ \citenamefont
  {Loss}}]{Golovach2006}%
  \BibitemOpen
  \bibfield  {author} {\bibinfo {author} {\bibfnamefont {V.~N.}\ \bibnamefont
  {Golovach}}, \bibinfo {author} {\bibfnamefont {M.}~\bibnamefont {Borhani}},\
  and\ \bibinfo {author} {\bibfnamefont {D.}~\bibnamefont {Loss}},\ }\bibfield
  {title} {\bibinfo {title} {Electric-dipole-induced spin resonance in quantum
  dots},\ }\bibfield  {journal} {\bibinfo  {journal} {Phys Rev B}\ }\textbf
  {\bibinfo {volume} {74}},\ \href {https://doi.org/10.1103/physrevb.74.165319}
  {10.1103/physrevb.74.165319} (\bibinfo {year} {2006})\BibitemShut {NoStop}%
\bibitem [{\citenamefont {van~den Berg}\ \emph {et~al.}(2013)\citenamefont
  {van~den Berg}, \citenamefont {Nadj-Perge}, \citenamefont {Pribiag},
  \citenamefont {Plissard}, \citenamefont {Bakkers}, \citenamefont {Frolov},\
  and\ \citenamefont {Kouwenhoven}}]{vandenBerg2013}%
  \BibitemOpen
  \bibfield  {author} {\bibinfo {author} {\bibfnamefont {J.~W.~G.}\
  \bibnamefont {van~den Berg}}, \bibinfo {author} {\bibfnamefont
  {S.}~\bibnamefont {Nadj-Perge}}, \bibinfo {author} {\bibfnamefont {V.~S.}\
  \bibnamefont {Pribiag}}, \bibinfo {author} {\bibfnamefont {S.~R.}\
  \bibnamefont {Plissard}}, \bibinfo {author} {\bibfnamefont {E.~P. A.~M.}\
  \bibnamefont {Bakkers}}, \bibinfo {author} {\bibfnamefont {S.~M.}\
  \bibnamefont {Frolov}},\ and\ \bibinfo {author} {\bibfnamefont {L.~P.}\
  \bibnamefont {Kouwenhoven}},\ }\bibfield  {title} {\bibinfo {title} {Fast
  spin-orbit qubit in an indium antimonide nanowire},\ }\bibfield  {journal}
  {\bibinfo  {journal} {Phys Rev Lett}\ }\textbf {\bibinfo {volume} {110}},\
  \href {https://doi.org/10.1103/PhysRevLett.110.066806}
  {10.1103/PhysRevLett.110.066806} (\bibinfo {year} {2013})\BibitemShut
  {NoStop}%
\bibitem [{\citenamefont {Metzger}\ \emph {et~al.}(2021)\citenamefont
  {Metzger}, \citenamefont {Park}, \citenamefont {Tosi}, \citenamefont
  {Janvier}, \citenamefont {Reynoso}, \citenamefont {Goffman}, \citenamefont
  {Urbina}, \citenamefont {Yeyati},\ and\ \citenamefont
  {Pothier}}]{Metzger2021}%
  \BibitemOpen
  \bibfield  {author} {\bibinfo {author} {\bibfnamefont {C.}~\bibnamefont
  {Metzger}}, \bibinfo {author} {\bibfnamefont {S.}~\bibnamefont {Park}},
  \bibinfo {author} {\bibfnamefont {L.}~\bibnamefont {Tosi}}, \bibinfo {author}
  {\bibfnamefont {C.}~\bibnamefont {Janvier}}, \bibinfo {author} {\bibfnamefont
  {A.~A.}\ \bibnamefont {Reynoso}}, \bibinfo {author} {\bibfnamefont {M.~F.}\
  \bibnamefont {Goffman}}, \bibinfo {author} {\bibfnamefont {C.}~\bibnamefont
  {Urbina}}, \bibinfo {author} {\bibfnamefont {A.~L.}\ \bibnamefont {Yeyati}},\
  and\ \bibinfo {author} {\bibfnamefont {H.}~\bibnamefont {Pothier}},\
  }\bibfield  {title} {\bibinfo {title} {Circuit-{QED} with phase-biased
  {Josephson} weak links},\ }\bibfield  {journal} {\bibinfo  {journal} {Phys.
  Rev. Research}\ }\textbf {\bibinfo {volume} {3}},\ \href
  {https://doi.org/10.1103/physrevresearch.3.013036}
  {10.1103/physrevresearch.3.013036} (\bibinfo {year} {2021})\BibitemShut
  {NoStop}%
\bibitem [{\citenamefont {{Wesdorp}}\ \emph {et~al.}(2022)\citenamefont
  {{Wesdorp}}, \citenamefont {{Matute-Ca{\v{n}}adas}}, \citenamefont
  {{Vaartjes}}, \citenamefont {{Gr{\"u}nhaupt}}, \citenamefont {{Laeven}},
  \citenamefont {{Roelofs}}, \citenamefont {{Splitthoff}}, \citenamefont
  {{Pita-Vidal}}, \citenamefont {{Bargerbos}}, \citenamefont {{van Woerkom}},
  \citenamefont {{Krogstrup}}, \citenamefont {{Kouwenhoven}}, \citenamefont
  {{Andersen}}, \citenamefont {{Levy Yeyati}}, \citenamefont {{van Heck}},\
  and\ \citenamefont {{de Lange}}}]{Wesdorp2022}%
  \BibitemOpen
  \bibfield  {author} {\bibinfo {author} {\bibfnamefont {J.~J.}\ \bibnamefont
  {{Wesdorp}}}, \bibinfo {author} {\bibfnamefont {F.~J.}\ \bibnamefont
  {{Matute-Ca{\v{n}}adas}}}, \bibinfo {author} {\bibfnamefont {A.}~\bibnamefont
  {{Vaartjes}}}, \bibinfo {author} {\bibfnamefont {L.}~\bibnamefont
  {{Gr{\"u}nhaupt}}}, \bibinfo {author} {\bibfnamefont {T.}~\bibnamefont
  {{Laeven}}}, \bibinfo {author} {\bibfnamefont {S.}~\bibnamefont {{Roelofs}}},
  \bibinfo {author} {\bibfnamefont {L.~J.}\ \bibnamefont {{Splitthoff}}},
  \bibinfo {author} {\bibfnamefont {M.}~\bibnamefont {{Pita-Vidal}}}, \bibinfo
  {author} {\bibfnamefont {A.}~\bibnamefont {{Bargerbos}}}, \bibinfo {author}
  {\bibfnamefont {D.~J.}\ \bibnamefont {{van Woerkom}}}, \bibinfo {author}
  {\bibfnamefont {P.}~\bibnamefont {{Krogstrup}}}, \bibinfo {author}
  {\bibfnamefont {L.~P.}\ \bibnamefont {{Kouwenhoven}}}, \bibinfo {author}
  {\bibfnamefont {C.~K.}\ \bibnamefont {{Andersen}}}, \bibinfo {author}
  {\bibfnamefont {A.}~\bibnamefont {{Levy Yeyati}}}, \bibinfo {author}
  {\bibfnamefont {B.}~\bibnamefont {{van Heck}}},\ and\ \bibinfo {author}
  {\bibfnamefont {G.}~\bibnamefont {{de Lange}}},\ }\bibfield  {title}
  {\bibinfo {title} {Microwave spectroscopy of interacting {A}ndreev spins},\
  }\href@noop {} {\bibfield  {journal} {\bibinfo  {journal} {arXiv e-prints}\ }
  (\bibinfo {year} {2022})},\ \Eprint {https://arxiv.org/abs/2208.11198}
  {arXiv:2208.11198} \BibitemShut {NoStop}%
\bibitem [{Sup()}]{Supplement}%
  \BibitemOpen
  \href@noop {} {}\bibinfo {note} {See Supplemental Material at [URL], which
  contains further details about theoretical modeling, fabrication and
  experimental setup, device tuneup, and additional data.}\BibitemShut {Stop}%
\bibitem [{\citenamefont {Han}\ \emph {et~al.}(2022)\citenamefont {Han},
  \citenamefont {Chan}, \citenamefont {de~Jong}, \citenamefont {Prosko},
  \citenamefont {Badawy}, \citenamefont {Gazibegovic}, \citenamefont {Bakkers},
  \citenamefont {Kouwenhoven}, \citenamefont {Malinowski},\ and\ \citenamefont
  {Pfaff}}]{Han2022}%
  \BibitemOpen
  \bibfield  {author} {\bibinfo {author} {\bibfnamefont {L.}~\bibnamefont
  {Han}}, \bibinfo {author} {\bibfnamefont {M.}~\bibnamefont {Chan}}, \bibinfo
  {author} {\bibfnamefont {D.}~\bibnamefont {de~Jong}}, \bibinfo {author}
  {\bibfnamefont {C.}~\bibnamefont {Prosko}}, \bibinfo {author} {\bibfnamefont
  {G.}~\bibnamefont {Badawy}}, \bibinfo {author} {\bibfnamefont
  {S.}~\bibnamefont {Gazibegovic}}, \bibinfo {author} {\bibfnamefont {E.~P.
  A.~M.}\ \bibnamefont {Bakkers}}, \bibinfo {author} {\bibfnamefont {L.~P.}\
  \bibnamefont {Kouwenhoven}}, \bibinfo {author} {\bibfnamefont {F.~K.}\
  \bibnamefont {Malinowski}},\ and\ \bibinfo {author} {\bibfnamefont
  {W.}~\bibnamefont {Pfaff}},\ }\bibfield  {title} {\bibinfo {title} {Variable
  and orbital-dependent spin-orbit field orientations in a {InSb} double
  quantum dot characterized via dispersive gate sensing},\ }\href@noop {}
  {\bibfield  {journal} {\bibinfo  {journal} {arXiv e-prints}\ } (\bibinfo
  {year} {2022})},\ \Eprint {https://arxiv.org/abs/2203.06047}
  {arXiv:2203.06047} \BibitemShut {NoStop}%
\bibitem [{\citenamefont {Werninghaus}\ \emph {et~al.}(2021)\citenamefont
  {Werninghaus}, \citenamefont {Egger}, \citenamefont {Roy}, \citenamefont
  {Machnes}, \citenamefont {Wilhelm},\ and\ \citenamefont
  {Filipp}}]{Werninghaus2021}%
  \BibitemOpen
  \bibfield  {author} {\bibinfo {author} {\bibfnamefont {M.}~\bibnamefont
  {Werninghaus}}, \bibinfo {author} {\bibfnamefont {D.~J.}\ \bibnamefont
  {Egger}}, \bibinfo {author} {\bibfnamefont {F.}~\bibnamefont {Roy}}, \bibinfo
  {author} {\bibfnamefont {S.}~\bibnamefont {Machnes}}, \bibinfo {author}
  {\bibfnamefont {F.~K.}\ \bibnamefont {Wilhelm}},\ and\ \bibinfo {author}
  {\bibfnamefont {S.}~\bibnamefont {Filipp}},\ }\bibfield  {title} {\bibinfo
  {title} {Leakage reduction in fast superconducting qubit gates via optimal
  control},\ }\bibfield  {journal} {\bibinfo  {journal} {npj Quantum Inf.}\
  }\textbf {\bibinfo {volume} {7}},\ \href
  {https://doi.org/10.1038/s41534-020-00346-2} {10.1038/s41534-020-00346-2}
  (\bibinfo {year} {2021})\BibitemShut {NoStop}%
\bibitem [{\citenamefont {Hays}\ \emph {et~al.}(2020)\citenamefont {Hays},
  \citenamefont {Fatemi}, \citenamefont {Serniak}, \citenamefont {Bouman},
  \citenamefont {Diamond}, \citenamefont {de~Lange}, \citenamefont {Krogstrup},
  \citenamefont {Nyg{\aa}rd}, \citenamefont {Geresdi},\ and\ \citenamefont
  {Devoret}}]{Hays2020}%
  \BibitemOpen
  \bibfield  {author} {\bibinfo {author} {\bibfnamefont {M.}~\bibnamefont
  {Hays}}, \bibinfo {author} {\bibfnamefont {V.}~\bibnamefont {Fatemi}},
  \bibinfo {author} {\bibfnamefont {K.}~\bibnamefont {Serniak}}, \bibinfo
  {author} {\bibfnamefont {D.}~\bibnamefont {Bouman}}, \bibinfo {author}
  {\bibfnamefont {S.}~\bibnamefont {Diamond}}, \bibinfo {author} {\bibfnamefont
  {G.}~\bibnamefont {de~Lange}}, \bibinfo {author} {\bibfnamefont
  {P.}~\bibnamefont {Krogstrup}}, \bibinfo {author} {\bibfnamefont
  {J.}~\bibnamefont {Nyg{\aa}rd}}, \bibinfo {author} {\bibfnamefont
  {A.}~\bibnamefont {Geresdi}},\ and\ \bibinfo {author} {\bibfnamefont {M.~H.}\
  \bibnamefont {Devoret}},\ }\bibfield  {title} {\bibinfo {title} {Continuous
  monitoring of a trapped superconducting spin},\ }\bibfield  {journal}
  {\bibinfo  {journal} {Nat Phys}\ }\textbf {\bibinfo {volume} {16}},\ \href
  {https://doi.org/10.1038/s41567-020-0952-3} {10.1038/s41567-020-0952-3}
  (\bibinfo {year} {2020})\BibitemShut {NoStop}%
\bibitem [{\citenamefont {Stockill}\ \emph {et~al.}(2016)\citenamefont
  {Stockill}, \citenamefont {Gall}, \citenamefont {Matthiesen}, \citenamefont
  {Huthmacher}, \citenamefont {Clarke}, \citenamefont {Hugues},\ and\
  \citenamefont {Atatüre}}]{Stockill2016}%
  \BibitemOpen
  \bibfield  {author} {\bibinfo {author} {\bibfnamefont {R.}~\bibnamefont
  {Stockill}}, \bibinfo {author} {\bibfnamefont {C.~L.}\ \bibnamefont {Gall}},
  \bibinfo {author} {\bibfnamefont {C.}~\bibnamefont {Matthiesen}}, \bibinfo
  {author} {\bibfnamefont {L.}~\bibnamefont {Huthmacher}}, \bibinfo {author}
  {\bibfnamefont {E.}~\bibnamefont {Clarke}}, \bibinfo {author} {\bibfnamefont
  {M.}~\bibnamefont {Hugues}},\ and\ \bibinfo {author} {\bibfnamefont
  {M.}~\bibnamefont {Atatüre}},\ }\bibfield  {title} {\bibinfo {title}
  {Quantum dot spin coherence governed by a strained nuclear environment},\
  }\bibfield  {journal} {\bibinfo  {journal} {Nat Commun}\ }\textbf {\bibinfo
  {volume} {7}},\ \href {https://doi.org/10.1038/ncomms12745}
  {10.1038/ncomms12745} (\bibinfo {year} {2016})\BibitemShut {NoStop}%
\bibitem [{\citenamefont {Hahn}(1950)}]{Hahn1950}%
  \BibitemOpen
  \bibfield  {author} {\bibinfo {author} {\bibfnamefont {E.~L.}\ \bibnamefont
  {Hahn}},\ }\bibfield  {title} {\bibinfo {title} {Spin echoes},\ }\bibfield
  {journal} {\bibinfo  {journal} {Phys Rev}\ }\textbf {\bibinfo {volume}
  {80}},\ \href {https://doi.org/10.1103/PhysRev.80.580}
  {10.1103/PhysRev.80.580} (\bibinfo {year} {1950})\BibitemShut {NoStop}%
\bibitem [{\citenamefont {Carr}\ and\ \citenamefont
  {Purcell}(1954)}]{Carr1954}%
  \BibitemOpen
  \bibfield  {author} {\bibinfo {author} {\bibfnamefont {H.~Y.}\ \bibnamefont
  {Carr}}\ and\ \bibinfo {author} {\bibfnamefont {E.~M.}\ \bibnamefont
  {Purcell}},\ }\bibfield  {title} {\bibinfo {title} {Effects of diffusion on
  free precession in nuclear magnetic resonance experiments},\ }\bibfield
  {journal} {\bibinfo  {journal} {Phys Rev}\ }\textbf {\bibinfo {volume}
  {94}},\ \href {https://doi.org/10.1103/PhysRev.94.630}
  {10.1103/PhysRev.94.630} (\bibinfo {year} {1954})\BibitemShut {NoStop}%
\bibitem [{\citenamefont {Barthel}\ \emph {et~al.}(2010)\citenamefont
  {Barthel}, \citenamefont {Medford}, \citenamefont {Marcus}, \citenamefont
  {Hanson},\ and\ \citenamefont {Gossard}}]{Barthel2010}%
  \BibitemOpen
  \bibfield  {author} {\bibinfo {author} {\bibfnamefont {C.}~\bibnamefont
  {Barthel}}, \bibinfo {author} {\bibfnamefont {J.}~\bibnamefont {Medford}},
  \bibinfo {author} {\bibfnamefont {C.~M.}\ \bibnamefont {Marcus}}, \bibinfo
  {author} {\bibfnamefont {M.~P.}\ \bibnamefont {Hanson}},\ and\ \bibinfo
  {author} {\bibfnamefont {A.~C.}\ \bibnamefont {Gossard}},\ }\bibfield
  {title} {\bibinfo {title} {Interlaced dynamical decoupling and coherent
  operation of a singlet-triplet qubit},\ }\bibfield  {journal} {\bibinfo
  {journal} {Phys Rev Lett}\ }\textbf {\bibinfo {volume} {105}},\ \href
  {https://doi.org/10.1103/PhysRevLett.105.266808}
  {10.1103/PhysRevLett.105.266808} (\bibinfo {year} {2010})\BibitemShut
  {NoStop}%
\bibitem [{\citenamefont {Bylander}\ \emph {et~al.}(2011)\citenamefont
  {Bylander}, \citenamefont {Gustavsson}, \citenamefont {Yan}, \citenamefont
  {Yoshihara}, \citenamefont {Harrabi}, \citenamefont {Fitch}, \citenamefont
  {Cory}, \citenamefont {Nakamura}, \citenamefont {Tsai},\ and\ \citenamefont
  {Oliver}}]{Bylander2011}%
  \BibitemOpen
  \bibfield  {author} {\bibinfo {author} {\bibfnamefont {J.}~\bibnamefont
  {Bylander}}, \bibinfo {author} {\bibfnamefont {S.}~\bibnamefont
  {Gustavsson}}, \bibinfo {author} {\bibfnamefont {F.}~\bibnamefont {Yan}},
  \bibinfo {author} {\bibfnamefont {F.}~\bibnamefont {Yoshihara}}, \bibinfo
  {author} {\bibfnamefont {K.}~\bibnamefont {Harrabi}}, \bibinfo {author}
  {\bibfnamefont {G.}~\bibnamefont {Fitch}}, \bibinfo {author} {\bibfnamefont
  {D.~G.}\ \bibnamefont {Cory}}, \bibinfo {author} {\bibfnamefont
  {Y.}~\bibnamefont {Nakamura}}, \bibinfo {author} {\bibfnamefont {J.-S.}\
  \bibnamefont {Tsai}},\ and\ \bibinfo {author} {\bibfnamefont {W.~D.}\
  \bibnamefont {Oliver}},\ }\bibfield  {title} {\bibinfo {title} {Noise
  spectroscopy through dynamical decoupling with a superconducting flux
  qubit},\ }\bibfield  {journal} {\bibinfo  {journal} {Nat Phys}\ }\textbf
  {\bibinfo {volume} {7}},\ \href {https://doi.org/10.1038/nphys1994}
  {10.1038/nphys1994} (\bibinfo {year} {2011})\BibitemShut {NoStop}%
\bibitem [{\citenamefont {Cywi\ifmmode~\acute{n}\else \'{n}\fi{}ski}\ \emph
  {et~al.}(2008)\citenamefont {Cywi\ifmmode~\acute{n}\else \'{n}\fi{}ski},
  \citenamefont {Lutchyn}, \citenamefont {Nave},\ and\ \citenamefont
  {Das~Sarma}}]{Cywinski2008}%
  \BibitemOpen
  \bibfield  {author} {\bibinfo {author} {\bibfnamefont {L.}~\bibnamefont
  {Cywi\ifmmode~\acute{n}\else \'{n}\fi{}ski}}, \bibinfo {author}
  {\bibfnamefont {R.~M.}\ \bibnamefont {Lutchyn}}, \bibinfo {author}
  {\bibfnamefont {C.~P.}\ \bibnamefont {Nave}},\ and\ \bibinfo {author}
  {\bibfnamefont {S.}~\bibnamefont {Das~Sarma}},\ }\bibfield  {title} {\bibinfo
  {title} {How to enhance dephasing time in superconducting qubits},\
  }\bibfield  {journal} {\bibinfo  {journal} {Phys Rev B}\ }\textbf {\bibinfo
  {volume} {77}},\ \href {https://doi.org/10.1103/PhysRevB.77.174509}
  {10.1103/PhysRevB.77.174509} (\bibinfo {year} {2008})\BibitemShut {NoStop}%
\bibitem [{\citenamefont {Medford}\ \emph {et~al.}(2012)\citenamefont
  {Medford}, \citenamefont {Cywi\ifmmode~\acute{n}\else \'{n}\fi{}ski},
  \citenamefont {Barthel}, \citenamefont {Marcus}, \citenamefont {Hanson},\
  and\ \citenamefont {Gossard}}]{Medford2012}%
  \BibitemOpen
  \bibfield  {author} {\bibinfo {author} {\bibfnamefont {J.}~\bibnamefont
  {Medford}}, \bibinfo {author} {\bibfnamefont {L.}~\bibnamefont
  {Cywi\ifmmode~\acute{n}\else \'{n}\fi{}ski}}, \bibinfo {author}
  {\bibfnamefont {C.}~\bibnamefont {Barthel}}, \bibinfo {author} {\bibfnamefont
  {C.~M.}\ \bibnamefont {Marcus}}, \bibinfo {author} {\bibfnamefont {M.~P.}\
  \bibnamefont {Hanson}},\ and\ \bibinfo {author} {\bibfnamefont {A.~C.}\
  \bibnamefont {Gossard}},\ }\bibfield  {title} {\bibinfo {title} {Scaling of
  dynamical decoupling for spin qubits},\ }\bibfield  {journal} {\bibinfo
  {journal} {Phys Rev Lett}\ }\textbf {\bibinfo {volume} {108}},\ \href
  {https://doi.org/10.1103/PhysRevLett.108.086802}
  {10.1103/PhysRevLett.108.086802} (\bibinfo {year} {2012})\BibitemShut
  {NoStop}%
\bibitem [{\citenamefont {Schreier}\ \emph {et~al.}(2008)\citenamefont
  {Schreier}, \citenamefont {Houck}, \citenamefont {Koch}, \citenamefont
  {Schuster}, \citenamefont {Johnson}, \citenamefont {Chow}, \citenamefont
  {Gambetta}, \citenamefont {Majer}, \citenamefont {Frunzio}, \citenamefont
  {Devoret}, \citenamefont {Girvin},\ and\ \citenamefont
  {Schoelkopf}}]{Schreier2008}%
  \BibitemOpen
  \bibfield  {author} {\bibinfo {author} {\bibfnamefont {J.~A.}\ \bibnamefont
  {Schreier}}, \bibinfo {author} {\bibfnamefont {A.~A.}\ \bibnamefont {Houck}},
  \bibinfo {author} {\bibfnamefont {J.}~\bibnamefont {Koch}}, \bibinfo {author}
  {\bibfnamefont {D.~I.}\ \bibnamefont {Schuster}}, \bibinfo {author}
  {\bibfnamefont {B.~R.}\ \bibnamefont {Johnson}}, \bibinfo {author}
  {\bibfnamefont {J.~M.}\ \bibnamefont {Chow}}, \bibinfo {author}
  {\bibfnamefont {J.~M.}\ \bibnamefont {Gambetta}}, \bibinfo {author}
  {\bibfnamefont {J.}~\bibnamefont {Majer}}, \bibinfo {author} {\bibfnamefont
  {L.}~\bibnamefont {Frunzio}}, \bibinfo {author} {\bibfnamefont {M.~H.}\
  \bibnamefont {Devoret}}, \bibinfo {author} {\bibfnamefont {S.~M.}\
  \bibnamefont {Girvin}},\ and\ \bibinfo {author} {\bibfnamefont {R.~J.}\
  \bibnamefont {Schoelkopf}},\ }\bibfield  {title} {\bibinfo {title}
  {Suppressing charge noise decoherence in superconducting charge qubits},\
  }\bibfield  {journal} {\bibinfo  {journal} {Phys Rev B}\ }\textbf {\bibinfo
  {volume} {77}},\ \href {https://doi.org/10.1103/physrevb.77.180502}
  {10.1103/physrevb.77.180502} (\bibinfo {year} {2008})\BibitemShut {NoStop}%
\bibitem [{\citenamefont {Braumüller}\ \emph {et~al.}(2020)\citenamefont
  {Braumüller}, \citenamefont {Ding}, \citenamefont {Vepsäläinen},
  \citenamefont {Sung}, \citenamefont {Kjaergaard}, \citenamefont {Menke},
  \citenamefont {Winik}, \citenamefont {Kim}, \citenamefont {Niedzielski},
  \citenamefont {Melville}, \citenamefont {Yoder}, \citenamefont
  {Hirjibehedin}, \citenamefont {Orlando}, \citenamefont {Gustavsson},\ and\
  \citenamefont {Oliver}}]{Braumueller2020}%
  \BibitemOpen
  \bibfield  {author} {\bibinfo {author} {\bibfnamefont {J.}~\bibnamefont
  {Braumüller}}, \bibinfo {author} {\bibfnamefont {L.}~\bibnamefont {Ding}},
  \bibinfo {author} {\bibfnamefont {A.~P.}\ \bibnamefont {Vepsäläinen}},
  \bibinfo {author} {\bibfnamefont {Y.}~\bibnamefont {Sung}}, \bibinfo {author}
  {\bibfnamefont {M.}~\bibnamefont {Kjaergaard}}, \bibinfo {author}
  {\bibfnamefont {T.}~\bibnamefont {Menke}}, \bibinfo {author} {\bibfnamefont
  {R.}~\bibnamefont {Winik}}, \bibinfo {author} {\bibfnamefont
  {D.}~\bibnamefont {Kim}}, \bibinfo {author} {\bibfnamefont {B.~M.}\
  \bibnamefont {Niedzielski}}, \bibinfo {author} {\bibfnamefont
  {A.}~\bibnamefont {Melville}}, \bibinfo {author} {\bibfnamefont {J.~L.}\
  \bibnamefont {Yoder}}, \bibinfo {author} {\bibfnamefont {C.~F.}\ \bibnamefont
  {Hirjibehedin}}, \bibinfo {author} {\bibfnamefont {T.~P.}\ \bibnamefont
  {Orlando}}, \bibinfo {author} {\bibfnamefont {S.}~\bibnamefont
  {Gustavsson}},\ and\ \bibinfo {author} {\bibfnamefont {W.~D.}\ \bibnamefont
  {Oliver}},\ }\bibfield  {title} {\bibinfo {title} {Characterizing and
  optimizing qubit coherence based on {SQUID} geometry},\ }\bibfield  {journal}
  {\bibinfo  {journal} {Phys. Rev. Applied}\ }\textbf {\bibinfo {volume}
  {13}},\ \href {https://doi.org/10.1103/physrevapplied.13.054079}
  {10.1103/physrevapplied.13.054079} (\bibinfo {year} {2020})\BibitemShut
  {NoStop}%
\bibitem [{\citenamefont {Malinowski}\ \emph
  {et~al.}(2017{\natexlab{a}})\citenamefont {Malinowski}, \citenamefont
  {Martins}, \citenamefont {Nissen}, \citenamefont {Fallahi}, \citenamefont
  {Gardner}, \citenamefont {Manfra}, \citenamefont {Marcus},\ and\
  \citenamefont {Kuemmeth}}]{Malinowski2017}%
  \BibitemOpen
  \bibfield  {author} {\bibinfo {author} {\bibfnamefont {F.~K.}\ \bibnamefont
  {Malinowski}}, \bibinfo {author} {\bibfnamefont {F.}~\bibnamefont {Martins}},
  \bibinfo {author} {\bibfnamefont {P.~D.}\ \bibnamefont {Nissen}}, \bibinfo
  {author} {\bibfnamefont {S.}~\bibnamefont {Fallahi}}, \bibinfo {author}
  {\bibfnamefont {G.~C.}\ \bibnamefont {Gardner}}, \bibinfo {author}
  {\bibfnamefont {M.~J.}\ \bibnamefont {Manfra}}, \bibinfo {author}
  {\bibfnamefont {C.~M.}\ \bibnamefont {Marcus}},\ and\ \bibinfo {author}
  {\bibfnamefont {F.}~\bibnamefont {Kuemmeth}},\ }\bibfield  {title} {\bibinfo
  {title} {Symmetric operation of the resonant exchange qubit},\ }\bibfield
  {journal} {\bibinfo  {journal} {Phys Rev B}\ }\textbf {\bibinfo {volume}
  {96}},\ \href {https://doi.org/10.1103/PhysRevB.96.045443}
  {10.1103/PhysRevB.96.045443} (\bibinfo {year}
  {2017}{\natexlab{a}})\BibitemShut {NoStop}%
\bibitem [{\citenamefont {Malinowski}\ \emph
  {et~al.}(2017{\natexlab{b}})\citenamefont {Malinowski}, \citenamefont
  {Martins}, \citenamefont {Cywi\ifmmode~\acute{n}\else \'{n}\fi{}ski},
  \citenamefont {Rudner}, \citenamefont {Nissen}, \citenamefont {Fallahi},
  \citenamefont {Gardner}, \citenamefont {Manfra}, \citenamefont {Marcus},\
  and\ \citenamefont {Kuemmeth}}]{Malinowski2017b}%
  \BibitemOpen
  \bibfield  {author} {\bibinfo {author} {\bibfnamefont {F.~K.}\ \bibnamefont
  {Malinowski}}, \bibinfo {author} {\bibfnamefont {F.}~\bibnamefont {Martins}},
  \bibinfo {author} {\bibfnamefont {L.}~\bibnamefont
  {Cywi\ifmmode~\acute{n}\else \'{n}\fi{}ski}}, \bibinfo {author}
  {\bibfnamefont {M.~S.}\ \bibnamefont {Rudner}}, \bibinfo {author}
  {\bibfnamefont {P.~D.}\ \bibnamefont {Nissen}}, \bibinfo {author}
  {\bibfnamefont {S.}~\bibnamefont {Fallahi}}, \bibinfo {author} {\bibfnamefont
  {G.~C.}\ \bibnamefont {Gardner}}, \bibinfo {author} {\bibfnamefont {M.~J.}\
  \bibnamefont {Manfra}}, \bibinfo {author} {\bibfnamefont {C.~M.}\
  \bibnamefont {Marcus}},\ and\ \bibinfo {author} {\bibfnamefont
  {F.}~\bibnamefont {Kuemmeth}},\ }\bibfield  {title} {\bibinfo {title}
  {Spectrum of the nuclear environment for {GaAs} spin qubits},\ }\bibfield
  {journal} {\bibinfo  {journal} {Phys Rev Lett}\ }\textbf {\bibinfo {volume}
  {118}},\ \href {https://doi.org/10.1103/PhysRevLett.118.177702}
  {10.1103/PhysRevLett.118.177702} (\bibinfo {year}
  {2017}{\natexlab{b}})\BibitemShut {NoStop}%
\bibitem [{\citenamefont {Landig}\ \emph {et~al.}(2019)\citenamefont {Landig},
  \citenamefont {Koski}, \citenamefont {Scarlino}, \citenamefont {M{\"u}ller},
  \citenamefont {Abadillo-Uriel}, \citenamefont {Kratochwil}, \citenamefont
  {Reichl}, \citenamefont {Wegscheider}, \citenamefont {Coppersmith},
  \citenamefont {Friesen}, \citenamefont {Wallraff}, \citenamefont {Ihn},\ and\
  \citenamefont {Ensslin}}]{Landig2019}%
  \BibitemOpen
  \bibfield  {author} {\bibinfo {author} {\bibfnamefont {A.~J.}\ \bibnamefont
  {Landig}}, \bibinfo {author} {\bibfnamefont {J.~V.}\ \bibnamefont {Koski}},
  \bibinfo {author} {\bibfnamefont {P.}~\bibnamefont {Scarlino}}, \bibinfo
  {author} {\bibfnamefont {C.}~\bibnamefont {M{\"u}ller}}, \bibinfo {author}
  {\bibfnamefont {J.~C.}\ \bibnamefont {Abadillo-Uriel}}, \bibinfo {author}
  {\bibfnamefont {B.}~\bibnamefont {Kratochwil}}, \bibinfo {author}
  {\bibfnamefont {C.}~\bibnamefont {Reichl}}, \bibinfo {author} {\bibfnamefont
  {W.}~\bibnamefont {Wegscheider}}, \bibinfo {author} {\bibfnamefont {S.~N.}\
  \bibnamefont {Coppersmith}}, \bibinfo {author} {\bibfnamefont
  {M.}~\bibnamefont {Friesen}}, \bibinfo {author} {\bibfnamefont
  {A.}~\bibnamefont {Wallraff}}, \bibinfo {author} {\bibfnamefont
  {T.}~\bibnamefont {Ihn}},\ and\ \bibinfo {author} {\bibfnamefont
  {K.}~\bibnamefont {Ensslin}},\ }\bibfield  {title} {\bibinfo {title}
  {Virtual-photon-mediated spin-qubit--transmon coupling},\ }\bibfield
  {journal} {\bibinfo  {journal} {Nat Commun}\ }\textbf {\bibinfo {volume}
  {10}},\ \href {https://doi.org/10.1038/s41467-019-13000-z}
  {10.1038/s41467-019-13000-z} (\bibinfo {year} {2019})\BibitemShut {NoStop}%
\bibitem [{\citenamefont {Forn-D{\'{\i}}az}\ \emph {et~al.}(2019)\citenamefont
  {Forn-D{\'{\i}}az}, \citenamefont {Lamata}, \citenamefont {Rico},
  \citenamefont {Kono},\ and\ \citenamefont {Solano}}]{FornDiaz2019}%
  \BibitemOpen
  \bibfield  {author} {\bibinfo {author} {\bibfnamefont {P.}~\bibnamefont
  {Forn-D{\'{\i}}az}}, \bibinfo {author} {\bibfnamefont {L.}~\bibnamefont
  {Lamata}}, \bibinfo {author} {\bibfnamefont {E.}~\bibnamefont {Rico}},
  \bibinfo {author} {\bibfnamefont {J.}~\bibnamefont {Kono}},\ and\ \bibinfo
  {author} {\bibfnamefont {E.}~\bibnamefont {Solano}},\ }\bibfield  {title}
  {\bibinfo {title} {Ultrastrong coupling regimes of light-matter
  interaction},\ }\bibfield  {journal} {\bibinfo  {journal} {Rev Mod Phys}\
  }\textbf {\bibinfo {volume} {91}},\ \href
  {https://doi.org/10.1103/revmodphys.91.025005} {10.1103/revmodphys.91.025005}
  (\bibinfo {year} {2019})\BibitemShut {NoStop}%
\bibitem [{\citenamefont {Scarlino}\ \emph {et~al.}(2022)\citenamefont
  {Scarlino}, \citenamefont {Ungerer}, \citenamefont {van Woerkom},
  \citenamefont {Mancini}, \citenamefont {Stano}, \citenamefont {Müller},
  \citenamefont {Landig}, \citenamefont {Koski}, \citenamefont {Reichl},
  \citenamefont {Wegscheider}, \citenamefont {Ihn}, \citenamefont {Ensslin},\
  and\ \citenamefont {Wallraff}}]{Scarlino2022}%
  \BibitemOpen
  \bibfield  {author} {\bibinfo {author} {\bibfnamefont {P.}~\bibnamefont
  {Scarlino}}, \bibinfo {author} {\bibfnamefont {J.}~\bibnamefont {Ungerer}},
  \bibinfo {author} {\bibfnamefont {D.}~\bibnamefont {van Woerkom}}, \bibinfo
  {author} {\bibfnamefont {M.}~\bibnamefont {Mancini}}, \bibinfo {author}
  {\bibfnamefont {P.}~\bibnamefont {Stano}}, \bibinfo {author} {\bibfnamefont
  {C.}~\bibnamefont {Müller}}, \bibinfo {author} {\bibfnamefont
  {A.}~\bibnamefont {Landig}}, \bibinfo {author} {\bibfnamefont
  {J.}~\bibnamefont {Koski}}, \bibinfo {author} {\bibfnamefont
  {C.}~\bibnamefont {Reichl}}, \bibinfo {author} {\bibfnamefont
  {W.}~\bibnamefont {Wegscheider}}, \bibinfo {author} {\bibfnamefont
  {T.}~\bibnamefont {Ihn}}, \bibinfo {author} {\bibfnamefont {K.}~\bibnamefont
  {Ensslin}},\ and\ \bibinfo {author} {\bibfnamefont {A.}~\bibnamefont
  {Wallraff}},\ }\bibfield  {title} {\bibinfo {title} {{In-situ tuning of the
  electric-dipole strength of a double-dot charge qubit: qharge-noise
  protection and ultrastrong coupling}},\ }\bibfield  {journal} {\bibinfo
  {journal} {Phys. Rev. X}\ }\textbf {\bibinfo {volume} {12}},\ \href
  {https://doi.org/10.1103/physrevx.12.031004} {10.1103/physrevx.12.031004}
  (\bibinfo {year} {2022})\BibitemShut {NoStop}%
\bibitem [{\citenamefont {de~Leon}\ \emph {et~al.}(2021)\citenamefont
  {de~Leon}, \citenamefont {Itoh}, \citenamefont {Kim}, \citenamefont {Mehta},
  \citenamefont {Northup}, \citenamefont {Paik}, \citenamefont {Palmer},
  \citenamefont {Samarth}, \citenamefont {Sangtawesin},\ and\ \citenamefont
  {Steuerman}}]{Leon2021}%
  \BibitemOpen
  \bibfield  {author} {\bibinfo {author} {\bibfnamefont {N.~P.}\ \bibnamefont
  {de~Leon}}, \bibinfo {author} {\bibfnamefont {K.~M.}\ \bibnamefont {Itoh}},
  \bibinfo {author} {\bibfnamefont {D.}~\bibnamefont {Kim}}, \bibinfo {author}
  {\bibfnamefont {K.~K.}\ \bibnamefont {Mehta}}, \bibinfo {author}
  {\bibfnamefont {T.~E.}\ \bibnamefont {Northup}}, \bibinfo {author}
  {\bibfnamefont {H.}~\bibnamefont {Paik}}, \bibinfo {author} {\bibfnamefont
  {B.~S.}\ \bibnamefont {Palmer}}, \bibinfo {author} {\bibfnamefont
  {N.}~\bibnamefont {Samarth}}, \bibinfo {author} {\bibfnamefont
  {S.}~\bibnamefont {Sangtawesin}},\ and\ \bibinfo {author} {\bibfnamefont
  {D.~W.}\ \bibnamefont {Steuerman}},\ }\bibfield  {title} {\bibinfo {title}
  {Materials challenges and opportunities for quantum computing hardware},\
  }\bibfield  {journal} {\bibinfo  {journal} {Science}\ }\textbf {\bibinfo
  {volume} {372}},\ \href {https://doi.org/10.1126/science.abb2823}
  {10.1126/science.abb2823} (\bibinfo {year} {2021})\BibitemShut {NoStop}%
\bibitem [{\citenamefont {Hendrickx}\ \emph {et~al.}(2018)\citenamefont
  {Hendrickx}, \citenamefont {Franke}, \citenamefont {Sammak}, \citenamefont
  {Kouwenhoven}, \citenamefont {Sabbagh}, \citenamefont {Yeoh}, \citenamefont
  {Li}, \citenamefont {Tagliaferri}, \citenamefont {Virgilio}, \citenamefont
  {Capellini}, \citenamefont {Scappucci},\ and\ \citenamefont
  {Veldhorst}}]{Hendrickx2018}%
  \BibitemOpen
  \bibfield  {author} {\bibinfo {author} {\bibfnamefont {N.~W.}\ \bibnamefont
  {Hendrickx}}, \bibinfo {author} {\bibfnamefont {D.~P.}\ \bibnamefont
  {Franke}}, \bibinfo {author} {\bibfnamefont {A.}~\bibnamefont {Sammak}},
  \bibinfo {author} {\bibfnamefont {M.}~\bibnamefont {Kouwenhoven}}, \bibinfo
  {author} {\bibfnamefont {D.}~\bibnamefont {Sabbagh}}, \bibinfo {author}
  {\bibfnamefont {L.}~\bibnamefont {Yeoh}}, \bibinfo {author} {\bibfnamefont
  {R.}~\bibnamefont {Li}}, \bibinfo {author} {\bibfnamefont {M.~L.~V.}\
  \bibnamefont {Tagliaferri}}, \bibinfo {author} {\bibfnamefont
  {M.}~\bibnamefont {Virgilio}}, \bibinfo {author} {\bibfnamefont
  {G.}~\bibnamefont {Capellini}}, \bibinfo {author} {\bibfnamefont
  {G.}~\bibnamefont {Scappucci}},\ and\ \bibinfo {author} {\bibfnamefont
  {M.}~\bibnamefont {Veldhorst}},\ }\bibfield  {title} {\bibinfo {title}
  {Gate-controlled quantum dots and superconductivity in planar germanium},\
  }\bibfield  {journal} {\bibinfo  {journal} {Nat Commun}\ }\textbf {\bibinfo
  {volume} {9}},\ \href {https://doi.org/10.1038/s41467-018-05299-x}
  {10.1038/s41467-018-05299-x} (\bibinfo {year} {2018})\BibitemShut {NoStop}%
\bibitem [{\citenamefont {Scappucci}\ \emph {et~al.}(2021)\citenamefont
  {Scappucci}, \citenamefont {Kloeffel}, \citenamefont {Zwanenburg},
  \citenamefont {Loss}, \citenamefont {Myronov}, \citenamefont {Zhang},
  \citenamefont {De~Franceschi}, \citenamefont {Katsaros},\ and\ \citenamefont
  {Veldhorst}}]{Scappucci2021}%
  \BibitemOpen
  \bibfield  {author} {\bibinfo {author} {\bibfnamefont {G.}~\bibnamefont
  {Scappucci}}, \bibinfo {author} {\bibfnamefont {C.}~\bibnamefont {Kloeffel}},
  \bibinfo {author} {\bibfnamefont {F.~A.}\ \bibnamefont {Zwanenburg}},
  \bibinfo {author} {\bibfnamefont {D.}~\bibnamefont {Loss}}, \bibinfo {author}
  {\bibfnamefont {M.}~\bibnamefont {Myronov}}, \bibinfo {author} {\bibfnamefont
  {J.-J.}\ \bibnamefont {Zhang}}, \bibinfo {author} {\bibfnamefont
  {S.}~\bibnamefont {De~Franceschi}}, \bibinfo {author} {\bibfnamefont
  {G.}~\bibnamefont {Katsaros}},\ and\ \bibinfo {author} {\bibfnamefont
  {M.}~\bibnamefont {Veldhorst}},\ }\bibfield  {title} {\bibinfo {title} {The
  germanium quantum information route},\ }\bibfield  {journal} {\bibinfo
  {journal} {Nat. Rev. Mater.}\ }\textbf {\bibinfo {volume} {6}},\ \href
  {https://doi.org/10.1038/s41578-020-00262-z} {10.1038/s41578-020-00262-z}
  (\bibinfo {year} {2021})\BibitemShut {NoStop}%
\bibitem [{\citenamefont {Tosato}\ \emph {et~al.}(2022)\citenamefont {Tosato},
  \citenamefont {Levajac}, \citenamefont {Wang}, \citenamefont {Boor},
  \citenamefont {Borsoi}, \citenamefont {Botifoll}, \citenamefont {Borja},
  \citenamefont {Martí-Sánchez}, \citenamefont {Arbiol}, \citenamefont
  {Sammak}, \citenamefont {Veldhorst},\ and\ \citenamefont
  {Scappucci}}]{Tosato2022}%
  \BibitemOpen
  \bibfield  {author} {\bibinfo {author} {\bibfnamefont {A.}~\bibnamefont
  {Tosato}}, \bibinfo {author} {\bibfnamefont {V.}~\bibnamefont {Levajac}},
  \bibinfo {author} {\bibfnamefont {J.-Y.}\ \bibnamefont {Wang}}, \bibinfo
  {author} {\bibfnamefont {C.~J.}\ \bibnamefont {Boor}}, \bibinfo {author}
  {\bibfnamefont {F.}~\bibnamefont {Borsoi}}, \bibinfo {author} {\bibfnamefont
  {M.}~\bibnamefont {Botifoll}}, \bibinfo {author} {\bibfnamefont {C.~N.}\
  \bibnamefont {Borja}}, \bibinfo {author} {\bibfnamefont {S.}~\bibnamefont
  {Martí-Sánchez}}, \bibinfo {author} {\bibfnamefont {J.}~\bibnamefont
  {Arbiol}}, \bibinfo {author} {\bibfnamefont {A.}~\bibnamefont {Sammak}},
  \bibinfo {author} {\bibfnamefont {M.}~\bibnamefont {Veldhorst}},\ and\
  \bibinfo {author} {\bibfnamefont {G.}~\bibnamefont {Scappucci}},\ }\bibfield
  {title} {\bibinfo {title} {Hard superconducting gap in a high-mobility
  semiconductor},\ }\href@noop {} {\bibfield  {journal} {\bibinfo  {journal}
  {arXiv e-prints}\ } (\bibinfo {year} {2022})},\ \Eprint
  {https://arxiv.org/abs/2206.00569} {arXiv:2206.00569} \BibitemShut {NoStop}%
\bibitem [{\citenamefont {{Spethmann}}\ \emph {et~al.}(2022)\citenamefont
  {{Spethmann}}, \citenamefont {{Zhang}}, \citenamefont {{Klinovaja}},\ and\
  \citenamefont {{Loss}}}]{Spethmann2022}%
  \BibitemOpen
  \bibfield  {author} {\bibinfo {author} {\bibfnamefont {M.}~\bibnamefont
  {{Spethmann}}}, \bibinfo {author} {\bibfnamefont {X.-P.}\ \bibnamefont
  {{Zhang}}}, \bibinfo {author} {\bibfnamefont {J.}~\bibnamefont
  {{Klinovaja}}},\ and\ \bibinfo {author} {\bibfnamefont {D.}~\bibnamefont
  {{Loss}}},\ }\bibfield  {title} {\bibinfo {title} {{Coupled superconducting
  spin qubits with spin-orbit interaction}},\ }\href@noop {} {\bibfield
  {journal} {\bibinfo  {journal} {arXiv e-prints}\ } (\bibinfo {year}
  {2022})},\ \Eprint {https://arxiv.org/abs/2205.03843} {arXiv:2205.03843}
  \BibitemShut {NoStop}%
\end{thebibliography}%


\begin{thebibliography}{18}%
\makeatletter
\providecommand \@ifxundefined [1]{%
 \@ifx{#1\undefined}
}%
\providecommand \@ifnum [1]{%
 \ifnum #1\expandafter \@firstoftwo
 \else \expandafter \@secondoftwo
 \fi
}%
\providecommand \@ifx [1]{%
 \ifx #1\expandafter \@firstoftwo
 \else \expandafter \@secondoftwo
 \fi
}%
\providecommand \natexlab [1]{#1}%
\providecommand \enquote  [1]{``#1''}%
\providecommand \bibnamefont  [1]{#1}%
\providecommand \bibfnamefont [1]{#1}%
\providecommand \citenamefont [1]{#1}%
\providecommand \href@noop [0]{\@secondoftwo}%
\providecommand \href [0]{\begingroup \@sanitize@url \@href}%
\providecommand \@href[1]{\@@startlink{#1}\@@href}%
\providecommand \@@href[1]{\endgroup#1\@@endlink}%
\providecommand \@sanitize@url [0]{\catcode `\\12\catcode `\$12\catcode
  `\&12\catcode `\#12\catcode `\^12\catcode `\_12\catcode `\%12\relax}%
\providecommand \@@startlink[1]{}%
\providecommand \@@endlink[0]{}%
\providecommand \url  [0]{\begingroup\@sanitize@url \@url }%
\providecommand \@url [1]{\endgroup\@href {#1}{\urlprefix }}%
\providecommand \urlprefix  [0]{URL }%
\providecommand \Eprint [0]{\href }%
\providecommand \doibase [0]{https://doi.org/}%
\providecommand \selectlanguage [0]{\@gobble}%
\providecommand \bibinfo  [0]{\@secondoftwo}%
\providecommand \bibfield  [0]{\@secondoftwo}%
\providecommand \translation [1]{[#1]}%
\providecommand \BibitemOpen [0]{}%
\providecommand \bibitemStop [0]{}%
\providecommand \bibitemNoStop [0]{.\EOS\space}%
\providecommand \EOS [0]{\spacefactor3000\relax}%
\providecommand \BibitemShut  [1]{\csname bibitem#1\endcsname}%
\let\auto@bib@innerbib\@empty
\bibitem [{\citenamefont {Bargerbos}\ \emph {et~al.}(2020)\citenamefont
  {Bargerbos}, \citenamefont {Uilhoorn}, \citenamefont {Yang}, \citenamefont
  {Krogstrup}, \citenamefont {Kouwenhoven}, \citenamefont {de~Lange},
  \citenamefont {van Heck},\ and\ \citenamefont {Kou}}]{Bargerbos2020}%
  \BibitemOpen
  \bibfield  {author} {\bibinfo {author} {\bibfnamefont {A.}~\bibnamefont
  {Bargerbos}}, \bibinfo {author} {\bibfnamefont {W.}~\bibnamefont {Uilhoorn}},
  \bibinfo {author} {\bibfnamefont {C.-K.}\ \bibnamefont {Yang}}, \bibinfo
  {author} {\bibfnamefont {P.}~\bibnamefont {Krogstrup}}, \bibinfo {author}
  {\bibfnamefont {L.~P.}\ \bibnamefont {Kouwenhoven}}, \bibinfo {author}
  {\bibfnamefont {G.}~\bibnamefont {de~Lange}}, \bibinfo {author}
  {\bibfnamefont {B.}~\bibnamefont {van Heck}},\ and\ \bibinfo {author}
  {\bibfnamefont {A.}~\bibnamefont {Kou}},\ }\bibfield  {title} {\bibinfo
  {title} {Observation of vanishing charge dispersion of a nearly open
  superconducting island},\ }\bibfield  {journal} {\bibinfo  {journal} {Phys
  Rev Lett}\ }\textbf {\bibinfo {volume} {124}},\ \href
  {https://doi.org/10.1103/PhysRevLett.124.246802}
  {10.1103/PhysRevLett.124.246802} (\bibinfo {year} {2020})\BibitemShut
  {NoStop}%
\bibitem [{\citenamefont {Kringh\o{}j}\ \emph {et~al.}(2020)\citenamefont
  {Kringh\o{}j}, \citenamefont {Larsen}, \citenamefont {van Heck},
  \citenamefont {Sabonis}, \citenamefont {Erlandsson}, \citenamefont
  {Petkovic}, \citenamefont {Pikulin}, \citenamefont {Krogstrup}, \citenamefont
  {Petersson},\ and\ \citenamefont {Marcus}}]{Kringhoj2020b}%
  \BibitemOpen
  \bibfield  {author} {\bibinfo {author} {\bibfnamefont {A.}~\bibnamefont
  {Kringh\o{}j}}, \bibinfo {author} {\bibfnamefont {T.~W.}\ \bibnamefont
  {Larsen}}, \bibinfo {author} {\bibfnamefont {B.}~\bibnamefont {van Heck}},
  \bibinfo {author} {\bibfnamefont {D.}~\bibnamefont {Sabonis}}, \bibinfo
  {author} {\bibfnamefont {O.}~\bibnamefont {Erlandsson}}, \bibinfo {author}
  {\bibfnamefont {I.}~\bibnamefont {Petkovic}}, \bibinfo {author}
  {\bibfnamefont {D.~I.}\ \bibnamefont {Pikulin}}, \bibinfo {author}
  {\bibfnamefont {P.}~\bibnamefont {Krogstrup}}, \bibinfo {author}
  {\bibfnamefont {K.~D.}\ \bibnamefont {Petersson}},\ and\ \bibinfo {author}
  {\bibfnamefont {C.~M.}\ \bibnamefont {Marcus}},\ }\bibfield  {title}
  {\bibinfo {title} {Controlled {DC} monitoring of a superconducting qubit},\
  }\bibfield  {journal} {\bibinfo  {journal} {Phys Rev Lett}\ }\textbf
  {\bibinfo {volume} {124}},\ \href
  {https://doi.org/10.1103/physrevlett.124.056801}
  {10.1103/physrevlett.124.056801} (\bibinfo {year} {2020})\BibitemShut
  {NoStop}%
\bibitem [{\citenamefont {{Bargerbos}}\ \emph {et~al.}(2022)\citenamefont
  {{Bargerbos}}, \citenamefont {{Pita-Vidal}}, \citenamefont {{{\v{Z}}itko}},
  \citenamefont {{Splitthoff}}, \citenamefont {{Gr{\"u}nhaupt}}, \citenamefont
  {{Wesdorp}}, \citenamefont {{Liu}}, \citenamefont {{Kouwenhoven}},
  \citenamefont {{Aguado}}, \citenamefont {{Kraglund Andersen}}, \citenamefont
  {{Kou}},\ and\ \citenamefont {{van Heck}}}]{Bargerbos2022b}%
  \BibitemOpen
  \bibfield  {author} {\bibinfo {author} {\bibfnamefont {A.}~\bibnamefont
  {{Bargerbos}}}, \bibinfo {author} {\bibfnamefont {M.}~\bibnamefont
  {{Pita-Vidal}}}, \bibinfo {author} {\bibfnamefont {R.}~\bibnamefont
  {{{\v{Z}}itko}}}, \bibinfo {author} {\bibfnamefont {L.~J.}\ \bibnamefont
  {{Splitthoff}}}, \bibinfo {author} {\bibfnamefont {L.}~\bibnamefont
  {{Gr{\"u}nhaupt}}}, \bibinfo {author} {\bibfnamefont {J.~J.}\ \bibnamefont
  {{Wesdorp}}}, \bibinfo {author} {\bibfnamefont {Y.}~\bibnamefont {{Liu}}},
  \bibinfo {author} {\bibfnamefont {L.~P.}\ \bibnamefont {{Kouwenhoven}}},
  \bibinfo {author} {\bibfnamefont {R.}~\bibnamefont {{Aguado}}}, \bibinfo
  {author} {\bibfnamefont {C.}~\bibnamefont {{Kraglund Andersen}}}, \bibinfo
  {author} {\bibfnamefont {A.}~\bibnamefont {{Kou}}},\ and\ \bibinfo {author}
  {\bibfnamefont {B.}~\bibnamefont {{van Heck}}},\ }\bibfield  {title}
  {\bibinfo {title} {{Spectroscopy of spin-split Andreev levels in a quantum
  dot with superconducting leads}},\ }\href@noop {} {\bibfield  {journal}
  {\bibinfo  {journal} {arXiv e-prints}\ ,\ \bibinfo {eid} {arXiv:2208.09314}}
  (\bibinfo {year} {2022})},\ \Eprint {https://arxiv.org/abs/2208.09314}
  {arXiv:2208.09314} \BibitemShut {NoStop}%
\bibitem [{\citenamefont {{Wesdorp}}\ \emph {et~al.}(2022)\citenamefont
  {{Wesdorp}}, \citenamefont {{Matute-Ca{\v{n}}adas}}, \citenamefont
  {{Vaartjes}}, \citenamefont {{Gr{\"u}nhaupt}}, \citenamefont {{Laeven}},
  \citenamefont {{Roelofs}}, \citenamefont {{Splitthoff}}, \citenamefont
  {{Pita-Vidal}}, \citenamefont {{Bargerbos}}, \citenamefont {{van Woerkom}},
  \citenamefont {{Krogstrup}}, \citenamefont {{Kouwenhoven}}, \citenamefont
  {{Andersen}}, \citenamefont {{Levy Yeyati}}, \citenamefont {{van Heck}},\
  and\ \citenamefont {{de Lange}}}]{Wesdorp2022}%
  \BibitemOpen
  \bibfield  {author} {\bibinfo {author} {\bibfnamefont {J.~J.}\ \bibnamefont
  {{Wesdorp}}}, \bibinfo {author} {\bibfnamefont {F.~J.}\ \bibnamefont
  {{Matute-Ca{\v{n}}adas}}}, \bibinfo {author} {\bibfnamefont {A.}~\bibnamefont
  {{Vaartjes}}}, \bibinfo {author} {\bibfnamefont {L.}~\bibnamefont
  {{Gr{\"u}nhaupt}}}, \bibinfo {author} {\bibfnamefont {T.}~\bibnamefont
  {{Laeven}}}, \bibinfo {author} {\bibfnamefont {S.}~\bibnamefont {{Roelofs}}},
  \bibinfo {author} {\bibfnamefont {L.~J.}\ \bibnamefont {{Splitthoff}}},
  \bibinfo {author} {\bibfnamefont {M.}~\bibnamefont {{Pita-Vidal}}}, \bibinfo
  {author} {\bibfnamefont {A.}~\bibnamefont {{Bargerbos}}}, \bibinfo {author}
  {\bibfnamefont {D.~J.}\ \bibnamefont {{van Woerkom}}}, \bibinfo {author}
  {\bibfnamefont {P.}~\bibnamefont {{Krogstrup}}}, \bibinfo {author}
  {\bibfnamefont {L.~P.}\ \bibnamefont {{Kouwenhoven}}}, \bibinfo {author}
  {\bibfnamefont {C.~K.}\ \bibnamefont {{Andersen}}}, \bibinfo {author}
  {\bibfnamefont {A.}~\bibnamefont {{Levy Yeyati}}}, \bibinfo {author}
  {\bibfnamefont {B.}~\bibnamefont {{van Heck}}},\ and\ \bibinfo {author}
  {\bibfnamefont {G.}~\bibnamefont {{de Lange}}},\ }\bibfield  {title}
  {\bibinfo {title} {Microwave spectroscopy of interacting {A}ndreev spins},\
  }\href@noop {} {\bibfield  {journal} {\bibinfo  {journal} {arXiv e-prints}\ }
  (\bibinfo {year} {2022})},\ \Eprint {https://arxiv.org/abs/2208.11198}
  {arXiv:2208.11198} \BibitemShut {NoStop}%
\bibitem [{\citenamefont {Bargerbos}\ \emph {et~al.}(2022)\citenamefont
  {Bargerbos}, \citenamefont {Pita-Vidal}, \citenamefont
  {\ifmmode~\check{Z}\else \v{Z}\fi{}itko}, \citenamefont {\'Avila},
  \citenamefont {Splitthoff}, \citenamefont {Gr\"unhaupt}, \citenamefont
  {Wesdorp}, \citenamefont {Andersen}, \citenamefont {Liu}, \citenamefont
  {Kouwenhoven}, \citenamefont {Aguado}, \citenamefont {Kou},\ and\
  \citenamefont {van Heck}}]{Bargerbos2022}%
  \BibitemOpen
  \bibfield  {author} {\bibinfo {author} {\bibfnamefont {A.}~\bibnamefont
  {Bargerbos}}, \bibinfo {author} {\bibfnamefont {M.}~\bibnamefont
  {Pita-Vidal}}, \bibinfo {author} {\bibfnamefont {R.}~\bibnamefont
  {\ifmmode~\check{Z}\else \v{Z}\fi{}itko}}, \bibinfo {author} {\bibfnamefont
  {J.}~\bibnamefont {\'Avila}}, \bibinfo {author} {\bibfnamefont {L.~J.}\
  \bibnamefont {Splitthoff}}, \bibinfo {author} {\bibfnamefont
  {L.}~\bibnamefont {Gr\"unhaupt}}, \bibinfo {author} {\bibfnamefont {J.~J.}\
  \bibnamefont {Wesdorp}}, \bibinfo {author} {\bibfnamefont {C.~K.}\
  \bibnamefont {Andersen}}, \bibinfo {author} {\bibfnamefont {Y.}~\bibnamefont
  {Liu}}, \bibinfo {author} {\bibfnamefont {L.~P.}\ \bibnamefont
  {Kouwenhoven}}, \bibinfo {author} {\bibfnamefont {R.}~\bibnamefont {Aguado}},
  \bibinfo {author} {\bibfnamefont {A.}~\bibnamefont {Kou}},\ and\ \bibinfo
  {author} {\bibfnamefont {B.}~\bibnamefont {van Heck}},\ }\bibfield  {title}
  {\bibinfo {title} {Singlet-doublet transitions of a quantum dot josephson
  junction detected in a transmon circuit},\ }\href
  {https://doi.org/10.1103/PRXQuantum.3.030311} {\bibfield  {journal} {\bibinfo
   {journal} {PRX Quantum}\ }\textbf {\bibinfo {volume} {3}},\ \bibinfo {pages}
  {030311} (\bibinfo {year} {2022})}\BibitemShut {NoStop}%
\bibitem [{\citenamefont {Padurariu}\ and\ \citenamefont
  {Nazarov}(2010)}]{Padurariu2010}%
  \BibitemOpen
  \bibfield  {author} {\bibinfo {author} {\bibfnamefont {C.}~\bibnamefont
  {Padurariu}}\ and\ \bibinfo {author} {\bibfnamefont {Y.~V.}\ \bibnamefont
  {Nazarov}},\ }\bibfield  {title} {\bibinfo {title} {Theoretical proposal for
  superconducting spin qubits},\ }\bibfield  {journal} {\bibinfo  {journal}
  {Phys Rev B}\ }\textbf {\bibinfo {volume} {81}},\ \href
  {https://doi.org/10.1103/PhysRevB.81.144519} {10.1103/PhysRevB.81.144519}
  (\bibinfo {year} {2010})\BibitemShut {NoStop}%
\bibitem [{\citenamefont {Hays}\ \emph {et~al.}(2020)\citenamefont {Hays},
  \citenamefont {Fatemi}, \citenamefont {Serniak}, \citenamefont {Bouman},
  \citenamefont {Diamond}, \citenamefont {de~Lange}, \citenamefont {Krogstrup},
  \citenamefont {Nyg{\aa}rd}, \citenamefont {Geresdi},\ and\ \citenamefont
  {Devoret}}]{Hays2020}%
  \BibitemOpen
  \bibfield  {author} {\bibinfo {author} {\bibfnamefont {M.}~\bibnamefont
  {Hays}}, \bibinfo {author} {\bibfnamefont {V.}~\bibnamefont {Fatemi}},
  \bibinfo {author} {\bibfnamefont {K.}~\bibnamefont {Serniak}}, \bibinfo
  {author} {\bibfnamefont {D.}~\bibnamefont {Bouman}}, \bibinfo {author}
  {\bibfnamefont {S.}~\bibnamefont {Diamond}}, \bibinfo {author} {\bibfnamefont
  {G.}~\bibnamefont {de~Lange}}, \bibinfo {author} {\bibfnamefont
  {P.}~\bibnamefont {Krogstrup}}, \bibinfo {author} {\bibfnamefont
  {J.}~\bibnamefont {Nyg{\aa}rd}}, \bibinfo {author} {\bibfnamefont
  {A.}~\bibnamefont {Geresdi}},\ and\ \bibinfo {author} {\bibfnamefont {M.~H.}\
  \bibnamefont {Devoret}},\ }\bibfield  {title} {\bibinfo {title} {Continuous
  monitoring of a trapped superconducting spin},\ }\bibfield  {journal}
  {\bibinfo  {journal} {Nat Phys}\ }\textbf {\bibinfo {volume} {16}},\ \href
  {https://doi.org/10.1038/s41567-020-0952-3} {10.1038/s41567-020-0952-3}
  (\bibinfo {year} {2020})\BibitemShut {NoStop}%
\bibitem [{\citenamefont {Janvier}\ \emph {et~al.}(2015)\citenamefont
  {Janvier}, \citenamefont {Tosi}, \citenamefont {Bretheau}, \citenamefont
  {Girit}, \citenamefont {Stern}, \citenamefont {Bertet}, \citenamefont
  {Joyez}, \citenamefont {Vion}, \citenamefont {Esteve}, \citenamefont
  {Goffman}, \citenamefont {Pothier},\ and\ \citenamefont
  {Urbina}}]{Janvier2015}%
  \BibitemOpen
  \bibfield  {author} {\bibinfo {author} {\bibfnamefont {C.}~\bibnamefont
  {Janvier}}, \bibinfo {author} {\bibfnamefont {L.}~\bibnamefont {Tosi}},
  \bibinfo {author} {\bibfnamefont {L.}~\bibnamefont {Bretheau}}, \bibinfo
  {author} {\bibfnamefont {{\c{C}}.}~\bibnamefont {Girit}}, \bibinfo {author}
  {\bibfnamefont {M.}~\bibnamefont {Stern}}, \bibinfo {author} {\bibfnamefont
  {P.}~\bibnamefont {Bertet}}, \bibinfo {author} {\bibfnamefont
  {P.}~\bibnamefont {Joyez}}, \bibinfo {author} {\bibfnamefont
  {D.}~\bibnamefont {Vion}}, \bibinfo {author} {\bibfnamefont {D.}~\bibnamefont
  {Esteve}}, \bibinfo {author} {\bibfnamefont {M.~F.}\ \bibnamefont {Goffman}},
  \bibinfo {author} {\bibfnamefont {H.}~\bibnamefont {Pothier}},\ and\ \bibinfo
  {author} {\bibfnamefont {C.}~\bibnamefont {Urbina}},\ }\bibfield  {title}
  {\bibinfo {title} {Coherent manipulation of {A}ndreev states in
  superconducting atomic contacts},\ }\bibfield  {journal} {\bibinfo  {journal}
  {Science}\ }\textbf {\bibinfo {volume} {349}},\ \href
  {https://doi.org/10.1126/science.aab2179} {10.1126/science.aab2179} (\bibinfo
  {year} {2015})\BibitemShut {NoStop}%
\bibitem [{\citenamefont {Hays}\ \emph {et~al.}(2018)\citenamefont {Hays},
  \citenamefont {de~Lange}, \citenamefont {Serniak}, \citenamefont {van
  Woerkom}, \citenamefont {Bouman}, \citenamefont {Krogstrup}, \citenamefont
  {Nyg\aa{}rd}, \citenamefont {Geresdi},\ and\ \citenamefont
  {Devoret}}]{Hays2018}%
  \BibitemOpen
  \bibfield  {author} {\bibinfo {author} {\bibfnamefont {M.}~\bibnamefont
  {Hays}}, \bibinfo {author} {\bibfnamefont {G.}~\bibnamefont {de~Lange}},
  \bibinfo {author} {\bibfnamefont {K.}~\bibnamefont {Serniak}}, \bibinfo
  {author} {\bibfnamefont {D.~J.}\ \bibnamefont {van Woerkom}}, \bibinfo
  {author} {\bibfnamefont {D.}~\bibnamefont {Bouman}}, \bibinfo {author}
  {\bibfnamefont {P.}~\bibnamefont {Krogstrup}}, \bibinfo {author}
  {\bibfnamefont {J.}~\bibnamefont {Nyg\aa{}rd}}, \bibinfo {author}
  {\bibfnamefont {A.}~\bibnamefont {Geresdi}},\ and\ \bibinfo {author}
  {\bibfnamefont {M.~H.}\ \bibnamefont {Devoret}},\ }\bibfield  {title}
  {\bibinfo {title} {Direct microwave measurement of {A}ndreev-bound-state
  dynamics in a semiconductor-nanowire {J}osephson junction},\ }\bibfield
  {journal} {\bibinfo  {journal} {Phys Rev Lett}\ }\textbf {\bibinfo {volume}
  {121}},\ \href {https://doi.org/10.1103/PhysRevLett.121.047001}
  {10.1103/PhysRevLett.121.047001} (\bibinfo {year} {2018})\BibitemShut
  {NoStop}%
\bibitem [{\citenamefont {Hays}\ \emph {et~al.}(2021)\citenamefont {Hays},
  \citenamefont {Fatemi}, \citenamefont {Bouman}, \citenamefont {Cerrillo},
  \citenamefont {Diamond}, \citenamefont {Serniak}, \citenamefont {Connolly},
  \citenamefont {Krogstrup}, \citenamefont {Nyg{\aa}rd}, \citenamefont
  {Levy~Yeyati}, \citenamefont {Geresdi},\ and\ \citenamefont
  {Devoret}}]{Hays2021}%
  \BibitemOpen
  \bibfield  {author} {\bibinfo {author} {\bibfnamefont {M.}~\bibnamefont
  {Hays}}, \bibinfo {author} {\bibfnamefont {V.}~\bibnamefont {Fatemi}},
  \bibinfo {author} {\bibfnamefont {D.}~\bibnamefont {Bouman}}, \bibinfo
  {author} {\bibfnamefont {J.}~\bibnamefont {Cerrillo}}, \bibinfo {author}
  {\bibfnamefont {S.}~\bibnamefont {Diamond}}, \bibinfo {author} {\bibfnamefont
  {K.}~\bibnamefont {Serniak}}, \bibinfo {author} {\bibfnamefont
  {T.}~\bibnamefont {Connolly}}, \bibinfo {author} {\bibfnamefont
  {P.}~\bibnamefont {Krogstrup}}, \bibinfo {author} {\bibfnamefont
  {J.}~\bibnamefont {Nyg{\aa}rd}}, \bibinfo {author} {\bibfnamefont
  {A.}~\bibnamefont {Levy~Yeyati}}, \bibinfo {author} {\bibfnamefont
  {A.}~\bibnamefont {Geresdi}},\ and\ \bibinfo {author} {\bibfnamefont {M.~H.}\
  \bibnamefont {Devoret}},\ }\bibfield  {title} {\bibinfo {title} {Coherent
  manipulation of an {A}ndreev spin qubit},\ }\bibfield  {journal} {\bibinfo
  {journal} {Science}\ }\textbf {\bibinfo {volume} {373}},\ \href
  {https://doi.org/10.1126/science.abf0345} {10.1126/science.abf0345} (\bibinfo
  {year} {2021})\BibitemShut {NoStop}%
\bibitem [{\citenamefont {Wesdorp}\ \emph {et~al.}(2021)\citenamefont
  {Wesdorp}, \citenamefont {Gr{\"u}nhaupt}, \citenamefont {Vaartjes},
  \citenamefont {Pita-Vidal}, \citenamefont {Bargerbos}, \citenamefont
  {Splitthoff}, \citenamefont {Krogstrup}, \citenamefont {van Heck},\ and\
  \citenamefont {de~Lange}}]{Wesdorp2021}%
  \BibitemOpen
  \bibfield  {author} {\bibinfo {author} {\bibfnamefont {J.~J.}\ \bibnamefont
  {Wesdorp}}, \bibinfo {author} {\bibfnamefont {L.}~\bibnamefont
  {Gr{\"u}nhaupt}}, \bibinfo {author} {\bibfnamefont {A.}~\bibnamefont
  {Vaartjes}}, \bibinfo {author} {\bibfnamefont {M.}~\bibnamefont
  {Pita-Vidal}}, \bibinfo {author} {\bibfnamefont {A.}~\bibnamefont
  {Bargerbos}}, \bibinfo {author} {\bibfnamefont {L.~J.}\ \bibnamefont
  {Splitthoff}}, \bibinfo {author} {\bibfnamefont {P.}~\bibnamefont
  {Krogstrup}}, \bibinfo {author} {\bibfnamefont {B.}~\bibnamefont {van
  Heck}},\ and\ \bibinfo {author} {\bibfnamefont {G.}~\bibnamefont
  {de~Lange}},\ }\bibfield  {title} {\bibinfo {title} {Dynamical polarization
  of the fermion parity in a nanowire {J}osephson junction},\ }\href@noop {}
  {\bibfield  {journal} {\bibinfo  {journal} {arXiv e-prints}\ } (\bibinfo
  {year} {2021})},\ \Eprint {https://arxiv.org/abs/2112.01936} {2112.01936}
  \BibitemShut {NoStop}%
\bibitem [{\citenamefont {Jin}\ \emph {et~al.}(2015)\citenamefont {Jin},
  \citenamefont {Kamal}, \citenamefont {Sears}, \citenamefont {Gudmundsen},
  \citenamefont {Hover}, \citenamefont {Miloshi}, \citenamefont {Slattery},
  \citenamefont {Yan}, \citenamefont {Yoder}, \citenamefont {Orlando},
  \citenamefont {Gustavsson},\ and\ \citenamefont {Oliver}}]{Jin2015}%
  \BibitemOpen
  \bibfield  {author} {\bibinfo {author} {\bibfnamefont {X.~Y.}\ \bibnamefont
  {Jin}}, \bibinfo {author} {\bibfnamefont {A.}~\bibnamefont {Kamal}}, \bibinfo
  {author} {\bibfnamefont {A.~P.}\ \bibnamefont {Sears}}, \bibinfo {author}
  {\bibfnamefont {T.}~\bibnamefont {Gudmundsen}}, \bibinfo {author}
  {\bibfnamefont {D.}~\bibnamefont {Hover}}, \bibinfo {author} {\bibfnamefont
  {J.}~\bibnamefont {Miloshi}}, \bibinfo {author} {\bibfnamefont
  {R.}~\bibnamefont {Slattery}}, \bibinfo {author} {\bibfnamefont
  {F.}~\bibnamefont {Yan}}, \bibinfo {author} {\bibfnamefont {J.}~\bibnamefont
  {Yoder}}, \bibinfo {author} {\bibfnamefont {T.~P.}\ \bibnamefont {Orlando}},
  \bibinfo {author} {\bibfnamefont {S.}~\bibnamefont {Gustavsson}},\ and\
  \bibinfo {author} {\bibfnamefont {W.~D.}\ \bibnamefont {Oliver}},\ }\bibfield
   {title} {\bibinfo {title} {Thermal and residual excited-state population in
  a 3d transmon qubit},\ }\href
  {https://doi.org/10.1103/PhysRevLett.114.240501} {\bibfield  {journal}
  {\bibinfo  {journal} {Phys. Rev. Lett.}\ }\textbf {\bibinfo {volume} {114}},\
  \bibinfo {pages} {240501} (\bibinfo {year} {2015})}\BibitemShut {NoStop}%
\bibitem [{\citenamefont {Stockill}\ \emph {et~al.}(2016)\citenamefont
  {Stockill}, \citenamefont {Gall}, \citenamefont {Matthiesen}, \citenamefont
  {Huthmacher}, \citenamefont {Clarke}, \citenamefont {Hugues},\ and\
  \citenamefont {Atatüre}}]{Stockill2016}%
  \BibitemOpen
  \bibfield  {author} {\bibinfo {author} {\bibfnamefont {R.}~\bibnamefont
  {Stockill}}, \bibinfo {author} {\bibfnamefont {C.~L.}\ \bibnamefont {Gall}},
  \bibinfo {author} {\bibfnamefont {C.}~\bibnamefont {Matthiesen}}, \bibinfo
  {author} {\bibfnamefont {L.}~\bibnamefont {Huthmacher}}, \bibinfo {author}
  {\bibfnamefont {E.}~\bibnamefont {Clarke}}, \bibinfo {author} {\bibfnamefont
  {M.}~\bibnamefont {Hugues}},\ and\ \bibinfo {author} {\bibfnamefont
  {M.}~\bibnamefont {Atatüre}},\ }\bibfield  {title} {\bibinfo {title}
  {Quantum dot spin coherence governed by a strained nuclear environment},\
  }\bibfield  {journal} {\bibinfo  {journal} {Nat Commun}\ }\textbf {\bibinfo
  {volume} {7}},\ \href {https://doi.org/10.1038/ncomms12745}
  {10.1038/ncomms12745} (\bibinfo {year} {2016})\BibitemShut {NoStop}%
\bibitem [{\citenamefont {Krogstrup}\ \emph {et~al.}(2015)\citenamefont
  {Krogstrup}, \citenamefont {Ziino}, \citenamefont {Chang}, \citenamefont
  {Albrecht}, \citenamefont {Madsen}, \citenamefont {Johnson}, \citenamefont
  {Nyg{\aa}rd}, \citenamefont {Marcus},\ and\ \citenamefont
  {Jespersen}}]{Krogstrup2015}%
  \BibitemOpen
  \bibfield  {author} {\bibinfo {author} {\bibfnamefont {P.}~\bibnamefont
  {Krogstrup}}, \bibinfo {author} {\bibfnamefont {N.~L.~B.}\ \bibnamefont
  {Ziino}}, \bibinfo {author} {\bibfnamefont {W.}~\bibnamefont {Chang}},
  \bibinfo {author} {\bibfnamefont {S.~M.}\ \bibnamefont {Albrecht}}, \bibinfo
  {author} {\bibfnamefont {M.~H.}\ \bibnamefont {Madsen}}, \bibinfo {author}
  {\bibfnamefont {E.}~\bibnamefont {Johnson}}, \bibinfo {author} {\bibfnamefont
  {J.}~\bibnamefont {Nyg{\aa}rd}}, \bibinfo {author} {\bibfnamefont
  {C.}~\bibnamefont {Marcus}},\ and\ \bibinfo {author} {\bibfnamefont {T.~S.}\
  \bibnamefont {Jespersen}},\ }\bibfield  {title} {\bibinfo {title} {Epitaxy of
  semiconductor-superconductor nanowires},\ }\bibfield  {journal} {\bibinfo
  {journal} {Nat. Mater.}\ }\textbf {\bibinfo {volume} {14}},\ \href
  {https://doi.org/10.1038/nmat4176} {10.1038/nmat4176} (\bibinfo {year}
  {2015})\BibitemShut {NoStop}%
\bibitem [{\citenamefont {Malinowski}\ \emph {et~al.}(2017)\citenamefont
  {Malinowski}, \citenamefont {Martins}, \citenamefont {Nissen}, \citenamefont
  {Fallahi}, \citenamefont {Gardner}, \citenamefont {Manfra}, \citenamefont
  {Marcus},\ and\ \citenamefont {Kuemmeth}}]{Malinowski2017}%
  \BibitemOpen
  \bibfield  {author} {\bibinfo {author} {\bibfnamefont {F.~K.}\ \bibnamefont
  {Malinowski}}, \bibinfo {author} {\bibfnamefont {F.}~\bibnamefont {Martins}},
  \bibinfo {author} {\bibfnamefont {P.~D.}\ \bibnamefont {Nissen}}, \bibinfo
  {author} {\bibfnamefont {S.}~\bibnamefont {Fallahi}}, \bibinfo {author}
  {\bibfnamefont {G.~C.}\ \bibnamefont {Gardner}}, \bibinfo {author}
  {\bibfnamefont {M.~J.}\ \bibnamefont {Manfra}}, \bibinfo {author}
  {\bibfnamefont {C.~M.}\ \bibnamefont {Marcus}},\ and\ \bibinfo {author}
  {\bibfnamefont {F.}~\bibnamefont {Kuemmeth}},\ }\bibfield  {title} {\bibinfo
  {title} {Symmetric operation of the resonant exchange qubit},\ }\bibfield
  {journal} {\bibinfo  {journal} {Phys Rev B}\ }\textbf {\bibinfo {volume}
  {96}},\ \href {https://doi.org/10.1103/PhysRevB.96.045443}
  {10.1103/PhysRevB.96.045443} (\bibinfo {year} {2017})\BibitemShut {NoStop}%
\bibitem [{\citenamefont {Nadj-Perge}\ \emph {et~al.}(2010)\citenamefont
  {Nadj-Perge}, \citenamefont {Frolov}, \citenamefont {Bakkers},\ and\
  \citenamefont {Kouwenhoven}}]{NadjPerge2010}%
  \BibitemOpen
  \bibfield  {author} {\bibinfo {author} {\bibfnamefont {S.}~\bibnamefont
  {Nadj-Perge}}, \bibinfo {author} {\bibfnamefont {S.~M.}\ \bibnamefont
  {Frolov}}, \bibinfo {author} {\bibfnamefont {E.~P. A.~M.}\ \bibnamefont
  {Bakkers}},\ and\ \bibinfo {author} {\bibfnamefont {L.~P.}\ \bibnamefont
  {Kouwenhoven}},\ }\bibfield  {title} {\bibinfo {title}
  {Spin{\textendash}orbit qubit in a semiconductor nanowire},\ }\bibfield
  {journal} {\bibinfo  {journal} {Nature}\ }\textbf {\bibinfo {volume} {468}},\
  \href {https://doi.org/10.1038/nature09682} {10.1038/nature09682} (\bibinfo
  {year} {2010})\BibitemShut {NoStop}%
\bibitem [{\citenamefont {van~den Berg}\ \emph {et~al.}(2013)\citenamefont
  {van~den Berg}, \citenamefont {Nadj-Perge}, \citenamefont {Pribiag},
  \citenamefont {Plissard}, \citenamefont {Bakkers}, \citenamefont {Frolov},\
  and\ \citenamefont {Kouwenhoven}}]{vandenBerg2013}%
  \BibitemOpen
  \bibfield  {author} {\bibinfo {author} {\bibfnamefont {J.~W.~G.}\
  \bibnamefont {van~den Berg}}, \bibinfo {author} {\bibfnamefont
  {S.}~\bibnamefont {Nadj-Perge}}, \bibinfo {author} {\bibfnamefont {V.~S.}\
  \bibnamefont {Pribiag}}, \bibinfo {author} {\bibfnamefont {S.~R.}\
  \bibnamefont {Plissard}}, \bibinfo {author} {\bibfnamefont {E.~P. A.~M.}\
  \bibnamefont {Bakkers}}, \bibinfo {author} {\bibfnamefont {S.~M.}\
  \bibnamefont {Frolov}},\ and\ \bibinfo {author} {\bibfnamefont {L.~P.}\
  \bibnamefont {Kouwenhoven}},\ }\bibfield  {title} {\bibinfo {title} {Fast
  spin-orbit qubit in an indium antimonide nanowire},\ }\bibfield  {journal}
  {\bibinfo  {journal} {Phys Rev Lett}\ }\textbf {\bibinfo {volume} {110}},\
  \href {https://doi.org/10.1103/PhysRevLett.110.066806}
  {10.1103/PhysRevLett.110.066806} (\bibinfo {year} {2013})\BibitemShut
  {NoStop}%
\bibitem [{\citenamefont {Landig}\ \emph {et~al.}(2019)\citenamefont {Landig},
  \citenamefont {Koski}, \citenamefont {Scarlino}, \citenamefont {M{\"u}ller},
  \citenamefont {Abadillo-Uriel}, \citenamefont {Kratochwil}, \citenamefont
  {Reichl}, \citenamefont {Wegscheider}, \citenamefont {Coppersmith},
  \citenamefont {Friesen}, \citenamefont {Wallraff}, \citenamefont {Ihn},\ and\
  \citenamefont {Ensslin}}]{Landig2019}%
  \BibitemOpen
  \bibfield  {author} {\bibinfo {author} {\bibfnamefont {A.~J.}\ \bibnamefont
  {Landig}}, \bibinfo {author} {\bibfnamefont {J.~V.}\ \bibnamefont {Koski}},
  \bibinfo {author} {\bibfnamefont {P.}~\bibnamefont {Scarlino}}, \bibinfo
  {author} {\bibfnamefont {C.}~\bibnamefont {M{\"u}ller}}, \bibinfo {author}
  {\bibfnamefont {J.~C.}\ \bibnamefont {Abadillo-Uriel}}, \bibinfo {author}
  {\bibfnamefont {B.}~\bibnamefont {Kratochwil}}, \bibinfo {author}
  {\bibfnamefont {C.}~\bibnamefont {Reichl}}, \bibinfo {author} {\bibfnamefont
  {W.}~\bibnamefont {Wegscheider}}, \bibinfo {author} {\bibfnamefont {S.~N.}\
  \bibnamefont {Coppersmith}}, \bibinfo {author} {\bibfnamefont
  {M.}~\bibnamefont {Friesen}}, \bibinfo {author} {\bibfnamefont
  {A.}~\bibnamefont {Wallraff}}, \bibinfo {author} {\bibfnamefont
  {T.}~\bibnamefont {Ihn}},\ and\ \bibinfo {author} {\bibfnamefont
  {K.}~\bibnamefont {Ensslin}},\ }\bibfield  {title} {\bibinfo {title}
  {Virtual-photon-mediated spin-qubit--transmon coupling},\ }\bibfield
  {journal} {\bibinfo  {journal} {Nat Commun}\ }\textbf {\bibinfo {volume}
  {10}},\ \href {https://doi.org/10.1038/s41467-019-13000-z}
  {10.1038/s41467-019-13000-z} (\bibinfo {year} {2019})\BibitemShut {NoStop}%
\end{thebibliography}%

\end{document}


\beginsupplement

\title{Supplementary information for ``Direct manipulation of a superconducting spin qubit strongly coupled to a transmon qubit''}

\author{Marta Pita-Vidal}
\thanks{These two authors contributed equally.}
\affiliation{QuTech and Kavli Institute of Nanoscience, Delft University of Technology, 2600 GA Delft, The Netherlands}

\author{Arno Bargerbos}
\thanks{These two authors contributed equally.}
\affiliation{QuTech and Kavli Institute of Nanoscience, Delft University of Technology, 2600 GA Delft, The Netherlands}

\author{Rok Žitko}
\affiliation{Jožef Stefan Institute, Jamova 39, SI-1000 Ljubljana, Slovenia}
\affiliation{Faculty of Mathematics and Physics, University of Ljubljana, Jadranska 19, SI-1000 Ljubljana, Slovenia}

\author{Lukas J. Splitthoff}
\affiliation{QuTech and Kavli Institute of Nanoscience, Delft University of Technology, 2600 GA Delft, The Netherlands}

\author{Lukas Grünhaupt}
\affiliation{QuTech and Kavli Institute of Nanoscience, Delft University of Technology, 2600 GA Delft, The Netherlands}

\author{Jaap J. Wesdorp}
\affiliation{QuTech and Kavli Institute of Nanoscience, Delft University of Technology, 2600 GA Delft, The Netherlands}

\author{Yu Liu}
\affiliation{Center for Quantum Devices, Niels Bohr Institute, University of Copenhagen, 2100 Copenhagen, Denmark}

\author{Leo P. Kouwenhoven}
\affiliation{QuTech and Kavli Institute of Nanoscience, Delft University of Technology, 2600 GA Delft, The Netherlands}

\author{Ramón Aguado}
\affiliation{Instituto de Ciencia de Materiales de Madrid (ICMM),
Consejo Superior de Investigaciones Cientificas (CSIC), Sor Juana Ines de la Cruz 3, 28049 Madrid, Spain}

\author{Bernard van Heck}
\affiliation{Leiden Institute of Physics, Leiden University, Niels Bohrweg 2, 2333 CA Leiden, The Netherlands}

\author{Angela Kou}
\affiliation{Department of Physics and Frederick Seitz Materials Research Laboratory,
University of Illinois Urbana-Champaign, Urbana, IL 61801, USA}

\author{Christian Kraglund Andersen}
\affiliation{QuTech and Kavli Institute of Nanoscience, Delft University of Technology, 2600 GA Delft, The Netherlands}

\date{\today}

\maketitle

\tableofcontents

\vspace{2 cm}

\newpage

\section{Modeling of joint ASQ-transmon system} \label{Ss:ASQ-transmon}
\subsection{Numerical diagonalization}
\label{Ss:diagonalization}
In order to obtain the transition frequencies of the joint ASQ-transmon system, we combine the Hamiltonian of the ASQ [Eq.~(1) in the main text] with the Hamiltonian of the transmon, as indicated in Eq.~(2) of the main text. This combined Hamiltonian is numerically diagonalized in the phase basis following the procedure in Refs.~\cite{Bargerbos2020, Kringhoj2020b}. This results in the transmon and ASQ energy levels $E_n$, as well as the associated transition frequencies $f_{nm} = \left(E_m - E_n\right)/h$. These frequencies are used in Figs.~1 and 4 to fit the spectroscopy measurements.

\subsection{Estimate of qubit-qubit coupling strength}
As demonstrated in Fig.~4, we observe avoided crossings between the transmon and the ASQ transitions, which is indicative of strong coherent coupling. In this section we derive how the coupling strength depends on the model parameters. 

We start by combining Eq.~(1) and (2) of the main text into the effective Hamiltonian
\begin{equation}
H_{\rm tot} = H_{\rm tmon} + H_{\rm Z} + H_{\rm coupling},
\end{equation}
with the individual terms given as
\begin{align} 
H_{\rm tmon} &= -4 E_{\rm c} \partial_\phi^2 - E_{\rm J} \cos{ ( \phi )} - E_0 \cos{\left(\phi-\phi_{\rm ext}\right)}, \\
 H_{\rm Z}  &= \frac{1}{2} \begin{pmatrix} 
                                E_{\rm Z}^{\rm \perp} & E_{\rm Z}^{\rm \parallel} \\
                                E_{\rm Z}^{\rm \parallel} & - E_{\rm Z}^{\rm \perp}
                                \end{pmatrix} = \frac{|\vec{E}_{\rm Z}|}{2} \begin{pmatrix} 
                                \sin{(\theta)} & \cos{(\theta)} \\
                                \cos{(\theta)}  & - \sin{(\theta)} 
                                \end{pmatrix} \\
 H_{\rm coupling} &= - E_{\rm SO} \sin(\phi-\phi_{\rm ext}) \sigma_x .
\end{align}
Here, $\sigma_x$ is the $x$ Pauli matrix and $\theta$ is the angle between the Zeeman field $\vec{E}_{\rm Z}$ and the spin-orbit direction, such that $E_{\rm Z}^\parallel = |\vec{E}_{\rm Z}| \cos{\theta}$ and $E_{\rm Z}^\perp = |\vec{E}_{\rm Z}| \sin{\theta}$. Next, we write the coupling term $H_{\rm coupling}$ in the eigenbasis of $H_{\rm Z}$, which is given by the states $\ket{v_1} = \left( \cos{(\theta/2)}, \sin{(\theta/2)} \right)$ and $\ket{v_2} = \left( -\sin{(\theta/2)}, \cos{(\theta/2)} \right)$. We identify that
\begin{align}\bra{v_1} \sigma_x \ket{v_2} = \bra{v_2} \sigma_x \ket{v_1} = \cos{\theta}
\end{align}
and 
\begin{align}\bra{v_1} \sigma_x \ket{v_1} = - \bra{v_2} \sigma_x \ket{v_2} = \sin{\theta},
\end{align}
such that $\sigma_x$ becomes $\cos{(\theta)} \sigma_{\overline{x}} + \sin{(\theta)} \sigma_{\overline{z}}$ in the $\{ \ket{v_1}, \ket{v_2} \}$ spin basis. 

We rewrite $H_{\rm coupling}$ and expand to first order in $\phi$, valid in the transmon limit $E_{\rm J}\gg E_{\rm c}$, where $\langle \phi \rangle \ll 1$, which results in 
\begin{align}
H_{\rm coupling} &\,=  E_{\rm SO}\left[\cos{\left(\phi\right)}\sin{\left(\phi_{\rm ext}\right)} - \cos{\left(\phi_{\rm ext}\right)} \sin{\left(\phi\right)}\right] \sigma_x \\
&\,\approx E_{\rm SO} \left[ \sin{(\phi_{\rm ext})} - \phi \cos{(\phi_{\rm ext})}  \right] \sigma_x.
\end{align}
Therefore, in the spin eigenbasis, we obtain
\begin{equation}
H_{\rm coupling} \approx E_{\rm SO} \left[ \sin{(\phi_{\rm ext})} - \phi \cos{(\phi_{\rm ext})}  \right] \left[\cos{(\theta)} \sigma_{\overline{x}} + \sin{(\theta)} \sigma_{\overline{z}}\right]. \label{eq:Hcoupling}
\end{equation}
This term of the Hamiltonian couples the ASQ to the transmon via the phase operator $\phi$ of the transmon and is, thus, reminiscent of a dipole coupling. In the transmon regime, we can express the operator $\phi$ in terms of the zero point fluctuations of the phase, $\phi_{\rm zpf}$, and the bosonic creation and annihilation transmon operators, $c^\dagger$ and $c$ respectively:  $\phi =  \phi_{\rm zpf} \left(c^\dagger + c \right)$. Inserting this operator into Eq.~\eqref{eq:Hcoupling}, we obtain
\begin{align}
H_{\rm coupling} &\,\approx \left[ E_{\rm SO} \sin{(\phi_{\rm ext})} -E_{\rm SO}  \phi_{\rm zpf} \left(c^\dagger + c \right) \cos{(\phi_{\rm ext})} \right]  \left[\cos{(\theta)} \sigma_{\overline{x}} + \sin{(\theta)} \sigma_{\overline{z}} \right] \\
&\,=  E_{\rm SO} \sin{(\phi_{\rm ext})} \left[\cos{(\theta)} \sigma_{\overline{x}} + \sin{(\theta)} \sigma_{\overline{z}} \right]+ \hbar J_{\overline{x}}\left(c^\dagger + c \right) \sigma_{\overline{x}} + \hbar J_{\overline{z}}\left(c^\dagger + c \right) \sigma_{\overline{z}}.
\end{align}
In this expression, we have the transversal coupling with the strength $\hbar J_{\overline{x}} = E_{\rm SO} \cos{(\phi_{\rm ext})} \phi_{\rm zpf} \cos{(\theta)}$ and a longitudinal coupling with the strength $\hbar J_{\overline{z}} = E_{\rm SO} \cos{(\phi_{\rm ext})} \phi_{\rm zpf} \sin{(\theta)}$. 

From fitting the spectroscopy data we find a charging energy of $E_{\rm c}/h =$~\SI{284}{MHz} and a Josephson energy of $E_{\rm J}/h =$~\SI{13.1}{GHz}, which results in  $\phi_{\rm zpf} = [2E_{\rm c}/E_{\rm J, eff}(\phi_{\rm ext})]^{1/4} \le 0.46$ where 
\begin{equation}
    E_{\rm J, eff}(\phi_{\rm ext})=(E_{\rm J}+E_0)\sqrt{\cos^2(\phi_{\rm ext}) + \left(\frac{E_{\rm J}-E_0}{E_{\rm J}+E_0}\right)^2\sin^2(\phi_{\rm ext})}.
\end{equation} 
For $E_{\rm SO}/h=$~\SI{309}{MHz}, this results on a transverse coupling of up to  $J_{\overline{x}}/(2\pi) =$~\SI{145}{MHz} when \flux~=~0 and the magnetic field is applied perpendicular to the spin-orbit direction. In the fit of Fig.~4 we instead find an avoided crossing of $2J_{\overline{x}}/(2\pi)=2\times 52 $~MHz, corresponding to a Zeeman field at an angle of $\theta=$~\SI{35.6}{\degree} with respect to the spin-orbit direction.

\newpage

\section{Device and experimental setup}
\label{Sss:device-setup}

\subsection{Device overview}

\begin{figure}[h!]
    \center
        \includegraphics[scale=1.0]{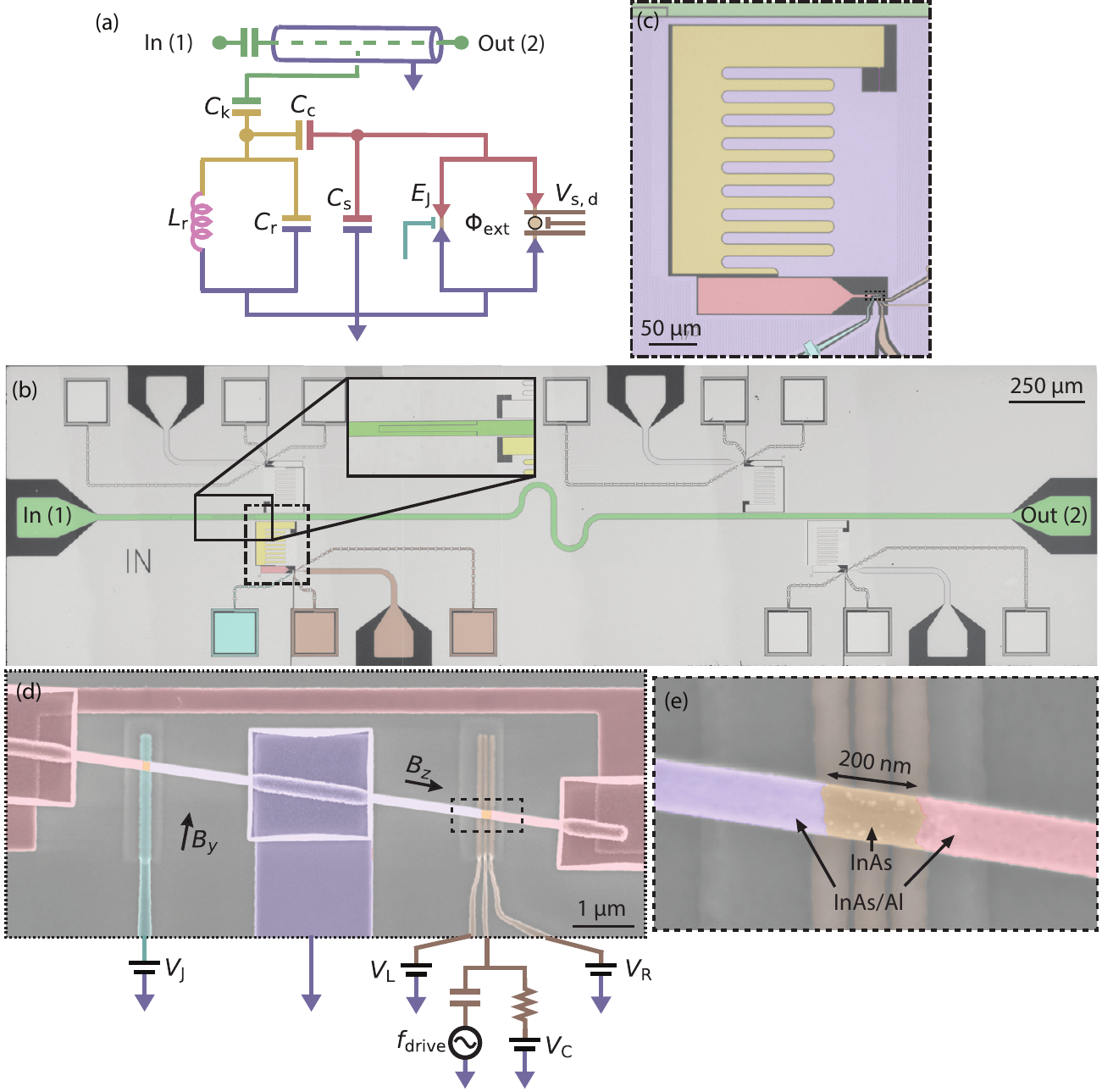}    
        \caption{{ \bf Device overview.}  (a) Diagram of the microwave circuit. A coplanar waveguide transmission line with an input capacitor (green center conductor) is capacitively coupled to a grounded LC resonator. The resonator consists of an island (yellow) capacitively and inductively  (pink) shunted to ground (blue). The resonator is in turn capacitively coupled to a transmon island (red), which is shunted to ground capacitively as well as via two parallel Josephson junctions.     
        (b) Chip containing four nearly identical devices coupled to the same transmission line, which has a capacitor at its input port, enlarged in the inset. 
        (c) False-colored optical microscope image of the device showing the qubit island, the resonator island, the resonator inductor, the transmission line, the electrostatic gates and ground. 
        (d) False-colored scanning electron micrograph (SEM) of a device comparable to that measured, showing the InAs/Al nanowire into which the junctions are defined. The $B_y$ component of the magnetic field is used to tune $\Phi_{\rm ext}$ \cite{Wesdorp2022}. $B_z$ is the magnetic field component parallel to the nanowire. 
        (e) False-colored SEM of a nearly identical device, showing the junction in which the quantum dot is gate defined. The three bottom gates have a width and spacing of \SI{40}{nm}, although this is obfuscated by the dielectric layer placed on top.
        }
        
    \label{fig:device}
\end{figure}

Fig.~\ref{fig:device} shows an overview of the device including the different elements forming the superconducting circuit used for readout and control of the qubits. The device under investigation in this work is the same as the one used in Ref.~\cite{Bargerbos2022b}, where further details about its physical implementation and fabrication can be found.

\subsection{Cryogenic and room temperature measurement setup}

\begin{figure}[h!]
    \center
    \includegraphics[scale=0.58]{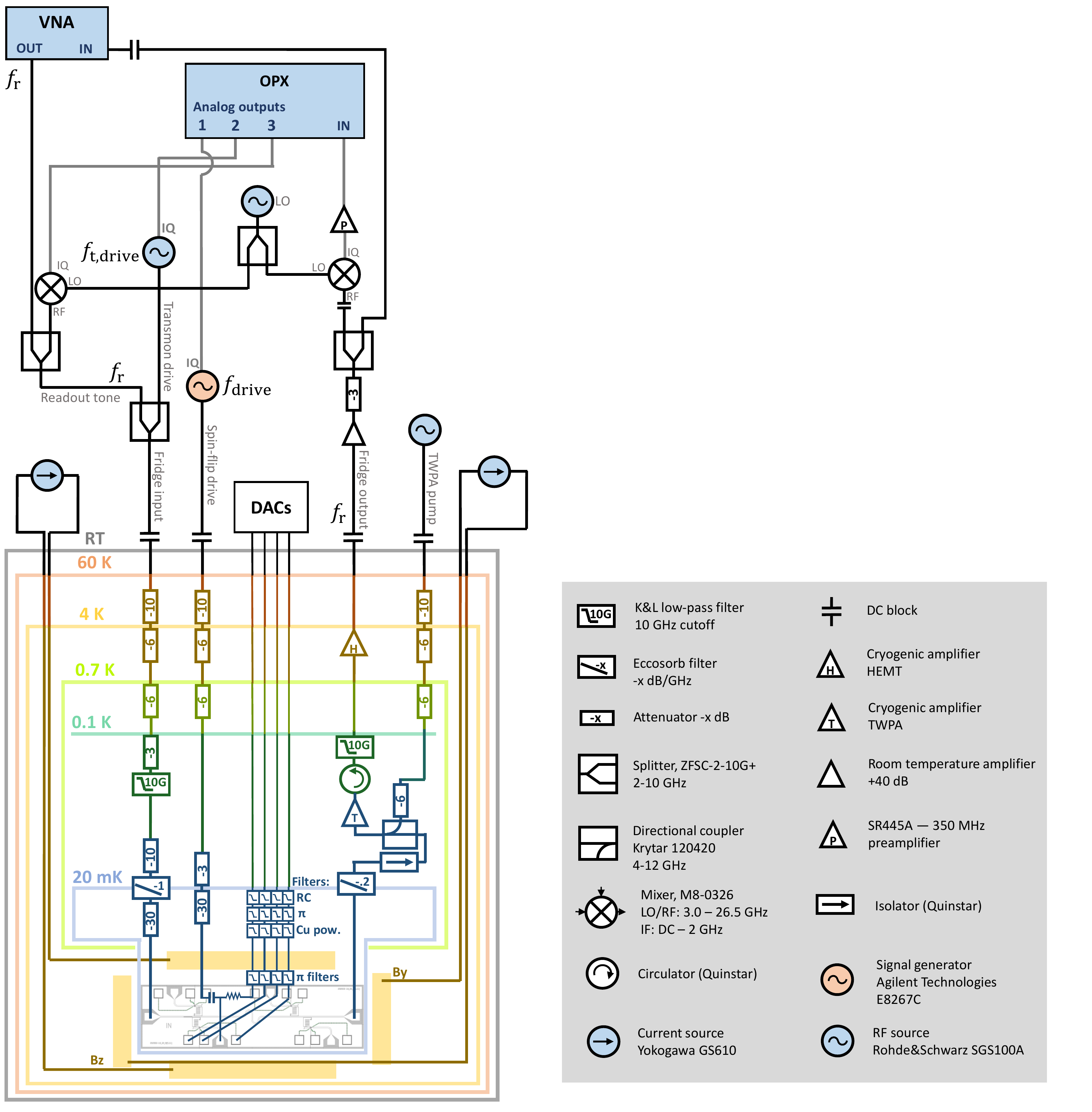}
    \caption{Measurement setup at cryogenic and room temperatures.}
    \label{fig:cryogenic_setup}
\end{figure}

The device was measured in a Triton dilution refrigerator with a base temperature of $\approx$~\SI{20}{mK}. Details of the wiring at room and cryogenic temperatures are shown in Fig.~\ref{fig:cryogenic_setup}. The setup contains an input radio-frequency (RF) line, an output RF line, an extra RF line for the spin-flip drive tone and multiple direct current (DC) lines, used to tune the electrostatic gate voltages. The DC gate lines are filtered at base temperature with multiple low-pass filters connected in series. 
The input and drive RF lines contain attenuators and low-pass filters at different temperature stages, as indicated. In turn, the output RF line contains amplifiers at different temperature stages: a travelling wave parametric amplifier (TWPA) at the mixing chamber plate ($\approx$~\SI{20}{mK}),  a high-electron-mobility transistor (HEMT) amplifier at the \SI{4}{K} stage, and an additional amplifier at room temperature. 
A three-axis vector magnet (x-axis not shown) is thermally anchored to the \SI{4}{K} temperature stage, with the device under study mounted at its center. The three magnet coils are  controlled with Yokogawa GS610 current sources. 
At room temperature, a vector network analyzer (VNA) is connected to the input and output RF lines for spectroscopy at frequency $f_{\rm r}$. On the input line, this signal is then combined with the IQ-modulated transmon drive tone at frequency $f_{\rm t, drive}$. A separate IQ-modulated tone at $f_{\rm r}$, only used for time-domain measurements, is also combined onto this line. The IQ-modulated spin-flip drive tone at frequency $f_{\rm drive}$ is sent through the drive line. For time-domain measurements the output signal is additionally split off into a separate branch and down-converted to be measured with a Quantum Machines OPX.

\subsection{Basic characterization and tune up}\label{Sss:tuneup}

The basic characterization and tune-up of the device proceeds such as detailed in Ref.~\cite{Bargerbos2022}, while the specific tune-up of the quantum dot resonance investigated in this device is detailed in the supplement of Ref.~\cite{Bargerbos2022b}, where it is labeled as resonance A. A brief summary is as follows: We first characterize the gate dependence of the reference junction with the dot fully closed, and fix \Vj~such that $E_{\rm J} \gg \sqrt{E_0^2+E_{\rm SO}^2}$, to ensure the phase drop set by \flux~happens mostly at the quantum dot junction. Furthermore, we choose \Ej~such that the transmon frequency $f_{\rm t}$ is close to the readout resonator frequency $ \approx 6.11$~GHz to obtain a large dispersive shift for two-tone spectroscopy and qubit readout. For the results shown in this work, we used \Vj~=~\SI{3860}{mV}. We then investigate the gate dependence of the quantum dot junction with the reference junction fully closed, determining the pinchoff voltages of the three quantum dot gates. Next, we open the reference junction to its gate set-point and explore the quantum dot junction gate space at both \flux~$=0$ and \flux~$=\pi$ to identify regions that show a $\pi$-shift in phase. For a given $\pi$-shifted region, we measure explicit \flux~dependence of the transmon to identify a resonance with a spin splitting comparable to the spin-independent Josephson energy. Finally, we choose a gate set-point in the selected resonance. For the results shown here, the setpoint chosen for the three quantum dot gates was \Vl~=~\SI{363}{mV},  \Vc~=~\SI{1000}{mV} and \Vc~=~\SI{81}{mV}, which corresponds to $V_{\rm T, A}$~=~\SI{-423.6}{mV},  $V_{\rm P, A}$~=~\SI{909.5}{mV} in the rotated gate frame shown in Fig.~S7 in Ref.~\cite{Bargerbos2022b}.

\newpage

\section{Extended dataset}

\subsection{Extended two-tone spectroscopy data}
\label{Ss:extended2tone}

\begin{figure}[h!]
    \center
    \includegraphics[scale=1]{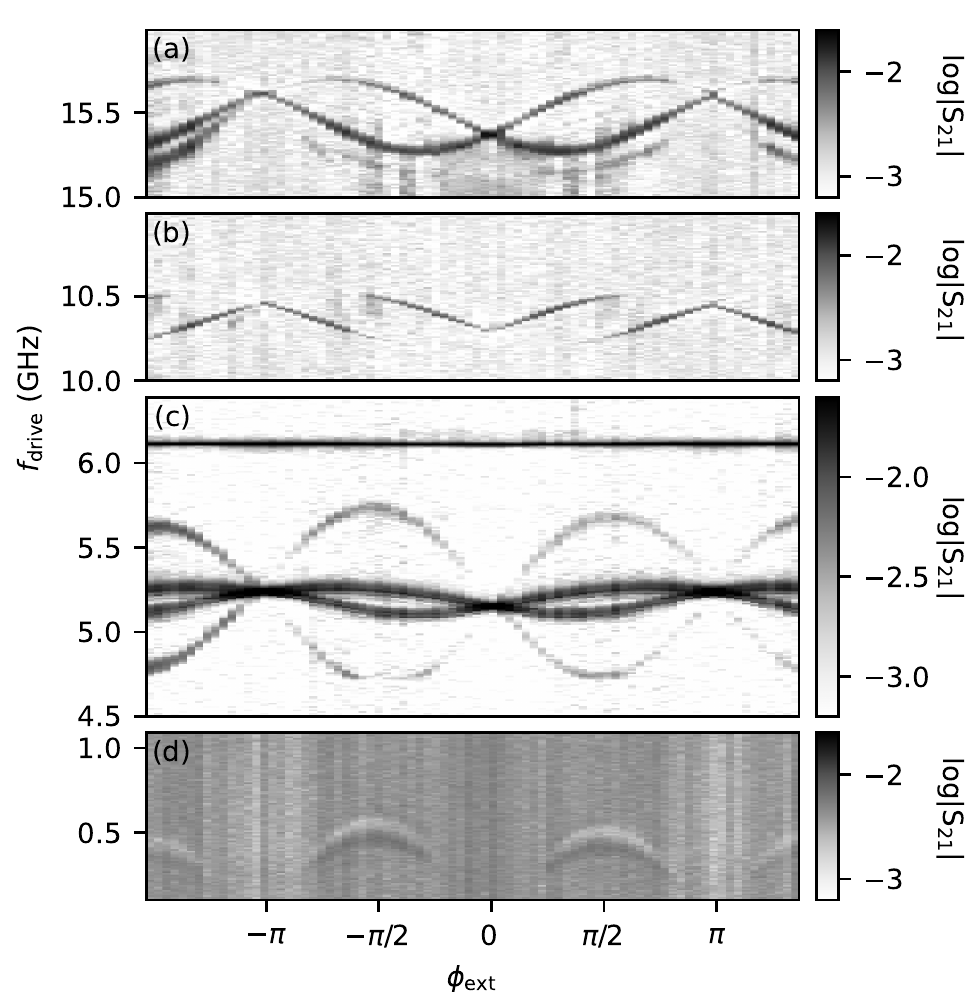}
    \caption{Normalized two-tone spectroscopy measurement of the transition spectrum versus external flux. Input power at the top of the spin-flip drive line is \SI{-36}{dBm} for (a-b), \SI{-46}{dBm} for (c) and \SI{-6}{dBm} for (d).}
    \label{fig:20GHz}
\end{figure}

Fig.~\ref{fig:20GHz} shows extended two-tone spectroscopy measurements at the setpoint of main text Fig.~4(b), performed over a range of \SI{20}{GHz}. It reveals several additional transition frequencies: panels (a) and (b) contain the higher-lying transmon transitions $f_{03}$ and $f_{02}$, respectively, while panel (c) shows five different transitions. These are the four transitions also shown in Fig.~4(b) and, above that, the resonator transition frequency. Panel (d) exhibits two low-frequency transitions: the bright top transition is the direct spin-flip transition with the transmon in its ground state, while the dark lower transition results from the direct spin-flip transition with the transmon in its excited state. The latter transition is visible as a result of a residual excited state population of the transmon. No other auxiliary transitions are found between 0 and \SI{20}{GHz}, nor does any transition develop for magnetic fields up to \SI{65}{mT}. We further note that the measurement of panel (d) requires a large drive power (31~dBm more than for the measurement shown in Fig.~1 of the main text), and that visibility is reduced compared to panel (c), which is expected since the matrix elements for the EDSR driving is suppressed in the absence of an external magnetic field.

\subsection{Single shot assignment fidelity}
\begin{figure}[h!]
    \center
    \includegraphics[scale=1]{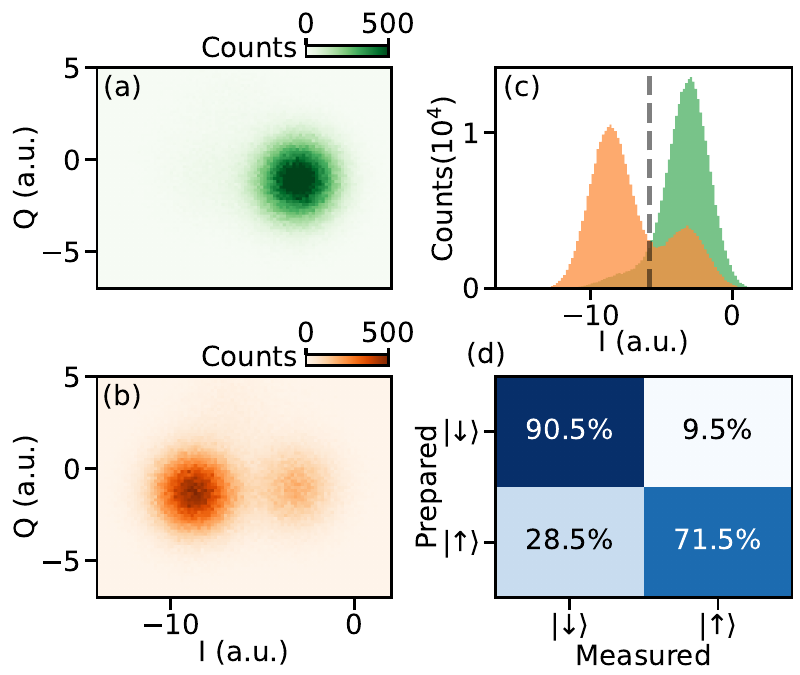}
    \caption{Single shot assignment fidelity. (a) Histogram in the complex plane of $3 \times 10^5$ sequential shots separated by \SI{200}{\micro s} and integrated for \SI{500}{ns} in the absence of an excitation pulse. (b) Same as (a) in the presence of a $\pi$-pulse with a FWHM of \SI{8}{ns} preceding each shot. 
    (c) Histograms of the I-quadrature response of the preceding panels. Green and orange colors correspond to panels (a) and (b), respectively. (d) Extracted single-shot fidelity's based on the threshold indicated in (c) with a gray dashed line. }
    \label{fig:fidelity}
\end{figure}

The time domain measurements in the main text are obtained by averaging over many shots. We now estimate the assignment fidelity of ASQ readout at the setpoint used for the coherence measurements in the main text ($B_z=$~\SI{65}{mT} and \flux~=~$3\pi/2$). To do so, we measure the IQ quadrature response of the readout resonator for the qubit prepared in the ground state [Fig.~\ref{fig:fidelity}(a)] and for the qubit prepared in the excited state [Fig.~\ref{fig:fidelity}(b)], after applying an 8-ns $\pi$-pulse. In both cases we read out for \SI{500}{ns}, more than 40 times shorter than $T_1$, and wait for $5 T_1$ between different measurements to let the qubit decay back to its ground state. We find an assignment fidelity of $F = 1 - (P(\downarrow|\uparrow) - P(\uparrow|\downarrow))/2=$~80\% [Fig.~\ref{fig:fidelity}(d)], where $P(a|b)$ denotes the probability of measuring the qubit to be in state $a$ after preparing it in state $b$. The fidelity is predominantly set by assignment errors for the excited state limited by decoherence during the excitation as the $\pi$-pulse duration is comparable to $T_2$. Longer readout times therefore do not significantly improve the assignment fidelity. However, shorter $\pi$ pulses would likely lead to improved performance although this experiment was not performed on the current device. 

\subsection{Parity lifetime}
\label{Ss:parity}
One of the advantages of using a quantum dot junction over a semiconducting weak link is that the charging energy of the quantum dot allows us to select an operational setpoint for which the doublet states are the lowest energy states of the system \cite{Padurariu2010, Bargerbos2022}. Therefore, the charging energy is expected to protect against qubit leakage via quasiparticle escape or recombination, which would take the junction outside of the computational space of the qubit. To confirm this protection, we measure the quasiparticle poisoning times of the junction around the gate setpoint used in the main text. 

\begin{figure}[h!]
    \center
    \includegraphics[scale=1]{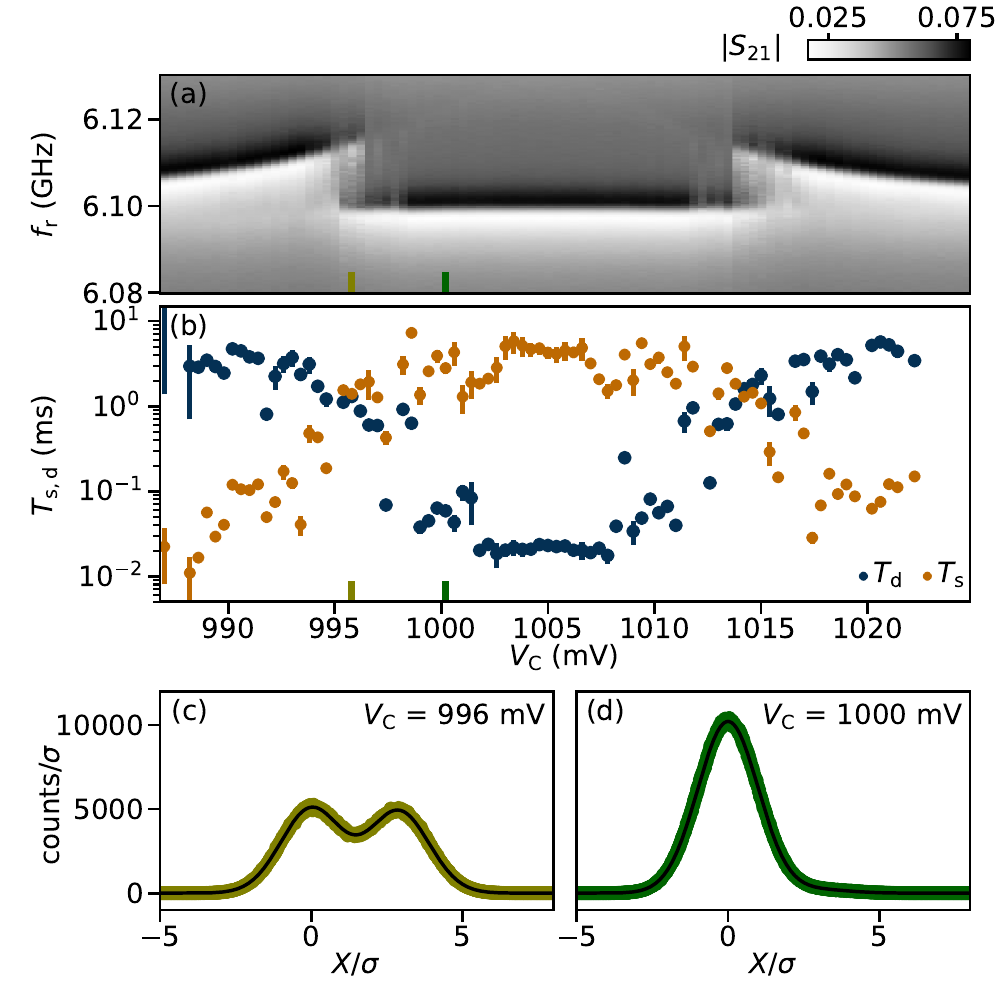}
    \caption{Gate dependence of parity lifetimes. 
    (a) \Vc~dependence of $|S_{21}|$ at \flux~=~0. 
    (b) \Vc~dependence of the extracted lifetimes. Markers indicate the mean while error bars indicate the maximum and minimum values of 10 consecutive time traces.
    (c) In yellow, 1D histogram of a continuously measured \SI{17}{s}-long time trace integrated in time bins of $t_{\rm int}$~=~\SI{4.3}{\micro s}, at \Vc~=~\SI{996}{mV}. In black, best fit to a double Gaussian shape.
    (d) Same as (c) but at \Vc~=~\SI{1000}{mV}.
    }
    \label{fig:QPP}
\end{figure}

Shown in Fig.~\ref{fig:QPP}(a), two resonances are visible as the central quantum dot gate \Vc~is varied around its setpoint \Vc~=~\SI{1000}{mV}, at \flux~=~0. Following the methods of Ref.~\cite{Bargerbos2022}, we identify the outer two \Vc~regions as having a singlet ground state (spin-zero) and the central region as having a doublet ground state (spin-$1/2$). For each gate point, we subsequently monitor the transmon circuit in real time and determine the switching time of the quantum dot junction parity. $T_{\rm s}$ and $T_{\rm d}$ denote the characteristic times for which the quantum dot maintains a singlet or doublet occupation, respectively. The extracted times are shown in Fig.~\ref{fig:QPP}(b). Note that this measurement is performed at \flux~=~0, where the $\ket{\uparrow}$ and $\ket{\downarrow}$  states result in equal transmon frequencies, thus becoming indistinguishable using our readout scheme \cite{Bargerbos2022b}. The spin-flip times $T_{\rm spin}$ are therefore not resolved here, as opposed to the experiments of Ref.~\cite{Hays2020}.

We observe that, for the outer two regions, where the ground state is the spin-0 state, the doublet switching time $T_{\rm d}$ ranges from a few $\upmu$s to hundreds of $\upmu$s, but is always much shorter than the singlet switching time $T_{\rm s}$. Close to the singlet-doublet ground state transition, both times become similar and of the order of \SI{1}{ms}, which can be seen in Fig.~\ref{fig:QPP}(c) for \Vc~=~\SI{996}{mV}, where the histogram of a continuous time trace, integrated in time bins of $t_{\rm int}$~=~\SI{4.3}{\micro \second}, shows two Gaussians with equal amplitudes. In the central region, where the doublet states are the lowest energy states, the situation is reversed and, away from the singlet-doublet transition, $T_{\rm d}$ is consistently above \SI{1}{ms}. The imbalance between average singlet and doublet occupation is shown in Fig.~\ref{fig:QPP}(d) for the setpoint used in the main text, \Vc~=~\SI{1000}{mV}. In this case we measure $T_{\rm s}=$~\SI{59}{\micro s} and $T_{\rm d}=$~\SI{2.8}{ms}. The latter is much larger than that of weak-link junctions, typically found to be in the range 10-500~$\upmu$s~\cite{Janvier2015, Hays2018, Hays2020, Hays2021, Wesdorp2021}, and thus demonstrates the advantage of using a quantum dot junction. In particular, for the weak-link ASQ \cite{Hays2021} the authors measured a parity lifetime $T_{\rm parity} =$~\SI{22}{\micro s} and a spin-flip time $T_{\rm spin} =$~\SI{17}{\micro s}, such that the parity lifetime was a relevant limitation to the qubit $T_1$. In contrast, we find that $T_{\rm d}\gg T_1$ such that the lifetime of the ASQ studied in this work is not limited by parity switches.

\subsection{Excited state population}
Similar to what is found in previous works investigating the doublet states of SNS junctions \cite{Hays2020, Hays2021, Wesdorp2022}, we observe that both $\ket{\uparrow}$ and $\ket{\downarrow}$ of the quantum dot junction are occupied at $B_z=$~\SI{0}{mT}, even in the absence of a drive. As such, we observe simultaneously both of the transmon branches corresponding to each spin state [see main text Fig.~4(b)]. We hypothesize that this residual excited state population is the result of excitations of either thermal or non-equilibrium origin, as the maximum zero-field ASQ transition frequency $f_{\rm s}\approx$~\SI{600}{MHz} corresponds to an effective temperature scale of $T_{\rm eff}\approx$~\SI{30}{\milli K}, below the typical electron temperatures found in transport and transmon \cite{Jin2015} experiments, \SIrange{35}{100}{\milli K}. 

\begin{figure}[h!]
    \center
    \includegraphics[scale=1]{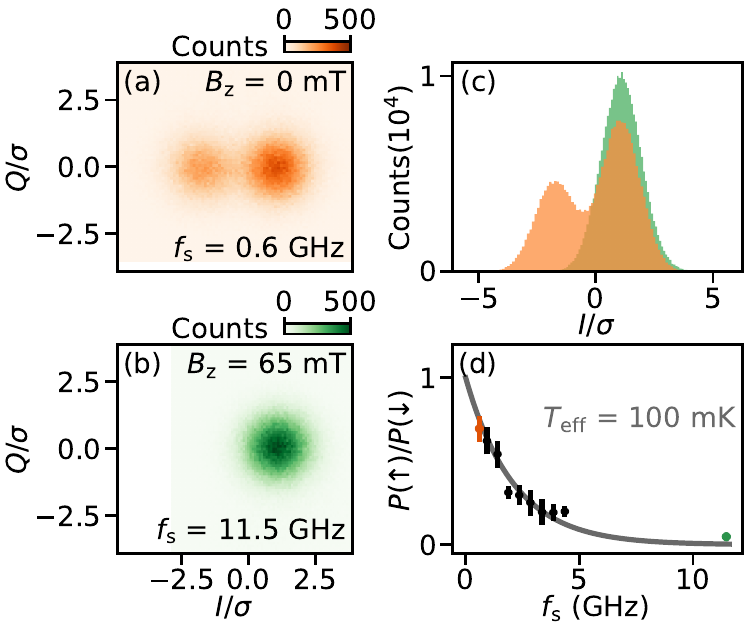}
    \caption{
    Excited state population of the spin states.
    (a) Histogram in the complex plane of $2.5\times10^5$ sequential shots, integrated for \SI{500}{ns} in the absence of an excitation pulse. Measured at $B_z = \SI{0}{mT}$ and $f_{\rm s}=$~\SI{0.6}{GHz}.
    (b) Same as (a) at $B_z = \SI{65}{mT}$ and $f_{\rm s}=$~\SI{11.5}{GHz}.
    (c) Histograms of the I-quadrature response of the preceding panels (a in orange, b in green). 
    (d) Extracted excited state population versus spin qubit frequency $f_{\rm s}$, as tuned with the magnetic field $B_z$. Data (markers) are fit with a Boltzmann equation (see text) resulting in an effective temperature of \SI{100}{mK}.
    }
    \label{fig:thermal}
\end{figure}

To investigate the residual population further, we monitor the transmon circuit in real-time, now at $\phi_{\rm ext}=3\pi/2$ so that we are maximally sensitive to changes in the spin state. At $B_z=$~\SI{0}{mT}, the IQ histogram of $2.5\times10^5$ sequential measurements confirms the presence of two populated states, as shown in Fig.~\ref{fig:thermal}(a). From a double Gaussian fit, we extract a ratio of state occupations of $P({\uparrow})/P({\downarrow})=$~0.7. Upon increasing the qubit frequency $f_{\rm s}$ with the magnetic field $B_z$, we find that the excited state population is strongly reduced, in line with expectation [Fig.~\ref{fig:thermal}(d)]. However, the ASQ frequency first crosses the transmon and then the resonator frequencies between 20 and \SI{30}{mT}, preventing the measurement of the spin states occupancy over a range of frequencies. Measuring again at $B_z=$~\SI{65}{mT}, where $f_{\rm s}=$~\SI{11.53}{GHz}, we find at most 4\% remaining excited state population, see Fig.~\ref{fig:thermal}(b). Here, the remaining excited state population is expected to be predominantly due to assignment errors, similar to those found in Fig.~\ref{fig:fidelity}(a). 

To extract the effective temperature of the ASQ, we subsequently fit the frequency dependence of the ratio of populations to a Boltzmann distribution, $P({\uparrow})/P({\downarrow})= \exp \left(-hf_{\rm s}/(k_{\rm B}T_{\rm eff}) \right)$, where $h$ and $k_{\rm B}$ are the Planck and Boltzmann constants, respectively. This leads to reasonable agreement with the data, resulting in an effective temperature of $T_{\rm eff}=100$~$\pm$~\SI{8}{mK} [see Fig.~\ref{fig:thermal}(d)].

\subsection{CP data}
 \label{Sss:CP}

\begin{figure}[h!]
    \center
    \includegraphics[scale=1]{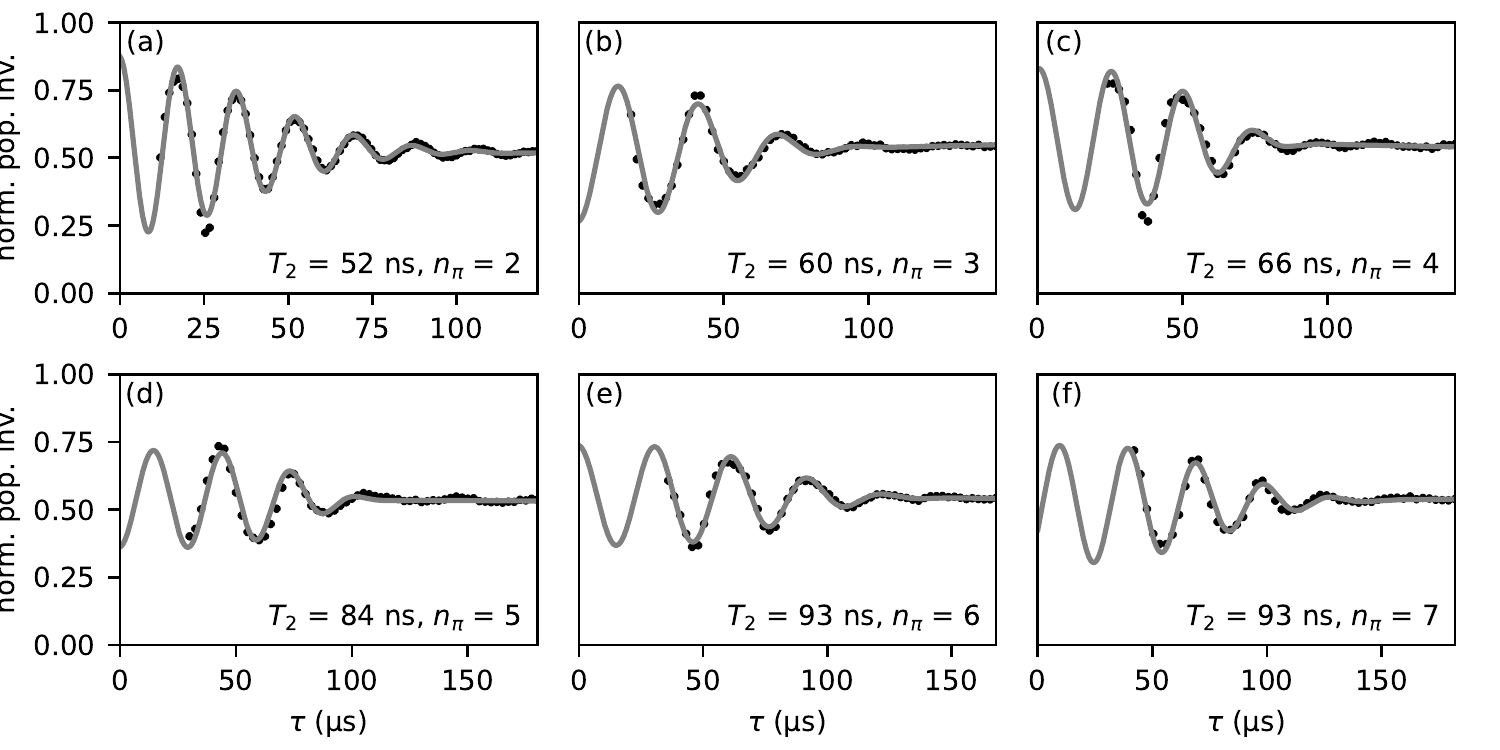}
    \caption{Extended CP experiment data. Solid lines indicate fits to the data (see text). All data is normalized to the visibility of a preceding Rabi oscillation measurement, and the data is obtained using a $\pi$-pulse ($\pi/2$-pulse) with a FWHM of \SI{4}{ns} (\SI{2}{ns}). The oscillations are introduced into the decay by adding a phase proportional to the delay time for the final $\pi/2$-pulse.}
    \label{fig:CP}
\end{figure}

In this section we provide further data for the CP measurements shown in Fig.~3(d) in the main text. As discussed, the CP sequence is constructed as follows: for each $n_\pi$, we apply a $\pi/2$-pulse, followed by $n_\pi$ equidistant $\pi$-pulses and a final $\pi/2$-pulse. All pulses are composed of a Gaussian envelope and have a FWHM of \SI{2}{ns} and \SI{4}{ns} for the $\pi/2$- and $\pi$-pulses, respectively. The separation between the centers of consecutive $\pi$-pulses is $\tau/n_\pi$ and the separation between a $\pi/2$ pulse and its nearest $\pi$ pulse is $\tau/(2n_\pi)$, resulting in a total delay time $\tau$ between the center of the two $\pi/2$ pulses. Fig.~\ref{fig:CP} shows CP measurements for $n_\pi$ values ranging from $2$ to $7$, accompanied by a fit to the expression
\begin{equation}\label{eq:CP}
a \cos{\left(\tau \Omega -\phi\right)} \exp{\left(-\left(\tau/T_{2}\right)^{d+1} \right)} + c + e\tau,
\end{equation} 
from which we extract the $T_2(n_\pi)$ values reported in Fig.~3(d). Note that the maximum waveform generator output power puts a limit on the minimum delay time $\tau$ for which the sequence can be generated, as the Gaussian pulses overlap for short delay times compared to the pulse width. This results in the absence of data for short $\tau$ in Fig.~\ref{fig:CP}.


\subsection{Transmon qubit coherence} \label{Sss:transmon}
\begin{figure}[H]
    \centering
    \includegraphics[scale=1.0]{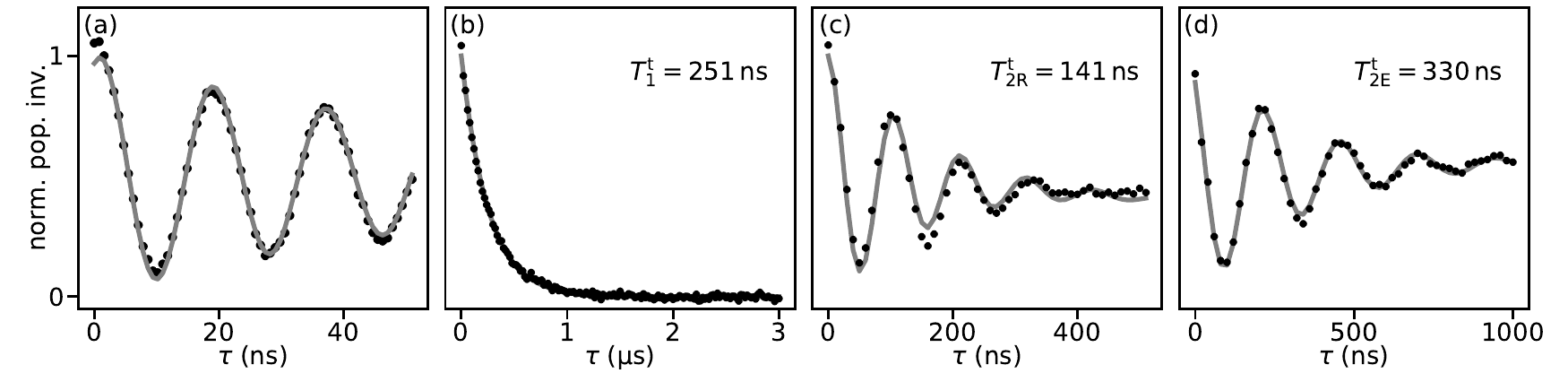} \caption{Coherence of the transmon qubit at $B_z=$~\SI{65}{mT}. (a) Rabi oscillations, (b) qubit lifetime, (c) Ramsey and (d) Hahn-echo experiments. Solid lines indicate fits to the data. For (c-d) oscillations are introduced into the decay by adding a phase proportional to the delay time for the final $\pi/2$-pulse. We plot the normalized population inversion, where each sub-panel is individually normalized to the resulting fit.}
    \label{fig:coherence_sup}
\end{figure}

We characterize the transmon performance at the flux and gate bias point used in the main text using standard time-domain techniques, see Fig.~\ref{fig:coherence_sup}. 

\subsection{ASQ coherence versus control parameters} \label{Sss:coherence-parameters}
In this section we provide additional data showing the dependence of the ASQ lifetime and coherence times on different control parameters. They are extracted by fitting their respective time evolutions using the same expressions employed in Fig.~3 of the main text: 
\begin{align} 
T_1: &\, \hspace{0.5 cm} a\exp{\left(t/T_{1}\right)} + c \\ 
T_{\rm 2R}: &\,  \hspace{0.5 cm} a \cos{\left(t \Omega -\phi\right)} \exp{\left(-\left(t / T_{\rm 2 R}\right)^{d+1}\right)} + c \\
T_{\rm 2E}: &\, \hspace{0.5  cm} a \cos{\left(t \Omega -\phi\right)} \exp{\left(-\left(t / T_{\rm 2 E}\right)^{d+1}\right)} + c + et
\end{align} 
Here, $a$, $c$, $d$, $e$, $\phi$, $\Gamma$, $T_1$, $T_{\rm 2R}$ and $T_{\rm 2E}$ are fit parameters. For $T_{\rm 2R}$ and $T_{\rm 2E}$, $\Omega$ accounts for the combination of detuning and the oscillations introduced by adding a phase proportional to the delay time for the final $\pi/2$-pulse.


\subsubsection{ASQ lifetime versus magnetic field} 
\begin{figure}[h!]
    \center
    \includegraphics[scale=1]{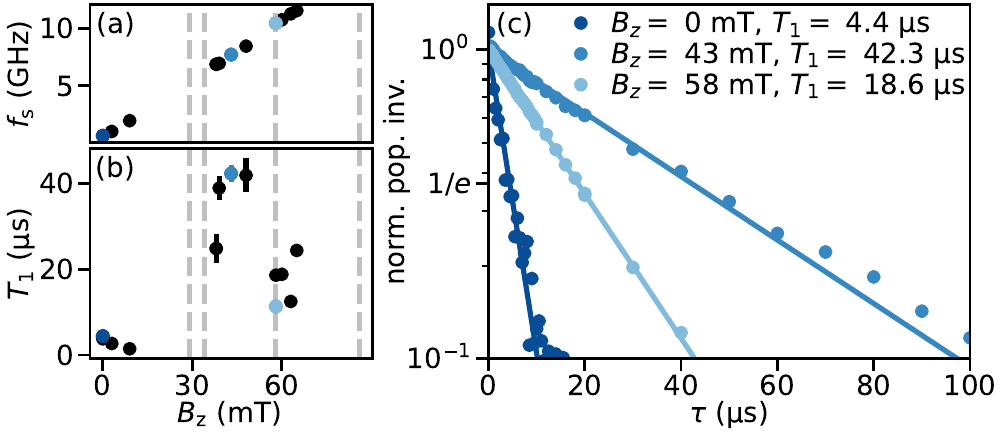}
    \caption{(a) Spin qubit frequency, $f_{\rm s}$, as a function of magnetic field, $B_z$. (b) Spin qubit lifetime, $T_1$ as a function of magnetic field. Dashed lines in (a-b) indicate the magnetic fields at which $f_{\rm s}$ crosses the first three transmon frequencies $f_{0j}$ and the resonator frequency [c.f. Fig.~\ref{fig:20GHz}]. (c) Representative qubit lifetime measurements, fit with an exponential decay.}
    \label{fig:T1_field}
\end{figure}
We start by investigating the evolution of the ASQ lifetime $T_1$ versus magnetic fields between 0 and \SI{65}{mT}. As shown in Fig.~\ref{fig:T1_field}(b), the qubit lifetime varies strongly, from around \SI{1}{\micro s} close to zero magnetic field and up to \SI{40}{\micro s} at intermediate fields, before once-more decreasing to approximately \SI{20}{\micro s}. For intermediate magnetic fields between \SI{15}{mT} and \SI{35}{mT}, the measurement of the qubit lifetime is hindered by the vicinity to the transmon and resonator transition frequencies. In this region it is not possible to drive the ASQ independently as, due to the capacitance between the gate drive line and the transmon island,  the transmon qubit is also excited.  This simultaneous driving of both qubits impedes the distinction of the response coming from each of them. 

The strong reduction of $T_1$ at low fields is potentially due to resonant exchange with the nuclear spins in InAs \cite{Stockill2016}; given the large $g$-factor of the ASQ, this process only takes places at low magnetic fields.  This is supported by the finding that at elevated magnetic fields, in the range \SIrange{45}{50}{mT}, we find the ASQ lifetime to exceed \SI{40}{\micro  s}. At even higher fields we observe a drop of the lifetime to around \SI{20}{\micro s}. As discussed in the main text, we conjecture the ASQ lifetime found in these regimes is limited by Purcell-like decay from coupling to the transmon, given the short transmon lifetime of around \SI{250}{ns} [Fig.~\ref{fig:coherence_sup}(b)].

\begin{figure}[h!]
    \center
    \includegraphics[scale=1]{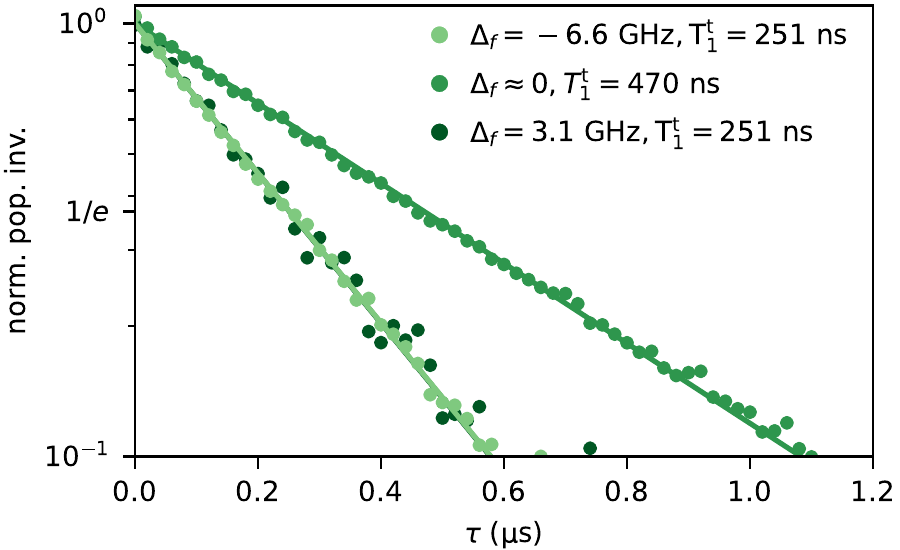}
    \caption{Transmon qubit lifetime $T_1^{\rm t}$ as a function of detuning $\Delta_f = f_{\rm{t}}-f_{\rm s}$ from the spin qubit frequency as tuned with the magnetic field. Detunings $\Delta_f = $~3.1, 0 and \SI{-6.6}{GHz} correspond to  $B_z =$~9, 28 and \SI{65}{mT}, respectively.
    }
    \label{fig:trasmonT1}
\end{figure}

To support the assertion that the reduction in the ASQ lifetime for qubit frequencies in the proximity of the transmon transitions is due to Purcell-like decay, we investigate whether the transmon lifetime is enhanced by proximity to the ASQ. Fig.~\ref{fig:trasmonT1} shows the transmon lifetime $T_1^{\rm t}$ for three different detunings between transmon and ASQ. When the qubits are detuned from each other, we measure $T_1^{\rm t} \approx$~\SI{250}{ns}. However, when the transmon is resonant with the ASQ, its lifetime is enhanced by almost factor of two, reaching 470~$\pm$~\SI{5}{ns}. This is consistent with hybridization of the two qubits, given that $T_1^{\rm s} \gg T_1^{\rm t}$, and supports that the lifetime of the ASQ can be decreased by vicinity to the transmon modes. These findings furthermore compliment the the observations discussed surrounding main text Fig.~4, serving as an additional signature of coherent coupling.

\subsubsection{Independence of ASQ coherence on gate voltages, magnetic field and flux}
We investigate the effect of different sources of noise by measuring the dependence of the $T_{2 \rm R}$ and $T_{2 \rm E}$ coherence times on gate voltage, magnetic field, and flux.

\begin{figure}[h!]
    \center
    \includegraphics[scale=1]{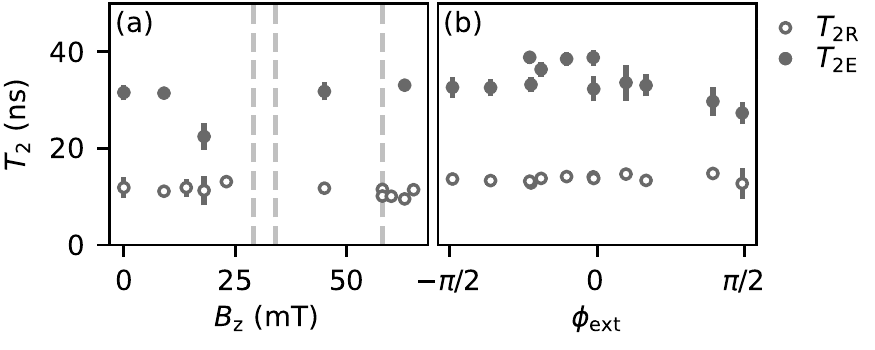}
    \caption{Dependence of the spin qubit coherence on the external magnetic field (a) and the external flux (b). The dashed lines in (a) indicate the magnetic fields at which $f_{\rm s}$ crosses the  transmon transition frequencies $f_{01}$ and $f_{02}$ as well as the resonator frequency.}
    \label{fig:coherence_flux}
\end{figure}

The $B_z$ dependence of coherence times is shown in Fig.~\ref{fig:coherence_flux}(a), for which we do not observe a measurable dependence over the $B_z$ range investigated. Therefore charge noise is likely not the dominant contribution to qubit dephasing since, if it was the case, an increase in $B_z$ would increase the effectiveness of EDSR at coupling charge noise to the qubit, which would result on a reduction of the decoherence times. In contrast, this $B_z$-independence of coherence times is compatible with nuclear magnetic noise being a strong contribution to qubit dephasing; due to the small magnetic moment of the nuclei spin, a magnetic fields of \SI{65}{mT} do not yet lead to a significant nuclear splitting. As a result of this we do not reach the regime of strong nuclear spin polarization, such that the precession of the nuclear bath in the external fields still leads to a significant Overhauser field for the range of fields explored. Additionally, the Overhauser field could have a field-independent component originating from the quadrupolar coupling of the nuclei to electric field gradients, induced by strain in the nanowire \cite{Krogstrup2015, Stockill2016}. A more complete understanding of the system will require further investigation. 

Next, we consider the dependence of coherence times on the external flux $\phi_{\rm ext}$. As shown in Fig.~\ref{fig:coherence_flux}(b), we again do not find a pronounced dependence of the coherence times. In particular, we do not observe an increase of the $T_2$ times  near the sweet spots at \flux~$=\pm \pi/2$. From this we conclude that flux noise does not strongly contribute to dephasing.

\begin{figure}[h!]
    \center
    \includegraphics[scale=1]{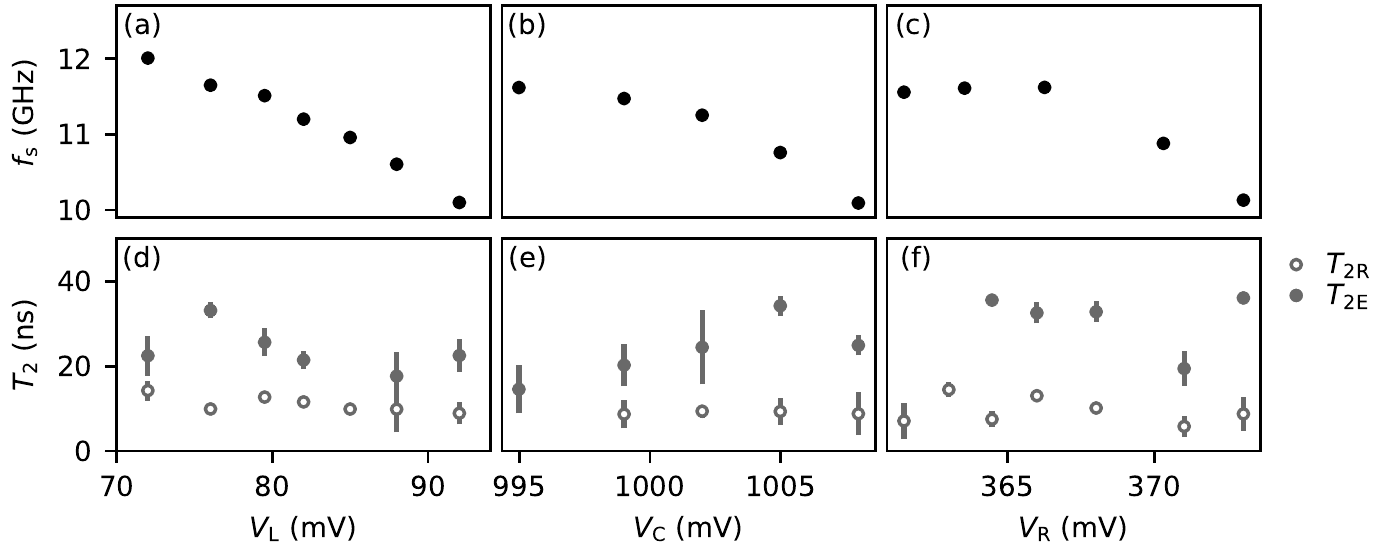}
    \caption{Dependence of the spin qubit coherence on the three quantum dot gates. (a-c) Spin qubit frequency versus gate voltage. (d-f) Ramsey and Hahn echo $T_2$ times versus gate voltage.}
    \label{fig:coherence_gate}
\end{figure}

Finally we investigate the dependence of coherence times on the voltages applied to the three gate electrodes situated underneath the quantum dot junction [see Fig.~\ref{fig:device}(e)]. As shown in Fig.~\ref{fig:coherence_gate}, we do not find a clear correlation between $T_{2 \rm R}$ or $T_{2 \rm E}$ and the slope of the qubit frequency versus any of the three gate voltages. This indicates that  voltage noise also does not provide a large contribution to the dephasing rate. However, although we measure $T_2$ in the vicinity of the available sweet spots of the individual gate electrodes, we did not find a simultaneous sweet spot for all three quantum dot gates, and the effect of voltage noise cannot be entirely ruled out. Further investigation of the qubit's susceptibility to voltage and magnetic noise based on the Rabi decay times are discussed in the next section, Sec.~\ref{Sss:noise}.

\subsection{Estimating the amplitude of charge and magnetic noise fluctuations}
\label{Sss:noise}
\begin{figure}[b!]
    \center
    \includegraphics[scale=1]{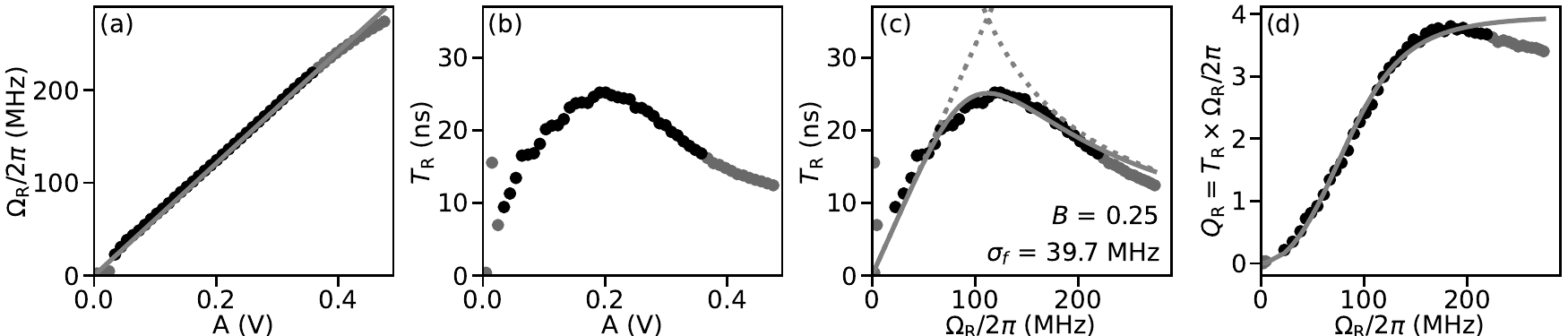}
    \caption{
    (a) Rabi frequency versus pulse amplitude (markers) and fit to a linear dependence (line). Same data as in Fig.~2.
    (b) Rabi decay time versus pulse amplitude (markers). Grey markers denote the points for which the data deviates from the linear dependence in (a).
    (c) Rabi decay time versus Rabi frequency (markers) an result of a fit of the black markers to Eq.~\ref{eq:TrfR} (continuous line). The dotted lines show the individual contributions of the two summands in Eq.~\ref{eq:TrfR}. 
    (d) Rabi quality factor $Q=T_{\rm R} f_{\rm R}$ versus  Rabi frequency (markers) and result of the same fit as in (c) (line). 
    }
    \label{fig:TRfR}
\end{figure}

A method for estimating upper bounds on the amplitude of fluctuations originating from different noise sources is provided in Ref.~\cite{Malinowski2017}, where the authors study the relation between the Rabi frequency, $f_{\rm R} = \Omega_{\rm R}/2\pi$, and the Rabi decay time, $T_{\rm R}$. These quantities, 
respectively shown in Figs.~\ref{fig:TRfR}(a) and (b), can be extracted from a fit to the Rabi signal with the expression $a \cos{\left(t \Omega_R\right)} \exp{\left(\rm{t}/T_{\rm R}\right)} + c$, where $t$ denotes the full-width half maximum of the applied Gaussian pulse, see Fig.~2 in the main text. We fit the extracted decay times to the model of Ref.~\cite{Malinowski2017} 
\begin{equation}\label{eq:TrfR}
\left( \frac{1}{T_{\rm R}} \right)^2 = \frac{\sigma_f^4}{4f_{\rm R}^2} + C^2 f_{\rm R}^2,
\end{equation} 
where $\sigma_f$ is the standard deviation of the fluctuations of the qubit frequency $f_{\rm s}$ due to noise in the control and model parameters and $C$ is a measure of noise of the drive field. The data is fitted up to the region where the Rabi frequency stops being linear as a function of the pulse amplitude $A$, indicated with grey markers in Fig.~\ref{fig:TRfR}, and extract  $\sigma_f=$~\SI{39.7}{MHz} and $C=0.25$.

If we assume that the dominating contribution to $\sigma_f$ originates from noise in just one control parameter, we can obtain upper bounds on the noise amplitude for various types of noise. Since the coherence time is mostly independent on the external flux [Fig.~\ref{fig:coherence_flux}], we focus only on two possible origins of decoherence: voltage noise and nuclear magnetic noise. We first determine the susceptibility of the qubit frequency with respect to the external parameters (\Vl, \Vc, \Vr, $B_\parallel$ and $B_\perp$) at the ASQ operational setpoint, calculated as the partial derivatives of the qubit frequency with respect to each parameter. From two-tone spectroscopy measurements, we find the susceptibilities with respect to the left, central and right quantum dot gates of $S_{\rm L} \approx$~\SI{0.16}{GHz/mV}, $S_{\rm C} \approx$~\SI{0.07}{GHz/mV} and $S_{\rm R} \approx$~\SI{0.08}{GHz/mV}, respectively, and the susceptibilities to the parallel and perpendicular magnetic fields of $S_{\parallel} \approx$~\SI{0.18}{GHz/mT} and  $S_\perp \approx$~\SI{0.05}{GHz/mT}, respectively.

We start by evaluating the contribution of voltage noise on the DC lines. Considering noise from the gate with highest susceptibility we obtain an upper bound of $\sigma_{\rm L} < \sigma_f / S_{\rm L}=$~\SI{0.25}{mV} for the standard deviation of the gate voltage fluctuations. While this agrees with the gate noise observed in Ref.~\cite{Hays2021}, where the estimated standard deviation of the voltage gate fluctuations was $\sigma_V=$~\SI{0.24}{mV}, we do not expect fluctuations of this magnitude to be present in our system. Previous experiments measured in the same experimental setup [Fig.~\ref{fig:cryogenic_setup}] observed gate stability below \SI{60}{\micro eV} for similar device geometries \cite{Bargerbos2020}. Furthermore, the DC lines used to control the gate electrodes are strongly filtered with a sequence of \SI{9}{kHz} RC filters, \SI{80}{MHz} to \SI{5}{GHz} $\pi$ filters and, finally, custom made copper powder filters, all mounted at the mixing chamber stage. and an additional set of \SI{80}{MHz} $\pi$ filter on the printed circuit board. The left and right gates additionally have first order LC filters on-chip, with an expected cutoff frequency of \SI{200}{MHz}. We therefore suspect that the dominant contribution to $\sigma_f$ does not arise from gate voltage fluctuations on the DC lines. However, charge fluctuations on the device, unrelated to the gate control, could still limit the coherence time.

Alternatively, the gate voltage noise could originate from the RF drive line connected to the central gate electrode. This would result in an upper bound to gate voltage noise of $\sigma_{\rm C} < \sigma_f / S_{\rm C} =$~\SI{0.57}{mV}, which corresponds to an effective power of \SI{-53}{dBm} at the sample. Given the \SI{-55}{dB} attenuation of the drive line [Fig.~\ref{fig:cryogenic_setup}], this would correspond to a noise power of \SI{2}{dBm} at the fridge input, which we consider implausible. Furthermore, the RF line is connected via both a DC block and a bias tee, providing strong high-pass filtering.

Next, we consider the contribution of nuclear magnetic noise. We estimate upper bounds to the longitudinal and transverse magnetic fluctuations of $\sigma_\parallel < \sigma_f / S_{\rm \parallel} =$~\SI{0.22}{mT} and $\sigma_\perp < \sigma_f / S_{\rm \perp} =$~\SI{0.80}{mT}, respectively. These estimates are comparable to the values obtained for InAs and InSb spin-orbit qubits in previous works: $\sigma_{B} =$~\SI{0.66}{mT} \cite{NadjPerge2010} and $\sigma_{B} =$~\SI{0.16}{mT} \cite{vandenBerg2013}, respectively. Nuclear magnetic noise is therefore a plausible dominating contribution to the dephasing observed in the ASQ. However, we emphasize that these calculations are only an estimate and that further investigation is needed to discern between the different possible causes of dephasing.

\subsection{Virtual-photon-mediated ASQ–resonator coupling}
\label{Ss:resonator}
\begin{figure}[h!]
    \center
    \includegraphics[scale=1]{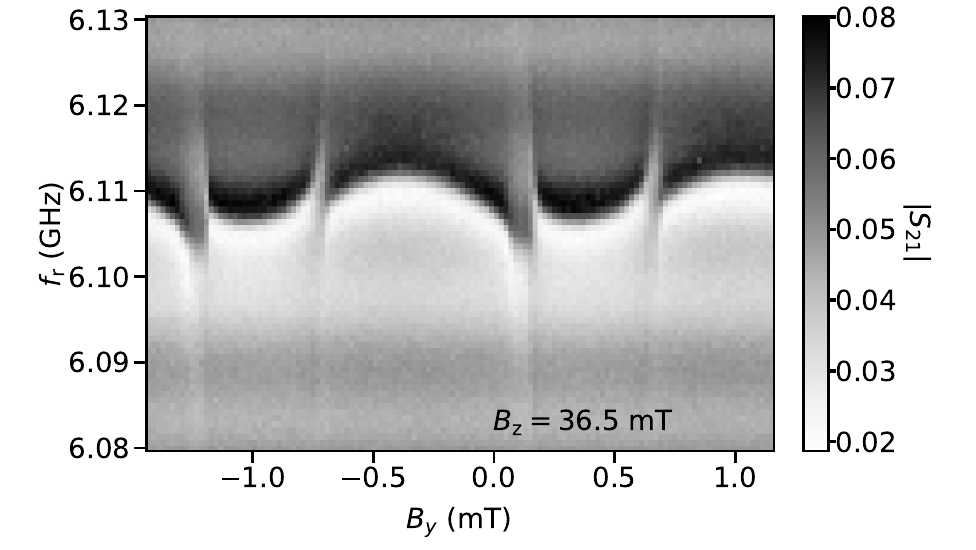}
    \caption{Single-tone spectroscopy of the readout resonator versus the magnetic field in the chip plane and perpendicular to the nanowire, $B_y$, for $B_z$ = \SI{36.5}{mT}.}
    \label{fig:qubit-res}
\end{figure}

In this section we provide additional data showing coherent coupling between the readout resonator and the Andreev spin qubit. As shown in Fig.~\ref{fig:qubit-res}, we observe avoided crossings between the ASQ and resonator transitions when they are on resonance, at $B_z=$~\SI{36.5}{mT}. This coherent coupling is of note, as the ASQ and readout resonator are not directly coupled. However, both are directly and strongly coupled to the transmon qubit, detuned by \SI{900}{MHz} in this case, which mediates a strong virtual coupling. This effect is analogous to the work of Ref.~\cite{Landig2019} where, instead, a resonator mediated virtual coupling between a transmon and a spin qubit.

\bibliography{ms.bib}